\documentclass[epj]{svjour}

\usepackage{amsfonts}
\usepackage{amsmath}
\usepackage{amssymb}
\usepackage{graphicx}
\usepackage{xspace}

\newcommand{\ie}{{i.e.}\xspace} % 
\newcommand{\eg}{{e.g.}\xspace} % 
\newcommand{\etc}{{etc.}\ }     % 
\newcommand{\dd}{\mathrm{d}}
\newcommand{\Psuccess}{P_\mathrm{success}}
\newcommand{\psuccess}{p_\mathrm{no-contr}}
\newcommand{\Psat}{P_\mathrm{sat}}
\newcommand{\imag}{\hat \imath}
\newcommand{\alphaH}{\alpha_H}
\newcommand{\alphaR}{\alpha_\mathrm{R}}
\newcommand{\alphainit}{\alpha}
\newcommand{\alphaflat}{\alpha_\mathrm{flat}}

\newcommand{\expproba}{\lambda}

\newcommand{\expcorraprobcond}{\zeta}
\newcommand{\expdeviationcjs}{\delta}
\newcommand{\expprobadeuxsat}{\gamma}
\newcommand{\Ai}{\mathrm{Ai}}
\newcommand{\adeux}{d_2}
\newcommand{\tst}{t^*}
\newcommand{\expDeltatemps}{a}
\newcommand{\pib}{\overline{\pi}}
\newcommand{\cbarre}{\overline{c}}
\newcommand{\epsalpha}{\epsilon_\alpha}
\newcommand{\epsp}{\epsilon_p}
\newcommand{\tvt}{t_2}
\newcommand{\tvtinit}{t_{2\mathrm{init}}}
\newcommand{\tcs}{t_3}
\newcommand{\rhot}{\tilde{\rho}}
\newcommand{\rhob}{\overline{\rho}}
\newcommand{\erf}{\mathrm{erf}}
\newcommand{\recdifflnai}{\mathcal{A}}
\def\tendsto{\,\mathop{\relbar\joinrel\relbar\joinrel\relbar\joinrel\relbar\joinrel\rightarrow}}
\newcommand{\gG}{\mathcal{G}}
\newcommand{\beq}{\begin{equation}}
\newcommand{\eeq}{\end{equation}}
\newcommand{\bea}{\begin{eqnarray}}
\newcommand{\eea}{\end{eqnarray}}
\newcommand{\largeurgraphes}{0.47\textwidth}

\begin{document}

\title{Criticality and Universality in the Unit-Propagation Search Rule.%
  \thanks{Preprint LPTENS-05/24.}}

\vskip .5cm
\author{C. Deroulers and R. Monasson}
\institute{CNRS-Laboratoire de Physique Th{\'e}orique de l'ENS,
24 rue Lhomond, 75231 Paris CEDEX 05, France.}
\date{\today}

\abstract{
The probability $\Psuccess(\alpha, N)$ that stochastic greedy
algorithms successfully solve the random SATisfiability problem is
studied as a function of the ratio $\alpha$ of constraints per variable
and the number $N$ of variables. These algorithms assign variables
according to the unit-propagation (UP) rule in presence of constraints
involving a unique variable (1-clauses), to some heuristic (H)
prescription otherwise. In the infinite $N$ limit, $\Psuccess$ vanishes
at some critical ratio $\alpha_H$ which depends on the heuristic $H$. We
show that the critical behaviour is determined by the UP rule only. In
the case where only constraints with 2 and 3 variables are present, we
give the phase diagram and identify two universality classes: the power
law class, where $\Psuccess[\alpha_H (1+\epsilon N^{-1/3}), N] \sim
A(\epsilon) / N^\gamma$; the stretched exponential class, where
$\Psuccess[\alpha_H (1+\epsilon N^{-1/3}), N] \sim \exp[-N^{1/6} \,
\Phi(\epsilon)]$. Which class is selected depends on the characteristic
parameters of input data. The critical exponent $\gamma$ is universal
and calculated; the scaling functions $A$ and $\Phi$ weakly depend on
the heuristic H and are obtained from the solutions of
reaction-diffusion equations for 1-clauses. Computation of some
non-universal corrections allows us to match numerical results with good
precision. The critical behaviour for constraints with $>3$ variables is
given. Our results are interpreted in terms of dynamical graph
percolation and we argue that they should apply to more general
situations where UP is used.
}

\maketitle  

\section{Introduction}

Many computational problems rooted in practical applications or issued
from theoretical considerations are considered to be very difficult in
that all exact algorithms designed so far have (worst-case) running
times growing exponentially with the size $N$ of the input. To tackle
such problems one is thus enticed to look for randomized (stochastic)
polynomial-time algorithms, guaranteed to run fast but not to find a 
solution. The key estimator of the performances of
these search procedures is the probability of success, $\Psuccess
(N)$, that is, the probability over the stochastic choices carried
out by the algorithm that a solution is found (when it exists) 
in polynomial time, say, 
less than  $N$. Roughly speaking, depending on the problem to be
solved and the nature of the input, two cases may arise in the
large $N$ limit~\footnote{Recall that $f(x)=\Theta (x)$ means that there
exist three strictly positive real numbers $X,a,b$ such that, 
$\forall x >X$, $a x < f(x)<b x$.} 
\begin{equation} \label{proba1}
\Psuccess (N) = \left\{ \begin{array}{c c}
\Theta (1) & \hbox{\rm (success case)} \\
\exp( - \Theta (N)) & \hbox{\rm (failure case).}
\end{array} \right.
\end{equation} 
When $\Psuccess$ is bounded from below by a strictly positive number 
at large $N$, a few runs will be
sufficient to provide a solution or a (probabilistic) proof of absence
of solution; this case is referred to as {\em success case} in the 
following. Unfortunately, in many cases, $\Psuccess$ is
exponentially small in the system's size and an exponentially large
number of runs is necessary to reach a conclusion about the existence
or not of a solution; this situation is hereafter 
denoted by {\em failure case}. 

An example is provided  by the K-SAT problem, informally defined as follows.
An instance of K-SAT is a set of constraints (called clauses) over
a set of Boolean variables. Each constraint
is the logical OR of $K$ variables or their negations. 
Solving an instance means either finding an assignment of the
variables that satisfies all clauses, or
proving that no such assignment exists. 
While 2-SAT can be solved in a time linear in the instance 
size~\cite{aspvall-plass-tarjan-temps-lineaire}, K-SAT with $K\ge 3$ is
NP-complete~\cite{papadimitriou-steiglitz-livre}.
The worst-case running time may be quite different
from the running time once averaged over some distribution of inputs.
A simple and theoretically-motivated distribution consists in
choosing, independently for each clause, a $K$-uplet 
of distinct variables, and negate each of them with probability one
half. The K-SAT problem, together with the flat distribution of instances
is called random K-SAT. 

In the past twenty years, various randomized algorithms were designed to
solve the random 3-SAT problem in linear time with a strictly positive
probability $\Psuccess$~\cite{franco-pure-literal} when the numbers $M$
of clauses and $N$ of variables tend to infinity at a fixed
ratio~\footnote{Other scalings for $M,N$ were investigated too and
appear to be easier to handle, see
reference~\cite{franco-swaminathan-temps-polynomial-partout}.}
$\alpha=M/N$. These algorithms are based on the recognition of the
special role played by clauses involving a unique variable, called
unit-clauses (unit-clauses are initially absent from the instance but
may result from the operation of the algorithm), and obey the following
\vskip .3cm
\noindent \textsc{Unit-Propagation (UP) rule: 
When the instance includes (at least) a unit-clause, sa\-tis\-fy 
this\linebreak clau\-se by as\-sign\-ing its variable appropriately, 
prior to any other action.}
\vskip .3cm

\noindent
UP merely asserts that drawing obvious logical consequences from the
constraints is better done first. Its \linebreak strength lies in its
recursive character: assignment of a variable from a 1-clause may create
further 1-clauses, and lead to further assignments. Hence, the
propagation of logical constraints is very efficient to reduce the size
of the instance to be treated. In the absence of unit-clause, some
choice has to be made for the variable to assign and the truth value to
give. This choice is usually made according to some heuristic, the
simplest one being

\vskip .3cm
\noindent {\sc  Random (R) heuristic: 
When the instance in-\linebreak{}cludes no unit-clause, pick any unset  
variable
uniformly at random, and set it to True or False with probabilities
one half.
}
\vskip .3cm

\noindent
The specification of the heuristic e.g. R, together with UP, entirely
defines the randomized search algorithm~\footnote{This procedure is
called the Unit-Clause algorithm in the computer science
literature~\cite{chao-franco-uc-guc,achlioptas-bornes-inferieures-via-equadiffs}.}.
The output of the procedure is either `Solution' if a solution is found,
or `Don't know' if a contradiction (two opposite unit-clauses) is
encountered. The probability that the procedure finds a solution,
averaged over the choices of instances with ratio $\alpha$, was studied
by Chao and Franco~\cite{chao-franco-sc1,chao-franco-uc-guc}, Frieze
and Suen~\cite{frieze-suen-guc-sc}, with the result
\begin{equation} \label{proba2}
\lim _{N\to\infty} \Psuccess (\alpha, N) = \left\{ \begin{array}{c c c}
\Theta (1) & \hbox{\rm if} &\alpha <\alphaR = \frac 83 \\
0 &\hbox{\rm if} & \alpha > \alphaR
\end{array} \right.
\end{equation} 
for Random 3-SAT, R heuristic \& UP. The above study was then extended 
to more sophisticated and powerful heuristic rules H, defined in
Section~\ref{section_definition_algorithmes}. It appears that identity
(\ref{proba2}) holds for all the randomized algorithms based on UP and a
heuristic rule H, with a critical ratio value $\alpha_H$ that depends on
H. Cocco and Monasson then showed that, for ratios above $\alpha _H$,
the probability of success is indeed exponentially small in $N$, in
agreement with the generally expected behaviour (\ref{proba1}).

\begin{figure}
\begin{center}
\includegraphics[angle=-90,width=\largeurgraphes]{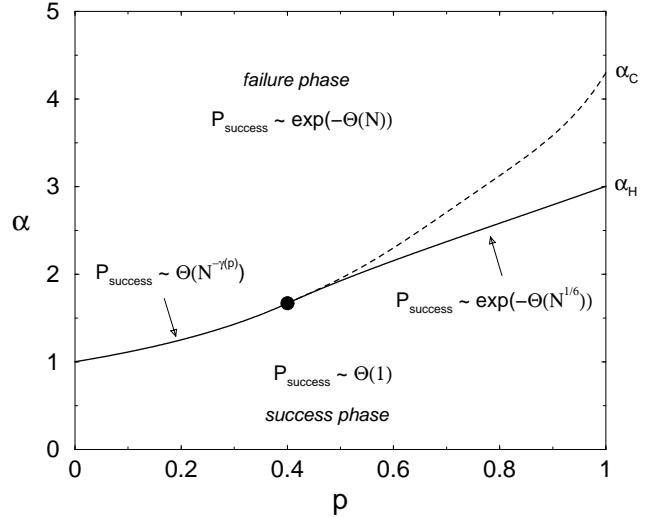}
\end{center}
\caption{Scaling of the probability of success of randomized search
algorithm based on a heuristic H and the UP rule for the 2+$p$-SAT model
with instances of size $N$ and having $\alpha$ clauses per variable. The
dynamic (or kinetic) critical line is represented for the R heuristic,
$\alphaR(p) = 1/(1-p)$ for $p<\frac 25$, $24 p/(2+p)^2$ for $p>\frac 25$
separates the failure phase (upper region) from the success phase (lower
region). Along the critical line, $\Psuccess$ decays as a power law for
$p<\frac 25$ with a $p$-dependent exponent $\expprobadeuxsat (p)=
(1-p)/(2-5p)/6$, and as a stretched exponential with exponent $\frac 16$
when $p>\frac 25$. The static critical line (dashed curve) coincide with
the dynamic one for $p<\frac 25$ and lies above for larger values of
$p$, see Section~\ref{section_definition_probleme_ksat}.}
\label{schema}
\end{figure}   

The transition from success to failure in random 3-SAT was quantitatively 
studied in a recent 
letter~\cite{deroulers-monasson-universalite-up-lettre}. 
We found that the width of the transition
scales as $N^{-\frac 13}$, and that the probability of success in the 
critical regime behaves as a stretched exponential of $N$ with 
exponent $\frac 16$. More precisely, 
\begin{equation} \label{proba3}
\Psuccess \left[\alpha _H (1+\epsilon N^{-\frac 13}) , N\right] = 
\exp \left( - N^\frac 16 \; \Phi(\epsilon)\; [1 + o(1) ]\right)
\end{equation} 
for Random 3-SAT, and UP. The calculation of the scaling function $\Phi$
relies on an accurate characterization of the critical distribution of
the number of unit clauses, for an exact expression see
eq.~(\ref{definition_Phi}). The important statement here is that the
above expression holds {\em independently} of the heuristic H, provided
the randomized procedure obeys UP~\footnote{Strictly speaking, $\Phi$
depends on $H$ through two global magnification ratios along the X and Y
axis: $\Phi_H (\epsilon)= y_H \Phi ( x_H \, \epsilon)$, where $x_H,y_H$
are heuristic-dependent real numbers and $\Phi$ is universal.}. This
result allowed us to evoke the existence of a universality class related
to UP. 

In the present paper, we provide all the calculations which led
us to equation (\ref{proba3}).
We also calculate subdominant corrections to the $N^{\frac 16}$
scaling of $\ln \Psuccess$ in (\ref{proba3}), which allow us to
account for numerical experiments in a more accurate manner than 
in reference~\cite{deroulers-monasson-universalite-up-lettre} (see 
Section~\ref{section_classe_3sat}). In addition, we study the
robustness of the stretched exponential behaviour with respect
to variations in the problem to be solved. We argue that the class
of problems to be considered for this purpose is random 
2+$p$-SAT, where instances are mixed 2- and 3-SAT instances with relative
fractions $p$ and $1-p$  (for fixed $0 \le p \le 
1$)~\cite{monasson-zecchina-kirkpatrick-selman-troyansky-2plusp-sat}. 
Our results are sketched on Figure \ref{schema}. It is found that
the stretched exponential behaviour holds for a whole set of critical
random 2+$p$-SAT problems, but not all of them. More precisely,
equation (\ref{proba3}) remains true for $p \ge 2/5$. For $p<2/5$, the
probability of success at criticality decreases as a power law only, 
\begin{equation} \label{proba4}
\Psuccess \left[\alphaR (p) \,  (1+\epsilon N^{-\frac 13}) , N\right] = 
\frac{A_H(\epsilon, p)}{N^{\expprobadeuxsat(p)}} [1+o(1)]
\end{equation}
for Random 2+$p$-SAT, $p<2/5$, and UP.
The value of the universal decay exponent is $\expprobadeuxsat(p) =
(1-p)/(2-5p)/6$. The calculation of the prefactor $A_H(p,\epsilon)$
shows similarities with the one of $\Phi(\epsilon )$ above, but is more
difficult and shown in Section~\ref{section_famille_2sat}.

\section{Definitions and brief overview of known results}
\subsection{Random $K$-SAT and 2+$p$-SAT problems}
\label{section_definition_probleme_ksat}

 In the random $K$-SAT problem
\cite{kirkpatrick-selman-science,selman-kirkpatrick-ai}, where $K$ is an
integer no less than 2, one wants to find a solution to a set of
$M=\alphainit N$ randomly drawn constraints (called \emph{clauses}) over
a set of $N$ Boolean variables $x_i$ ($i=1\ldots N$): $x_i$ can take
only two values, namely True ($T$) and False ($F$). Each constraint reads
$z_{i_1} \vee z_{i_2} \vee \ldots \vee z_{i_K}$, where $\vee$ denotes
the logical OR; $z_\ell$ is called a \emph{literal}: it is either a
variable $x_{i_\ell}$ or its negation $\bar x_{i_\ell}$ with equal
probabilities ($=\frac 12$), and $(i_1, i_2, \ldots , i_K)$ is a
$K$-uplet of distinct integers unbiasedly drawn from the set of the
$\binom{N}{K}$ $K$-uplets. Such a clause with $K$ literals is called a
$K$-clause, or clause of length $K$. Such a set of $M$ clauses involving
$N$ variables is named an \emph{instance} or \emph{formula} of the 
$K$-SAT problem. An instance is either
\emph{satisfiable} (there exists at least one \emph{satisfying
assignment}) or \emph{unsatisfiable} (contrary case). We will be mainly
interested in the large $N$, large $M$ limit with fixed $\alpha=M/N$ and
$K$. Notice that the results presented in this paper
are, or have been obtained for this `flat' distribution only, and
do not hold for real-world, industrial instances of the SAT problem.

A distribution of constraints will appear naturally in the course of our 
study: the random 2+$p$-SAT
problem~\cite{monasson-zecchina-kirkpatrick-selman-troyansky-2plusp-sat}. 
For fixed $0 \le p \le 1$, each one of the $M = \alpha N$ clauses is of
length either 2 or 3, with respective probabilities $1-p$ and $p$. 
Parameter $p$ allows one to interpolate between 2-SAT ($p=0$) and 3-SAT
($p=1$). 

Experiments and theory show that the probability $\Psat$ that a randomly
drawn instance with parameters $p,\alpha$ be satisfiable is, in the
large $N$ limit, equal to 0 (respectively, 1) if the ratio $\alpha$ is
smaller (resp., larger) than some critical value $\alpha _C(p)$. 
Nowadays, the value of $\alpha_C(p)$ is rigorously known for $p \le
\frac25$ only, with the result
$\alpha_C(p)=1/(1-p)$~\cite{achlioptas-kirousis-kranakis-krizanc-2plusp-sat-rigoureux},
see Figure \ref{schema}. For 3-SAT the best current upper and lower
bounds to the threshold are 4.506 \cite{dubois-bornesup} and 3.52
\cite{kaporis-kirousis-lalas-hl-cl} respectively. For finite but large
$N$, the steep decrease of $\Psat$ with $\alpha$ (at fixed $p$) takes
place over a small change $\delta \alpha = \Theta (N^{-\frac 1\nu})$ in
the ratio of variables per clause, called width of the transition.
Wilson has shown that the width exponent $\nu$ is bounded from below by
$2$ (for all values of
$p$)~\cite{wilson-exposant-largeur-transition-statique-ksat}. For 2-SAT,
a detailed study by Bollob\`as {\em et al.} establishes that
$\nu=3$~\cite{bollobas-borgs-chayes-2sat-scaling-window}, and that
$\Psat$ is finite at the threshold $\alpha=1$. A numerical estimate of
this critical $\Psat$ may be found in reference
~\cite{shen-zhang-max-2-sat-etude-empirique} and we provide a more
precise one in Appendix~\ref{appendice_proba_2sat_au_seuil}: $\Psat(p=0,
\alpha=1) = 0.907 \pm 10^{-3}$.

\subsection{Greedy randomized search algorithms}
\label{section_definition_algorithmes}

In this paper, we are not interested in the probability of satisfaction
$\Psat$ (which is a property characteristic of the random SAT problem
only) but in the probability $\Psuccess$ ($\le \Psat$) 
that certain algorithms are
capable of finding a solution to randomly drawn instances. These
algorithms are defined as follows.

Initially
\cite{davis-putnam,kirkpatrick-selman-science,selman-kirkpatrick-ai},
all variables are unset and all clau\-ses have their initial length ($K$
in the $K$-SAT case, 2 or 3 in the $2+p$-SAT case). Then the algorithms
iteratively set variables to the value $T$ (true) or $F$ (false)
according to two well-defined rules (mentioned in the introduction and
detailed below), and update (reduce) the constraints accordingly. For
instance, the 3-clause ($\bar x_1 \vee \bar x_2 \vee x_3$) is turned
into the 2-clause ($\bar x_1 \vee x_3$) if $x_2$ is set to $T$, and is
satisfied, hence removed from the list of clauses, if $x_2$ is set to
$F$. A 1-clause (or unit-clause) like $\bar x_1$ may become a 0-clause
if its variable happens to be inappropriately assigned (here, to $T$);
this is called a \emph{contradiction}. In this case, the algorithms halt
and output `don't know', since it can not be decided whether the
contradiction results from a inadequate assignment of variables (while
the original instance is satisfiable) or from the fact that the instance
is not satisfiable. If no contradiction is ever produced, the process
ends up when all clauses have been satisfied and removed, and a solution
(or a set of solutions if some variable are not assigned yet) is found.
The output of the algorithms is then `Satisfiable'.

We now make explicit the aforementioned rules for variable assignment.
The first rule, UP (for Unit-\linebreak{}Propagation)
\cite{frieze-suen-guc-sc}, is common to all algorithms: if a clause with
a unique variable (a 1-clause), \eg $\bar x_1$, is produced at some
stage of the procedure, then this variable is assigned to satisfy the
clause, \eg $x_1 := F$. UP is a corner stone of practical SAT solvers.
Ignoring a 1-clause means taking the risk that it becomes a 0-clause
later on (and makes the whole search process fail), while making the
search uselessly longer in the case of an unsatisfiable
instance~\footnote{Another fundamental rule in SAT solvers that we do
not consider explicitly here, although it is contained in the CL
heuristic~\cite{kaporis-kirousis-lalas-hl-cl} to which our results
apply, is the Pure Literal rule \cite{franco-pure-literal}, where one
assigns only variables (called \emph{pure literals}) that appear always
the same way in clauses, \ie always negated or always not negated).
Removal of pure literals and of their attached clauses make the instance
shorter without affecting its logical status (satisfiable or not).}.

Therefore, as long as unit-clauses are present, the algorithms try to
satisfy them by proper variable assignment. New
1-clauses may be in turn produced by simplification of 2-clauses, 
and 0-clauses (contradictions) when several 1-clauses require the same 
variable to take opposite logical values. 

The second rule is a specific and arbitrary prescription 
for variable assignment taking over UP
when it cannot be used \ie in the absence of 1-clause. It is termed 
\emph{heuristic} rule because it does not rely on a logical
principle as the UP rule. In the simplest heuristic, referred to as random
(R) here, the prescription is to set
any unassigned variable to $T$ or $F$ with probability $\frac 12$
independently of the remaining clauses
\cite{chao-franco-uc-guc,frieze-suen-guc-sc}. 
More sophisticated heuristics are able to choose a variable that will 
satisfy (thus eliminate) the
largest number of clauses while minimizing the odds that a
contradiction is produced later. Some examples are:

\begin{enumerate}

\item GUC \cite{chao-franco-uc-guc} (for Generalized Unit Clause)
prescribes to take a variable in the shortest clause available, and to
assign this variable so as to satisfy this clause. In particular, when
there are no 1-clauses, 2-clauses are removed first, which decreases the
future production of 1-clauses and thus of contradictions. 

\item HL \cite{kaporis-kirousis-lalas-hl-cl} (for Heaviest Literal)
prescribes to take (in absence of 1-clauses, as always) the literal that
appears most in the reduced instance (at the time of choice), and to set
it to $T$ (by assigning accordingly its variable), disregarding the
number of occurrences of its negation or the length of the clauses it
appears in. 

\item CL \cite{kaporis-kirousis-lalas-hl-cl} (for Complementary
Literals) prescribes to take a variable according to the number of
occurrences of it and of its negation in a rather complex way, such that
the number of 2-clauses decreases maximally without making the number of
3-clauses too much decrease.

\item KCNFS \cite{dequen-dubois-kcnfs} is an even more complex
heuristic, specially designed to reduce the number of backtrackings
needed to prove that a given instance is unsatisfiable, on top of
standard tricks to improve the search. 

\end{enumerate}

\subsection{The success-to-failure transition}
\label{suctofaisec}

Chao and Franco have studied the probability $\Psuccess$ that the
randomized search process based on UP and the R heuristic, called UC (for
unit-clause) algorithm, successfully finds a solution to instances of
random 3-SAT with characteristic parameters
$\alpha,N$~\cite{chao-franco-uc-guc}, with the result (\ref{proba2}). This
study was extended by Achlioptas et al. to the case of random
2+$p$-SAT~\cite{achlioptas-kirousis-kranakis-krizanc-2plusp-sat-rigoureux},
with the following outcome:
\begin{equation} \label{proba2bis}
 \lim_{N\to\infty} \Psuccess (\alpha, p, N) \ \ 
 \begin{array}{lcl}
  >0 & & \text{  if } \alpha <\alphaR (p) \\
  =0 & & \text{  if } \alpha > \alphaR (p)
 \end{array}
\end{equation} 
for the R heuristic and UP, where
\begin{equation}
\alphaR(p) = \frac 1{1-p} \quad {\hbox{\rm if}}\quad p \le \frac 25
\ , \quad \frac {24\, p}{(2+p)^2} 
\quad {\hbox{\rm if}}\quad p \ge \frac 25 \ .
\end{equation}
Hence, as simple as is UC, this procedure is capable to reach 
the critical threshold $\alpha _C(p)$ separating satisfiable from 
unsatisfiable instances when $p \le \frac 25$. For $p>\frac 25$, a 
finite gap separates $\alphaR(p)$ from $\alpha _C(p)$ in which 
instances are (in the large $N$ limit) almost surely satisfiable 
but UC has a vanishingly small probability of success.
 
Similar results were obtained for the heuristics H listed above,
especially for $p=1$ (3-SAT). The smallest ratios, called
thresholds and denoted by $\alphaH$, at which $\Psuccess$ 
vanishes in the infinite $N$ limit
are: $\alpha _\mathrm{GUC} \simeq 3.003$~\cite{chao-franco-uc-guc}, 
$\alpha _\mathrm{HL} \simeq 3.42$~\cite{kaporis-kirousis-lalas-hl-cl}, 
$\alpha _\mathrm{CL} \simeq 3.52$~\cite{kaporis-kirousis-lalas-hl-cl}.
The 3-SAT threshold for the KCNFS heuristic is not known.

In this paper, we are interested in the critical scaling of $\Psuccess$ 
with $N$, that is, when $\alpha$ is chosen to be equal, or
very close to its critical and heuristic-dependent value $\alphaH$
(at fixed $p$). More precisely, we show that $\Psuccess$ may vanish
either as a stretched exponential (\ref{proba3}) or as an inverse
power law (\ref{proba4}). Strikingly, although the randomized search 
algorithms based on different heuristics exhibit quite different
performances e.g. values of $\alphaH$, we claim that the scaling of 
$\Psuccess$ at criticality is essentially unique. The mechanism that
monitors the transition from success to failure at $\alphaH$ of the 
corresponding algorithms is indeed UP.  For instance, for KCNFS, a
numerical study shows that the special, complex heuristic is never used
when $\alphainit$ is close to the $\alpha_\mathrm{KCNFS}$ threshold of
this algorithm.  

Hereafter, we show that universality holds for random K-SAT with $K\ge
3$ and for 2+$p$-SAT with $p\ge \frac 25$ one the one hand, and $p <
\frac{2}{5}$ on the other hand. In the $p<\frac 25$ case, for which
$\alpha _H$ and $\alpha _C$ coincide
(Sect.~\ref{section_famille_2sat}), there is strictly speaking an
infinite family of universality classes, depending on the parameter $p$
(in particular, the critical exponent $\gamma _H(p)$ --- see
(\ref{proba4}) --- varies continuously with $p$). These analytical
predictions are confirmed by numerical investigations.

\section{Generating function framework for the kinetics of search}
\label{section_fonctions_generatrices}

 This section is devoted to the analysis of the greedy UC= UP+R
algorithm, defined in the previous section, on instances of the random
$2+p$-SAT or $K$-SAT problems. We introduce a generating function
formalism to take into account the variety of instances which can be
produced in the course of the search process.
 We shall use $b(n;m,q)=\binom{m}{n} q^n(1-q)^{m-n}$ to denote the 
Binomial distribution, and $\delta_{n,m}$ to represent the Kronecker 
function over integers $n$: $\delta_{n,m}=\{1$ if $n=m$, $0$ 
otherwise\}.

\subsection{The evolution of the search process}
\label{section_cinetique_fonctions_generatrices}

 For the random $2+p$-SAT and $K$-SAT distributions of boolean formulas
(instances), it was shown by Chao and Franco
\cite{chao-franco-sc1,chao-franco-uc-guc} that, during the first descent
in the search tree \ie prior to any backtracking, the distribution of
residual formulas (instances reduced because of the partial assignment
of variables) keeps its uniformity conditioned to the numbers of
$\ell$-clauses ($0 \le \ell \le K$). This statement remains correct for
heuristics slightly more sophisticated than R \eg GUC, SC$_1$
\cite{chao-franco-uc-guc,achlioptas-bornes-inferieures-via-equadiffs,chao-franco-sc1},
and was recently extended to splitting heuristics based on literal
occurrences such as HL and CL \cite{kaporis-kirousis-lalas-hl-cl}.
Therefore, we have to keep track only of these numbers of $\ell$-clauses
instead of the full detailed residual formulas: our phase space has
dimension $K+1$ in the case of $K$-SAT (4 for $2+p$-SAT).
 Moreover, this makes random 2+$p$-SAT a natural problem. After 
partial reduction by the algorithm, a 3-SAT formula is
turned into a $2+p$-SAT formula, where $p$ depends on the steps
the algorithm has already performed.

 Call $P(\vec C;T)$ the probability that the search process leads, after
$T$ assignments, to a clause vector $\vec C=(C_0,$ $C_1,$ $C_2, \ldots, 
C_K)$. Then, we have
\begin{equation} \label{equation_evolution_tousCs}
 P(\vec C'; T+1) =\sum_{\vec C}  M  [\vec C' \gets \vec C; T] \ 
 P(\vec C; T)
\end{equation}
where the transition matrix $M$ is 
\bea
 M [\vec C' \gets \vec C; T] & = &
  \big( 1-\delta_{C_1,0} \big)  \, M _\mathrm{UP} [\vec C' \gets \vec C; T] +
   \nonumber \\ & &
  \delta_{C_1,0}\,  M_\mathrm{R} [\vec C' \gets \vec C; T] \ .
\eea
The transitions matrices corresponding to unit-pro\-pa\-ga\-tion (UP) and 
the random heuristic (R) are
\bea
 \label{definition_MX} & &
 M_X[\vec C' \gets \vec C; T] = \sum_{z_K=0}^{C_K} b[z_K; C_K,K\mu]
   \nonumber \\ & & \quad
 \times \sum_{r_{K-1}=0}^{z_K} b[r_{K-1};z_K,\frac{1}{2}] \ 
 \delta_{C'_K,C_K-z_K}
   \nonumber \\ & & \quad
 \times \sum_{z_{K-1}=0}^{C_{K-1}} b[z_{K-1}; C_{K-1}, (K-1)\mu]
   \nonumber \\ & & \quad
 \times \sum_{r_{K-2}=0}^{z_{K-1}} b[r_{K-2};z_{K-1},\frac{1}{2}]
  \  \delta_{C'_{K-1},C_{K-1}-z_{K-1}+r_{K-1}}
   \nonumber \\ & & \quad
 \times \ldots \times \sum_{z_2=0}^{C_2} b[z_2; C_2,2\mu]
   \\ & & \quad \nonumber
 \times \sum_{r_1=0}^{z_2} b[r_1;z_2,\frac 12]
  \  \delta_{C'_2,C_2-z_2+r_2} F_X[C'_1,C_1,r_1,C'_0,C_0,\mu]
\eea
where $\mu := \frac 1{N-T}$ and, for X=UP and R,
\bea
 F_\mathrm{UP} & := &
  \sum_{z_1=0}^{C_1-1} b[z_1; C_1-1,\mu] \ \delta_{C'_1 , C_1-1-z_1+r_1}
    \nonumber \\ & & \times
  \sum_{r_0=0}^{z_1} b\left[r_0; z_1, \frac{1}{2}\right]
  \delta_{C'_0,C_0+r_0} \ ,
    \nonumber \\
 F_\mathrm{R} & := & \delta_{C'_1, r_1} \delta_{C'_0,C_0} \ .
\eea

 The above expressions for the transition matrices can be understood as
follows. Let $A$ be the variable assignment after $T$ assignments, and
${\cal F}$ the residual formula. Call $\vec C$ the clause vector of
${\cal F}$. Assume first that $C_1\ge 1$. Pick up one 1-clause, say,
$\ell$. Call $z_j$ the number of $j$-clauses that contain $\bar \ell$ or
$\ell$ (for $j=1,2,3$). Due to uniformity, the $z_{j}$'s are binomial
variables with parameter $j/(N-T)$ among $C_j-\delta_{j,1}$ (the
1-clause that is satisfied through unit-propagation is removed).  Among
the $z_j$ clauses, $r_{j-1}$ contained $\bar \ell$ and are reduced to
$(j-1)$-clauses, while the remaining $z_j-r_{j-1}$ contained $\ell$ and
are satisfied and removed.  $r_{j-1}$ is a binomial variable with
parameter $1/2$ among $z_j$. 0-clauses are never destroyed and
accumulate. The new clause vector $\vec C'$ is expressed from $\vec C$
and the $z_{j+1}$'s, $r_j$'s using Kronecker functions; thus, $M_P[\vec
C',\vec C;T]$ expresses the probability that a residual formula at
step $T$ with clause vector $\vec C$ gives rise to a (non violated)
residual instance at step $T+1$ through unit-propagation. Assume now
$C_1=0$.  Then, a yet unset variable is chosen and set to T or F
uniformly at random. The calculation of the new vector $\vec C'$ is
identical to the unit-propagation case above, except that $z_1=r_0=0$
(absence of 1-clause).
 Hence, putting both $C_1\ge 1$ and $C_1=0$ contributions together,
$M[\vec C' \gets \vec C; T]$ expresses the probability to have an
instance after $T+1$ assignments and with clause vector $\vec
C'$ produced from an instance with $T$ assigned variables and
clause vector $\vec C$.

\subsection{Generating functions for the numbers of clauses}

It is convenient to introduce the generating function \linebreak $G 
(\,{\vec X}\,;T\,)$ of the probability $P(\,{\vec C}\,;T\,)$ where
\[ \vec X := (X_0, X_1, X_2, \ldots, X_K), \]
\[ G(\,{\vec X}\,;T\,) := \sum_{\vec C} \, X_0^{C_0} \,
X_1^{C_1} \ldots X_K^{\, C_K} \, P(\,{\vec C}\,,T\,). \]
Evolution equation (\ref{equation_evolution_tousCs}) for the $P$'s can
be rewritten in terms of a recurrence relation for the generating
function~$G$,
\bea
\label{equation_evolution_de_fonctgen_sous_UC}
 G(\,{\vec X}\,;T+1\,) & = & \frac 1{f_1}\;
  G \big(X_0, f_1, f_2, \ldots, f_K; T\,\big) +
  \\ & & \nonumber
 \bigg(1 - \frac 1{f_1} \bigg) \;
  G\big(X_0, 0, f_2, f_3, \ldots, f_K; T\, \big)
\eea
where $f_1, \ldots, f_K$ stand for the functions
\begin{equation} \label{deff}
f_j( \vec X; T) := X_j + \frac j{N-T}
\bigg( \frac{1+X_{j-1}}2 - X_j \bigg)
\end{equation}
($j=1,\ldots,K$). Notice that probability conservation is achieved in 
this equation: $G(1,1,\ldots,1;T)=1$ for all $T$.

Variants of the R heuristic will translate into additional contributions
to the recurrence relation 
(\ref{equation_evolution_de_fonctgen_sous_UC}). For instance, if
the algorithm stops as soon as there are no
reducible clauses left ($C_1=C_2=\ldots=C_K=0$) instead of assigning all
remaining variables at random (such a variation is closer to what is
used in a practical search algorithm), the transition
matrix is modified into
\bea & &
 M [\vec C' \gets \vec C; T] =
  \big(1-\delta_{C_1,0} \big)  \, M _\mathrm{UP} [\vec C' \gets \vec C; T] +
   \nonumber \\ & & \quad
  \delta_{C_1,0}
   \big(1 - \delta_{C_2,0}\delta_{C_3,0}\ldots\delta_{C_K,0} \big) \,
  M_\mathrm{R} [\vec C' \gets \vec C; T] \quad
\eea
and equation (\ref{equation_evolution_de_fonctgen_sous_UC}) becomes
\bea
\label{equation_evolution_de_fonctgen_sous_UC_arret_quand_plus_de_clauses}
 G(\,{\vec X}\,;T+1\,) & = & \frac 1{f_1}\;
  G \big(X_0, f_1, f_2, \ldots, f_K; T\,\big) +
   \nonumber \\ & &
  \bigg(1 - \frac 1{f_1} \bigg) \;
  G\big(X_0, 0, f_2, f_3, \ldots, f_K; T\, \big)
   \nonumber \\ & &
  - G\big(X_0, 0, 0, \ldots, 0 ; T) \ .
\eea
In this case, $G$ is not normalized any longer; \linebreak
$G(1,1,\ldots,1;T)$ is now the probability that search has not stopped 
after assignment of $T$ variables. One could also impose that the 
algorithm comes to a halt as soon as a contradiction is detected
\ie when $C_0$ gets larger than or equal to unity. 
This requirement is dealt with by setting
$X_0$ to 0 in the evolution equation for
$G$~(\ref{equation_evolution_de_fonctgen_sous_UC}), or
(\ref{equation_evolution_de_fonctgen_sous_UC_arret_quand_plus_de_clauses}).
All probabilities are now conditioned to the
absence of 0-clauses, and, again, $G$ is not normalized.

For the more complicated heuristic GUC (without stopping condition), the
recurrence relation reads
\begin{eqnarray}
\label{eqev_guc} & &
G(\,{\vec X}\,;T+1\,) = \frac 1{f_1}\;
  G \big(X_0, f_1, f_2, \ldots, f_K; T\,\big) \\ & & \quad
  + \bigg( \frac{1}{f_2} - \frac{1}{f_1} \bigg) \;
  G\big(X_0,  0, f_2, f_3, \ldots, f_K; T\, \big) \nonumber \\ & & \quad
  + \bigg( \frac{1}{f_3} - \frac{1}{f_2} \bigg) \;
  G\big(X_0, 0, 0, f_3, \ldots, f_K; T\, \big) \nonumber \\ & & \quad
  + \ldots + \bigg( \frac{1}{f_K} - \frac{1}{f_{K-1}} \bigg) \;
  G\big(X_0, 0, 0, \ldots, 0, f_K; T\, \big) \ . \nonumber
\end{eqnarray}

The above recurrence relations 
(\ref{equation_evolution_de_fonctgen_sous_UC}), 
(\ref{equation_evolution_de_fonctgen_sous_UC_arret_quand_plus_de_clauses}), 
(\ref{eqev_guc})... will be useful in subsection 3.4 to derive
the distribution of unit-clauses. As far as $j$-clauses are concerned
with $j\ge 2$, we shall see in subsection~\ref{secavecla} that, thanks to
self-a\-ve\-ra\-ge\-ness in the large $N$ limit, it is sufficient to know
their expectation values, $\langle C_j \rangle(T)$. The average number
of $j$-clauses is
the derivative, evaluated at the point $X_0 = X_1 = \ldots = X_K = 1$,
of the generating function $G$ with respect to $X_j$: $\langle C_j
\rangle(T) = \partial_{X_j} \ln G(1,1,\ldots,1;T)$~\footnote{The
logarithm plays no role when $G$ is normalized, \ie
$G(1,1,\ldots,1;T)=1$.}. Evaluating the derivative at another point is
used to take conditional average: for instance, the average of $C_1$
conditioned to the absence of 0-clauses is $\langle C_1 \rangle(T) =
\partial_{X_1} \ln G(0,1,\ldots,1;T)$ (here, $G(0,1,1,\ldots,1;T)$ may
be less than 1 as we have seen). Taking derivatives with respect to more
than one $X_j$ would give information about correlation functions and/or
higher order moments of the $C_j$'s.

 For evolution equation~(\ref{equation_evolution_de_fonctgen_sous_UC}),
the system of evolution equations for the $\langle C_j \rangle(T)$'s is
triangular:
\begin{eqnarray} & &
 \label{equation_evolution_C_j}
 \langle C_j \rangle(T+1) - \langle C_j \rangle(T) = \\ & & \quad
  -\frac{j}{N-T} \langle C_j \rangle (T)
  + \frac{1}{2} \frac{j+1}{N-T} \langle C_{j+1} \rangle(T)
 \quad \text{if} \ \, 2 \le j \le K \nonumber \\ & &
 \label{equation_evolution_C_1}
 \langle C_1 \rangle(T+1) - \langle C_1 \rangle(T) =
  -\frac{1}{N-T} \langle C_1 \rangle (T) \\ \nonumber & & \quad
  + \frac{1}{N-T} \langle C_2 \rangle(T)
  +(1-\frac{1}{N-T}) \Big( \langle \delta_{C_1,0}\rangle(T) - 1 \Big) \\ & &
 \label{equation_evolution_C_0}
  \langle C_0 \rangle(T+1) - \langle C_0 \rangle(T) = \\ & & \quad
  \frac{1}{2(N-T)} \Big(
  \langle C_1 \rangle (T) - 1 +
  \langle \delta_{C_1,0} \rangle(T) \Big) = \nonumber \\ & & \quad
  \frac{1}{2(N-T)} \langle \max(C_1-1,0) \rangle(T) \nonumber
\end{eqnarray}
(with $C_{K+1} := 0$) and it can be solved analytically, starting from
$\langle C_K \rangle(T)$ down to $\langle C_2 \rangle(T)$, with the
initial condition $C_j(0)= \alphainit N \delta_{j,K}$ ($\alphainit$ is
the initial clauses-per-variables ratio). However, the equations for
$\langle C_1 \rangle(T)$ and $\langle C_0 \rangle(T)$ involve more
information than the averages of the $C_j$'s, namely the probability
that there is no 1-clauses, and they can't be solved directly: we shall
study the full probability distribution of $C_1$ in the sequel, in order
to extract the finally useful information, that is the probability
$\langle \delta_{C_0,0} \rangle(T)$ that the search process doesn't fail
(doesn't detect any 0-clause or contradiction) up to step $T$ of the
algorithm.

 Before going on, let us point out that at least two strategies are at
hand to compute this finally useful quantity. We just explained one: we
set $X_0=X_1=1$ in $G$, compute the averages of the $C_j$'s ($j \ge 2$)
(these stochastic variables turn out to self-average in this case where
$C_0$ and $C_1$ are free --- see below), compute the full distribution
of $C_0$ and $C_1$ conditioned to the averages of the $C_j$'s ($j \ge
2$), and finally extract the probability that $C_0$ vanishes up to time
$T$. The other one starts with noticing that this last probability is
nothing other than $G(0,1,1,\ldots,1;T)$: thus, it seems more natural to
compute it through studying the generating function
$G(0,X_1,X_2,\ldots,X_K;T)$ with $X_0$ set to 0, \ie to condition all
probabilities and averages on the success of the search, or equivalently
to require the process to stop as soon as a contradiction appears. But
this would prevent us to solve the evolution equations for the $\langle
C_j \rangle(T)$'s and would finally lead to more complication: indeed,
in such a case, since $G$ is not normalized, the quantity
$G(0,1,1,\ldots,1;T)$, that expresses the probability that no
contradiction is found, and that can't be expressed without information
about $C_1$, appears in every equation --- or, put in another way, there
are correlations between all $C_j$'s and $C_1$. Therefore, we prefer to
take the seemingly less direct first route and study from now on only
the simplest kinetics~(\ref{equation_evolution_de_fonctgen_sous_UC}),
with UC heuristic and no stop condition.

 Another way of circumventing this problem could be to do a kind of
coarse-graining by grouping steps of the algorithm where 1-clauses are
present (and the Unit Propagation principle is used) into so-called
\emph{rounds}~\cite{achlioptas-sorkin-rounds,kaporis-kirousis-lalas-hl-cl},
and then do as if the rounds where the elementary steps: at the end of
each step, $C_1$ is always vanishing, so that one needs to keep track
only of the $C_j$'s, $j \ge 2$, in the coarse-grained process.

\subsection{On the self-averageness of clause numbers and resolution 
trajectories}
\label{secavecla}

 Is the knowledge of the sole averages $\langle C_j \rangle(T)$ enough,
at least for $2 \le j \le K$, to compute the success probability of the
search process? The answer is yes, in a large range of situations,
because the $C_j$'s are self-averaging (for $j \ge 2$).

It may be shown
rigorously~\cite{achlioptas-bornes-inferieures-via-equadiffs}, using
Wormald's theorem~\cite{wormald-theoreme}, that, with the kinetics
defined above and no constraints on the $C_j$'s (\ie with all $X_j$ set
to 1), $C_2$, $C_3$, ..., $C_K$ are self-averaging in the $T, N \to
+\infty$ with fixed $t :=T/N$ limit in such a way that we can
approximate them by continuous functions of the reduced parameter $t \in
[0,1]$:
\begin{equation}
 \label{forme_concentree_c_j}
 C_j(T) = N c_j(t) + o(N-T), \quad 2 \le j \le K
\end{equation}
where $o(N-T)$ is actually an asymptotically Gaussian fluctuation term,
\ie a stochastic variable with average $\mathcal{O}(1)$ and standard
deviation $\mathcal{O}(\sqrt{N-T})$ ($N-T$ is the number of
not-yet-assigned variables). The self-averageness of the $C_j$'s is a
consequence of the concentration of their
variations~\cite{achlioptas-bornes-inferieures-via-equadiffs}: given
$\vec{C}(T)$, the variation terms $\Delta C_j := C_j(T+1)-C_j(T)$ for $j
\ge 2$, eq.~(\ref{equation_evolution_C_j}) are concentrated around
constant averages, and these averages $\langle \Delta C_j \rangle$ may
be approximated by continuous functions $g_j(C_2/(N-T),$ $C_3/(N-T),$
$\ldots,$ $C_K/(N-T), t)$ with errors $o(1)$. However, the $\delta$ term
in eq.~(\ref{equation_evolution_C_1}) and the $\max$ term in
eq.~(\ref{equation_evolution_C_0}) are not smooth and prevent the
existence of continuous functions $g_1$ and $g_0$. This has deep
consequences, since the distribution of $C_1$ is found to be broad (in
the large $N$ limit, the standard deviation is not negligible w.r.t. the
average, but of the same order of magnitude). We conclude as in previous
subsection that we shall be obliged to study the full distribution of
$C_1$ and $C_0$.

 Eq.~(\ref{forme_concentree_c_j}) ensures that we can safely replace
the values of the $C_j$'s, $j \ge 2$, with their averages in the
large $N,T$ limits.  Let $\mathcal{E}(T)$ be a
probabilistic event at step $T$, such as: `the search detects no
contradiction up to step $T$'. We divide the space 
$C_2(T'), C_3(T'), \ldots, C_K(T')$, $0 \le T' \le T$ into a tube
$\mathcal{C}$ centered on the average trajectory $\langle C_2
\rangle(T'), \ldots, \langle C_K \rangle(T')$ and into its exterior
$\bar{\mathcal{C}}$. The probability of $\mathcal{E}(T)$ then reads:
\begin{equation}
 \label{partition_de_la_proba}
 \mathbb{E}[\mathcal{E}(T)] =
   \mathbb{E}[\mathcal{E}(T) \cap \mathcal{C}] +
   \mathbb{E}[\mathcal{E}(T) \cap \bar{\mathcal{C}}]
\end{equation}
and we shall choose the size of $\mathcal{C}$ so that the second term is
negligible with respect to the first one, \ie to the probability of
$\mathcal{E}(T)$ conditioned to the $C_j$'s lying close to their
averages at all times $0 \le T' \le T$.

 Fix $2 \le j \le K$ and $1/2 < \expdeviationcjs < 1$. At time $0 \le T'
\le T$, if the asymptotic standard deviation of $C_j(T')$ is
$\sigma_j(T') \sqrt{N-T}$, the probability that the discrete stochastic
variable $C_j(T')$ lies away from its average by more than $\sigma_j(T')
(N-T)^\expdeviationcjs$, or equivalently that the (asymptotically)
continuous stochastic variable $C_j(T')/[\sigma_j(T') \sqrt{N-T}]$ lies
away from its average by more than $(N-T)^{\expdeviationcjs-1/2}$, is
\bea
 \Delta(T') & := &
 2 \int_{(N-T)^{\expdeviationcjs-1/2}}^{+\infty} \frac{\dd x}{\sqrt{2\pi}}
   e^{-\frac{x^2}{2}} \nonumber \\
 & = & \frac{2}{\sqrt{2 \pi} (N-T)^{\expdeviationcjs-1/2}} 
   e^{-(N-T)^{2\expdeviationcjs-1}/2} \nonumber \\ & &
 + \mathcal{O} \left[ \frac{ e^{-(N-T)^{2\expdeviationcjs-1}/2} }
   {(N-T)^{3\expdeviationcjs-3/2}} \right] .
\eea
Although the value of $\sigma_j(T')$ depends on $T'$, it varies only
smoothly with the reduced parameter $t=T/N$ and it makes sense to use a
single exponent $\expdeviationcjs$ to define the region $\mathcal{C}$. 
The probability that $C_j(T')$ stays close to the average trajectory up
to $\sigma_j(T') (N-T)^\expdeviationcjs$ from $T'=0$ to $T$ is then
\beq
 \Delta := \prod_{T'=0}^T \left( 1 - \Delta(T') \right)
 = 1 - \mathcal{O}\Big( N^{3/2-\expdeviationcjs} 
         e^{-N^{2\expdeviationcjs-1}/2} \Big) \ .
\eeq
Generalizing this to the parallelepipedic region $\mathcal{C}$ with
boundaries such that each $C_j(T')$, $2 \le j \le K$, is always at the
distance at most $\sigma_j(T') (N-T)^\expdeviationcjs$ from its average, 
we find that the measure of $\mathcal{C}$ is
\beq
 \mathbb{E}[\mathcal{C}] = (1 - \Delta)^{K-1}
 = 1 - \mathcal{O}\Big( N^{3/2-\expdeviationcjs} 
         e^{-N^{2\expdeviationcjs-1}/2} \Big)
\eeq
and the complementary measure is $\mathbb{E}[\bar{\mathcal{C}}] = 1 -
\mathbb{E}[\mathcal{C}]$ so that
\begin{equation}
 \label{partition_de_la_proba_avec_borne_sur_Cbarre}
 \mathbb{E}[\mathcal{E}(T)] =
   \mathbb{E}[\mathcal E(T) \cap \mathcal{C}] +
   \mathcal{O}\Big( N^{3/2-\expdeviationcjs}
     e^{-N^{2\expdeviationcjs-1}/2} \Big)
\end{equation}
where the second term vanishes as $N$ gets large, as we wished.

 Finally, let us draw the scheme of the computations to follow: any
trajectory of $C_2(T), \ldots, C_K(T)$ inside $\mathcal{C}$ brings a
contribution to $\mathbb{E}[\mathcal E(T) \cap \mathcal{C}]$ that lies
close to the conditional average $\mathbb{E}[\mathcal E(T) | C_2(T)=N 
c_2(t), \ldots, C_K(T)=N c_K(t)]$ by an relative error at most
$\mathcal{O}(N^\expcorraprobcond)$ in any direction ($+C_2, -C_2,
\ldots$), $\expcorraprobcond$ being computed later, together with the
conditional average (it depends presumably on $\expdeviationcjs$). Thus,
we can approximate the total contribution $\mathbb{E}[\mathcal E(T) \cap
\mathcal{C}]$ with
\bea & &
 \left[1-\mathcal{O}\left( N^{3/2-\expdeviationcjs}
  e^{-N^{2\expdeviationcjs-1}/2} \right) \right]
   \times \nonumber \\ & & \nonumber \quad
 \mathbb{E}[\mathcal E(T) | C_2(T)=N c_2(t), \ldots, C_K(T)=N c_K(t)]
   \times \\ & & \quad
 \left[ 1 + \mathcal{O}\left( N^\expcorraprobcond \right) \right]
\eea
to get
\bea & &
 \label{principe_approximation_concentration_C_j}
 \mathbb{E}[\mathcal{E}(T)] =
     \nonumber \\ & & \quad
 \mathbb{E}\left[ \mathcal E(T) | C_2(T)=N c_2(t), \ldots,
   C_K(T)=N c_K(t) \right]
     \nonumber \\ & & \qquad
 \times \left[ 1 + \mathcal{O}\left( N^\expcorraprobcond \right) \right]
     \nonumber \\ & & \quad
 + \mathcal{O}\left( N^{3/2-\expdeviationcjs}
   e^{-N^{2\expdeviationcjs-1}/2} \right)
\eea
where we shall have to ensure (by a proper choice of $\expdeviationcjs$ if
possible) that the neglected terms are indeed negligible with respect to
the computed conditional expectation value: if $\expdeviationcjs$ is too
large, the weight of the region $\bar{\mathcal{C}}$ is very small, but we
allow deviations from the average and typical trajectory inside the (too
loose) region $\mathcal{C}$ that may bring contributions substantially
different from the typical one. Conversely, if we group into $\mathcal{C}$
near the typical trajectory only the most faithful trajectories, we have a
good control over the main contribution, but the weight of the
`treaters' in $\bar{\mathcal{C}}$ may not be negligible any more.

 The self-averaging of $C_1$ and $C_0$ (or its lack) has consequences
that may be observed numerically. Let us study the distribution over
instances of the probability $P$ that the UC=UP+R greedy, randomized
algorithm detects no contradiction during its run. That is, for each of
the $\approx 4000$ instances that we draw at random, we do $10^4$ to
$10^5$ runs of the algorithm (with different random choices of the
algorithm) and we estimate the probability of success of the algorithm
on this instance~\footnote{Alternatively, we could get the same result
by doing one run of the algorithm on each instance (and averaging over
many more instances) since the sequence $b$ of choices of the algorithm
on the instance $A$ is the same as the sequence $a$ of choices of the
algorithm on an instance $B$ obtained by relabeling the variables and
the clauses of instance $A$. However, this technique was slower in
practice because much time is spent building new instances.}.

\begin{figure}[htbp]
\begin{center}
\includegraphics[width=\largeurgraphes]{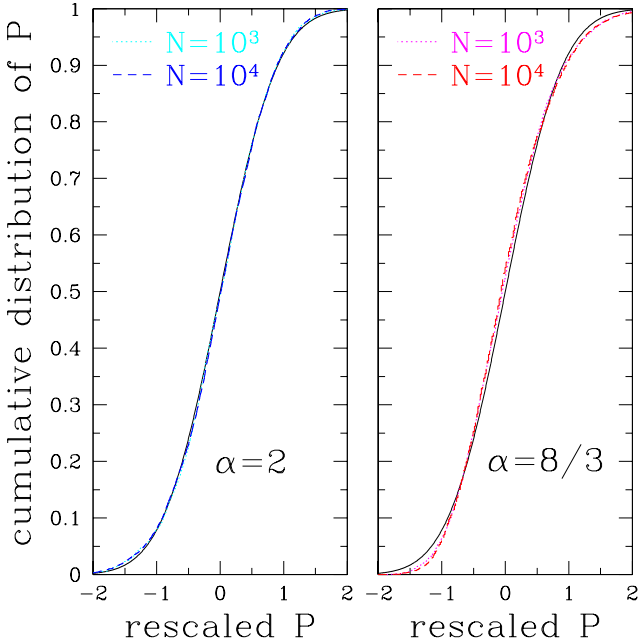}
\end{center}
\caption{\label{figure_concentration_psuccess} Numerical estimates for
the cumulative probability distributions of the probability $P$ of
success of the UC=R+UP algorithm on random 3-SAT instances with sizes
$N=10^3$ (dots) and $10^4$ (dashes) and initial clauses-per-variable
ratios $\alphainit=2$ (left) and $8/3$ (right). The $P$-axis for each
distribution (each value of $\alphainit$ and $N$) is chosen so that the
rescaled distributions have average 0 and standard deviation
$1/\sqrt{2}$. The solid lines show the Gaussian cumulative distribution
$[1+\mathrm{erf}(x)]/2$. For $\alphainit=2$, numerical distribution are
in good agreement with a Gaussian shape, but not for the critical ratio
$\alphainit=8/3$.}
\end{figure}

 The cumulative distribution function of $P$ is plotted in
figure~\ref{figure_concentration_psuccess} for instances of 3-SAT with
initial clauses-per-variable ratio $\alphainit=2$ (left curves) and
$8/3$ (right curves), for sizes of problems $N=1000$ and 10\,000. For
each size $N$, $P$ is rescaled to fix the average to 0 and the standard
deviation to $1/\sqrt{2}$. For $\alphainit<8/3$, $C_1$ has finite
average and standard deviation when $N \to +\infty$, and $\Delta C_1$
may be approximated by a continuous function $g_1$ like the $\Delta
C_j$'s for $j \ge 2$ (see Sect.~\ref{section_success_case}). The
numerical distributions of $P$ are successfully compared to a Gaussian
distribution (the average of $P$ for $\alphainit<8/3$ is computed in
Sect.~\ref{section_success_case}, see
Eq.~(\ref{expression_Psuccess_UC})). For $\alphainit=8/3$, things are
different. $C_1$ has average and standard deviation of the order of
$N^{1/3}$ (see Sect.~\ref{section_choix_echelles_critiques} and
following). As for $\alphainit=2$, the width of the finite-size
distributions of $P$ vanishes with $N$ --- they concentrate about their
average (computed in Section~\ref{section_classe_3sat}, see
Eq.~(\ref{moinslnPsuccess_famille_3sat})), and the rescaled finite-size
distributions of $P$ are numerically seen to converge to a well-defined
distribution. However, this distribution is \emph{not} Gaussian --- this
effect seems rather small, but significant.

 If we now plug the self-averaged form~(\ref{forme_concentree_c_j}) of
the $C_j$'s, $j \ge 2$, in their evolution
equations~(\ref{equation_evolution_C_j}), we get, using the reduced
parameter $t=T/N$,
\begin{equation}
 \label{equation_evolution_c_j}
 \frac{\dd c_j}{\dd t} =
 - \frac{j}{1-t} c_j(t) + \frac{j+1}{2(1-t)} c_{j+1}(t)
 \qquad \qquad 2 \le j \le K
\end{equation}
with $c_{K+1} := 0$. This triangular system of equations, with the
initial conditions $c_j(0)= \alphainit \delta_{j,K}$, is easily solved
for given $K$. For instance, for $2+p$-SAT and the R heuristic, the 
solution reads
\begin{eqnarray}
 \label{solution_c2c3_UC_c3}
 c_3(t) & = & \alphainit p (1-t)^3 \\
 \label{solution_c2c3_UC_c2}
 c_2(t) & = & \alphainit (3 p t / 2 + 1-p) (1-t)^2
\end{eqnarray}
whereas, for $2+p$-SAT and the GUC heuristic, 
\begin{eqnarray}
 \label{solution_c2c3_GUC_c3}
 c_3(t) & = & \alphainit p (1-t)^3 \\
 \label{solution_c2c3_GUC_c2}
 c_2(t) & = & (1-t) \left\{ \ln(1-t) +
                \alphainit [3 p t (2-t)/4 + (1-p)]\right\} .
  \nonumber \\ & &
\end{eqnarray}
For K-SAT with the R heuristics with initial ratios $\alpha_j = C_j/N$,
\beq
 \label{solution_c2_UC}
 c_2(t) = (1-t)^2 \sum_{j=2}^K \alpha_j j(j-1) 2^{1-j} t^{j-2} \ .
\eeq

 The parametric curve $[c_2(t), c_3(t), \ldots, c_K(t)]$, $0 \le t \le
1$, will be called \emph{resolution trajectory}. It describes,
forgetting about 0- and 1-clauses, the typical evolution of the reduced
formula under the operation of the algorithm. In the case of 3- or
$2+p$-SAT, a useful alternative phase space is the $p \times \alpha$
plane, where $\alpha = (c_2+c_3)/(1-t)$ is the (instantaneous)
2-or-3-clauses-per-variable ratio and $p$ is, as usual, the proportion
of 3-clauses amongst the 2- and 3-clauses. Some resolution trajectories
of the UC algorithm are shown on figure~\ref{figure_traj}. They all end
at the $c_1=c_2=c_3=0$ point with $t=1$: almost all variables have to be
assigned before no clauses are left (and a solution is found). More
clever algorithms such as GUC are able to find solutions with a finite
fraction of remaining unset variables, and give a family of solutions
with a finite entropy at once.

\subsection{Reduced generating functions for 0- and 1-clause numbers}

 From now on, we identify $C_2, C_3, \ldots, C_K$ with their
asymptotic averages $N C_j(t)$ as discussed
above, and study the kinetics of $C_0$ and $C_1$ as driven by $C_2,
C_3, \ldots, C_K$. Under these assumptions, it is easy to write the
evolution equation for $C_0$ and $C_1$ that corresponds
to~(\ref{equation_evolution_tousCs}): if \linebreak $P(C_0, C_1 ; T | 
C_2)$ is the probability that the search process leads, after $T$ 
assignments and with imposed values of $C_2, C_3, \ldots C_K$ at all 
steps 0 to $N$, to the numbers $C_0$, $C_1$ of 0-, 1-clauses,
\bea & &
 \label{equation_evolution_C1}
 P(C'_0, C'_1 ; T+1 | C_2) = \\ \nonumber & & \qquad
 \sum_{C_0} \sum_{C_1}  M_1[C'_0, C'_1 \gets C_0, C_1; T | C_2] \ 
   P(C_0, C_1 ; T | C_2)
\eea
where the expression of $M_1$ is deduced from that of $M$ by canceling
all what it written on the left of the sum over $z_2$ in
(\ref{definition_MX}). It is readily seen in that expression that,
actually, the transition matrix $M_1$ doesn't depend explicitly on
$C_3$, $C_4$, \ldots, $C_K$ but only on $C_2$; therefore, we dropped the
unnecessary dependence in the above equation. $M_1$ also depends
explicitly on time through $\mu = 1/(N-T)$. This evolution equation
translates into the following equation for the generating function
\[ G_{01}(X_0,X_1;T|C_2) := \sum_{C_0=0}^{M} \sum_{C_1=0}^{M} \!\!
X_0^{C_0} X_1^{C_1} P(C_0,C_1;T|C_2) : \]
\bea & &
\label{equation_evolution_fonctgen_c1}
 G_{01}(X_0,X_1;T+1|C_2) =
  \left(1+\frac{X_1-1}{N-T}\right)^{C_2(\alphainit,T)} \times
    \\ \nonumber & & \ 
  \left[
   \frac{1}{f_1} G_{01}(X_0,f_1;T|C_2) +
   \left(1 - \frac{1}{f_1} \right) G_{01}(X_0,0;T|C_2) \right]
\eea
where $f_1 = \frac{1+X_0}{2(N-T)} + X_1 (1-\frac{1}{N-T})$, see
(\ref{deff}). Since the $X_0$ argument of $G_{01}$ is the same in all
terms, we shall drop it when there is no ambiguity and use the lighter
notation $G_1(X_1;T|C_2) := G_{01}(X_0,X_1;T|C_2)$. The key equation
above yields the main results of the next Sections: in particular,
$G_{01}(0,1;T|C_2)$ is the probability that the search process
\emph{detected} no contradictions (\ie produced no 0-clauses, even if
there are already contradictory 1-clauses such as `$a$' and
`$\overline{a}$') while assigning the first $T$ variables, and
$G_{01}(0,1;T=N|C_2)$ is the probability that a solution has been found
by the search process, \ie that all $N$ variables were assigned without
production of a contradiction (if all variables were assigned, any
produced contradiction was necessarily detected).

\section{The probability of success}
In this section, the generating function formalism is used to study the
probability $\Psuccess (\alpha,p,N)$ that UC successfully finds a
solution to a random instance of the $2+p$-SAT problem with $N$
variables and $\alpha N$ clauses. We first consider the infinite size
limit, denoted by $\Psuccess (\alpha, p)$. As explained in Section
\ref{suctofaisec}, the probability of success vanishes everywhere but
for ratios smaller than a critical value $\alphaR(p)$. This threshold
line is re-derived, with a special emphasis on the critical behaviour of
$\Psuccess$ when $\alpha \to \alphaR(p)$.

We then turn to the critical behaviour at large but finite $N$, 
with the aim of making precise the scaling of $\Psuccess (\alpha,p,N)$
with $N$. A detailed analysis of the behaviour of the terms appearing 
in the evolution equation for the generating functions $G$ 
(\ref{equation_evolution_fonctgen_c1}) 
is performed. We show that the resulting scalings are largely common 
to all algorithms that use the Unit Propagation rule.

\subsection{The infinite size limit, and the success regime}
\label{section_success_case}

%As we showed in the previous section, we may evaluate the probability
%that the search process finds a solution (that is, does not produce any
%0-clause while assigning all $N$ variables) by computing
%$G_{01}(X_0,X_1;T=N|C_2)$, the generating function of $C_0$ and $C_1$
%conditioned to the average value of $C_2$. This uses the known typical
%evolution of $C_2$ (and $C_3$),
%(\ref{solution_c2c3_UC_c3})--(\ref{solution_c2c3_UC_c2}).

 For a sufficiently low initial clauses-per-variables ratio
$\alphainit$, the algorithm finds a solution with positive
probability when $N\to \infty$. It is natural to look for a solution of
equation~(\ref{equation_evolution_fonctgen_c1}) with the following 
scaling:
\begin{equation}
 \label{developpement_G1_succes_ordre0}
 G_1(X_1,T=tN|C_2) = \pi_0(X_1,t) + o(1)
\end{equation}
when $T,N \to +\infty$, $\pi_0$ being a smooth function of $X_1$ and $t$
($X_0$ is kept fixed to 0). $\pi_0(1,t)$ is therefore, in the $N \to
+\infty$ limit, the probability that the search process detected no
contradiction after a fraction $t$ of the variable has been assigned. The
probability of success we seek for is $\Psuccess = \pi _0(1,1)$.

 We furthermore know that $C_2$, which drives the evolution of $C_1$ in 
(\ref{equation_evolution_fonctgen_c1}), is concentrated around its 
average: we take
\begin{equation}
 \label{forme_de_C2_impose}
 C_2(\alphainit,T=t N) = (N-T) \adeux(\alphainit, t) +
   \mathcal{O}\left[(N-T)^\expdeviationcjs\right]
\end{equation}
with
\beq
 \label{definition_adeux}
 \adeux(t) := c_2(t)/(1-t)
\eeq
and $1/2 < \expdeviationcjs <1$ will be chosen later. Inserting the
above Ans\"atze into (\ref{equation_evolution_fonctgen_c1}) yields
\bea & &
 \label{eq1succ_avec_erreurs}
\pi_0(X_1,t) = \frac{e^{(X_1-1) \adeux(t)}}{X_1} \;
 \left[ \pi_0(X_1,t) + (X_1-1) \; \pi_0(0,t) \right]
  \nonumber \\ & & \qquad
 + \mathcal{O}\left[(1-X_1) (N-T)^{\delta-1}\right]
 + \mathcal{O}\left[(N-T)^{-1}\right]
\eea
hence, in the thermodynamic limit $N,T \to +\infty$, an equation for
$\pi_0$,
\beq
 \label{eq1suc}
\pi_0(X_1,t) = \frac{e^{(X_1-1) \adeux(t)}}{X_1} \;
 \left[ \pi_0(X_1,t) + (X_1-1) \; \pi_0(0,t) \right] .
\eeq
This equation does not suffice by itself to compute $\pi_0(1,t)$.  Yet it
yields two interesting results if we differentiate it w.r.t. $X_1$ to
the first and second orders in $X_1=1$:
\begin{eqnarray}
 \label{eq1asuc}
\pi_0 (0,t) &=& \big(1-\adeux(t) \big) \; \pi_0(1,t) \\
 \label{eq1bsuc}
\frac{\partial \pi_0}{\partial X_1}(1,t) &=&
 \frac{\adeux(t) \big( 2 - \adeux(t)\big)}{2 \big(1-\adeux(t) \big)} \;
 \pi_0(1,t) \ .
\end{eqnarray}
Under assumption (\ref{developpement_G1_succes_ordre0}),
$\pi_0(0,t)/\pi_0(1,t) = 1-\adeux(t)$ can be interpreted as the
probability $\rho_1(t)$ that there is no 1-clause at time $t$
conditioned to the survival of the search process. $\adeux(t)$ is then
the (conditional) probability that there is at least one
1-clause~\footnote{And the $\delta$ term that appears in
(\ref{equation_evolution_C_1}) is actually a continuous function of
$C_2/(N-T)$ so that (\ref{forme_concentree_c_j}) holds also for $j=1$.}.
As $\rho_1(t)$ has to be positive or null, $\adeux(t)$ cannot be larger
than 1. As long as this is ensured, $\rho_1(t)$ has a well-defined and
positive limit in the $N \to +\infty$ limit (at fixed reduced time $t$).
The conditional average of $C_1$, $\partial_{X_1} \pi_0(1,t) /
\pi_0(1,t)$, can be expressed from (\ref{eq1bsuc}) and is of the order
of one when $N \to +\infty$. The terms of the r.h.s. of
(\ref{equation_evolution_C_1}) compensate each other: 1-clauses are
produced from 2-clauses slower than they are eliminated, and do not
accumulate. Conversely, in the \emph{failure} regime
(Sect.~\ref{suctofaisec}), 1-clauses accumulate, and cause
contradictions.

 To complete the computation of $\pi_0(1,t)$, we consider higher orders
in the large $N$ expansion of $G_1$. In general, this would involve the
cumbersome fluctuation term $\mathcal{O}\left[(X_1-1)
(N-T)^{\delta-1}\right]$, but, at $X_1=1$, only the `deterministic'
$\mathcal{O}\left[(N-T)^{-1}\right]$ correction is left since $C_2$
disappears from equation (\ref{equation_evolution_fonctgen_c1}). Thus we
assume
\begin{equation}
 \label{developpement_G1_succes_ordre1enx1}
 G_1(1,T=tN|C_2) = \pi_0(1,t) + \frac{1}{N-T}\; \pi_1(1,t) + 
  o\left(\frac{1}{N-T}\right)
\end{equation}
which yields, when inserted into (\ref{equation_evolution_fonctgen_c1}),
\beq
 \label{eq2suc}
-(1-t) \frac{\partial \pi_0}{\partial t}(1,t) =
 \frac{1}{2}
 \bigg[ \frac{\partial \pi_0}{\partial X_1} (1,t) + 
 \pi_0(0,t)  -  \pi_0(1,t) \bigg] \ .
\eeq
 This equation (\ref{eq2suc}) can be turned into an ordinary
differential equation for $\pi_0(1,t)$ using (\ref{eq1asuc}) and
(\ref{eq1bsuc}); after integrating over $t$, with the initial condition
$\pi(1,t=0)=1$ \ie no contradiction can arise prior to any variable
setting, we find the central result of this
section~\cite{frieze-suen-guc-sc,achlioptas-kirousis-kranakis-krizanc-2plusp-sat-rigoureux}:
\begin{equation}
 \label{expr_proba_succes}
\pi_0 (1,t) = \exp \left( - \int_0 ^t \frac {d\tau}{4 (1-\tau)}
\, \frac{\adeux(\tau)^2}{1-\adeux(\tau)} \right)
\end{equation}
which is finite if, and only if, $\adeux(\tau)<1$ for all $0 \le \tau
\le 1$ (the apparent divergence at $\tau=1$ is in practice compensated
by the factors involving $\adeux$).

The above result can be used in (\ref{eq1asuc}) and (\ref{eq1suc}) 
to compute $\pi_0 (X_1,t)$, that is
the  generating function of the probability $P(C_1,t)$ that there
are $C_1$ 1-clauses and no contradiction has occurred 
after assignment of a fraction $t$ of the variables, 
\bea & &
 \label{expression_pi0_succes}
 \pi_0(X_1,t) := \sum _{C_1 \ge 0} P(C_1,t) \; X_1^{C_1}
    \nonumber \\ & & \qquad =
  [1-\adeux(t)]\, \frac{1-X_1}{1 - X_1 e^{-(X_1-1)\adeux(t)}}
  \; \pi_0(1,t) . \quad
\eea
As long as $\adeux(t)<1$, the average number of 1-clauses $\langle C_1
\rangle(t)$ is finite as $N \to +\infty$.  This sheds light on the
finiteness of $\Psuccess$. The probability of not detecting a
contradiction at the time step $T \to T+1$ is
$(1-\mu/2)^{\max(C_1-1,0)}\simeq 1-\max(C_1-1,0)/2/(N-T)$, and
$\Psuccess$ is the product of $\Theta(N)$ quantities of that
order~\footnote{We can't go further and compute a function
$\pi_1(X_1,t)$ corresponding to the order $1/N$ in
(\ref{developpement_G1_succes_ordre0}), since the `Gaussian
fluctuations' term $\mathcal{O}\left[(N-T)^\expdeviationcjs\right]$ in
(\ref{forme_de_C2_impose}) would dominate the $1/N$ introduced
correction --- only for $X_1=1$ is this $1/N$-term relevant, so that we
could write down (\ref{eq2suc}). It is also impossible to compute
$\pi_1(1,t)$ alone.}.

\begin{figure}[htbp]
\begin{center}
\includegraphics[width=\largeurgraphes]{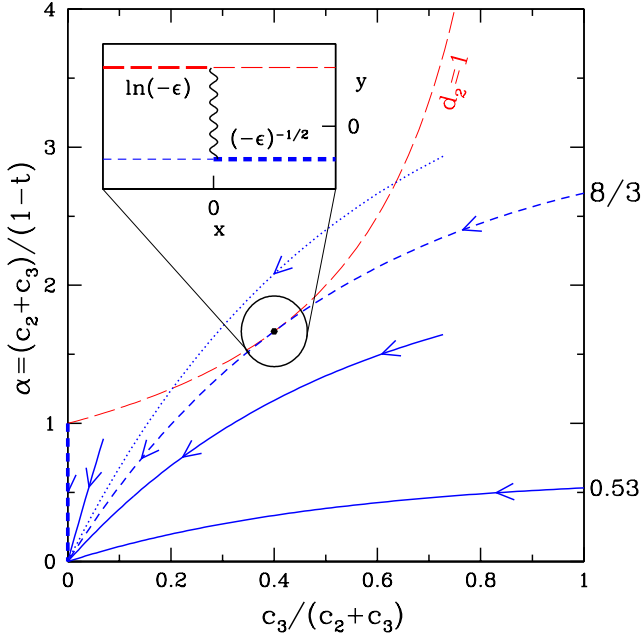}
\end{center}
\caption{\label{figure_traj} Some resolution trajectories of the UC=UP+R
algorithm. If the trajectory of some algorithm stays strictly below the
$\adeux=\alphainit (1-p)=1$ line (long-dashed line), 1-clauses don't
accumulate and the success probability $\Psuccess$ of the algorithm is
finite as $N \to +\infty$ (solid trajectories). Critical behaviour is
when the trajectory gets tangent to this line or starts on it
(short-dashed trajectories). When the trajectory spends time in the
region above this line (dotted trajectory), $\Psuccess$ gets
exponentially small in $N$. \textbf{Inset:} The singular behaviour of
$\Psuccess$ around the $(p=2/5, \alpha=5/3)$ point is better understood
if one uses local coordinates $(x,y)$. $x$ is the distance along the
common tangent to the $\adeux=1$ line and the critical trajectory. $y$
is such that lines of constant $y$ are, in the original coordinates,
parabolas, tangent to the $\adeux=1$ line. The point $(2/5,5/3)$ is
spread onto the waved line. Left and right limits of $\Psuccess$ for $x
\to 0$ on the waved line are well-defined but distinct. The $N \to
+\infty$ limit of $-\ln \Psuccess$ has the indicated singularities on
the thick lines.}
\end{figure}

 The validity condition $\adeux(t)=c_2(t)/(1-t)<1$ is fulfilled at all
steps $t$ if, and only if, the initial clauses-per-variable ratio
$\alphainit$ is smaller than a threshold, $\alphaR$, as can be seen from
the expression of $c_2(t)= \langle C_2 \rangle(T)/N$ that results from
equation (\ref{solution_c2c3_UC_c2}). Graphically, in the $(p, \alpha)$
plane, the resolution trajectory in Figure~\ref{figure_traj} stays below
the line $\alpha (1-p)=1$ iff. $\alphainit$ is small enough.
Finding the threshold value for $\alphainit$ and a given $p$ is an easy
ballistic problem:
\begin{itemize}

\item If $p<2/5$ (`2-SAT family'), whatever the initial
clauses-per-variable $\alphainit$, the resolution trajectory
(figure~\ref{figure_traj}) will always either be entirely below the
$\adeux=1$ line (success case, low $\alphainit$), or cut it (failure
case, high $\alphainit$). The threshold value of $\alphainit$ is reached
when the resolution trajectory starts exactly on it (critical case),
therefore
\beq
 \label{alpha_critique_famille2sat}
 \alphaR(p)=1/(1-p) \quad \text{if }p < 2/5.
\eeq

\item If $p \ge 2/5$ (`3-SAT class'), the resolution trajectory for low
$\alphainit$ is also entirely below the $\adeux=1$ line (success case).
This situation ends when the resolution trajectory gets \emph{tangent}
to the $\adeux=1$ line, whereas for $p<2/5$ it was \emph{secant}. 
All critical trajectories for $p \ge 2/5$ share the support of the
critical trajectory for $p=1$ (3-SAT) that starts at $(p=1,
\alphainit=8/3)$, and all become tangent to the $\adeux=1$ line at the
$(p=2/5, \alpha=5/3)$ point (reached after a finite time), whereas for
$p<2/5$ there are several critical trajectories. Here,
\beq
 \label{alpha_critique_famille3sat}
 \alphaR(p) = 24 p/(2+p)^2 \quad \text{if }p \ge 2/5.
\eeq

%\item If $p=2/5$, the critical resolution trajectory both starts on the 
%$\adeux=1$ line and is tangent to it. $\alphaR(p) = 5/3$.

\end{itemize}

The probability $\Psuccess = \pi_0(1,1)$ that the UC algorithm finds a
solution is obtained from equation (\ref{expr_proba_succes}) with
$\adeux(t) = \alphainit (1-t)(3/2 p t + 1-p)$,
equation~(\ref{solution_c2c3_UC_c2}):
\bea & &
 \label{expression_Psuccess_UC}
 -\ln \Psuccess(\alphainit,p) =
 \frac{1}{4 \sqrt{24 p/(2+p)^2 / \alphainit - 1}} \times
   \\ & & \qquad
 \left[ \mathrm{arctan} \frac{1}{\sqrt{24 p/(2+p)^2/\alphainit - 1}} +
   \right. \nonumber \\ & & \qquad \left. \ \,
        \mathrm{arctan} \frac{5(p-2/5)}
          {(2+p) \sqrt{24 p/(2+p)^2/\alphainit-1}} \right]
   \nonumber \\ & & \qquad \nonumber
 -\frac{1}{8} \ln \left( \frac{1}{1-p} - \alphainit \right)
 -\frac{1}{8} \ln(1-p) + \frac{\alphainit (p-4)}{16} \ .
\eea
%It shows especially that, when $\alphainit$ tends to $\alphaR(p)$, the
%success probability goes to 0, which indicates that $\alphaR$ is not an
%artefact of our computation: it lies on the frontier between the success
%case and a failure case, where $\Psuccess$ vanishes for large $N$.
Of particular interest is the singularity of $\Psuccess$ slightly below
the threshold ratio. At fixed $p$, as $\alphainit$ increases, the first
singularity is encountered when the resolution trajectory tangent to the
$\adeux=1$ line is crossed \ie for $\alphainit = 24 p/(2+p)^2$
(figure~\ref{figure_traj}, largest short-dashed line, and thick
short-dashed line in the inset). If $p>2/5$ (3-SAT class), the two
$\mathrm{arctan}$ in equation (\ref{expression_Psuccess_UC}) tend to 
$\pi/2$ and $\Psuccess$ vanishes as ($\epsilon<0$):
\beq
 \label{singularite_psuccess_alphafixe_famille3sat}
-\ln \Psuccess \left[ (1 + \epsilon) \alphaR(p),p \right] =
 \frac \pi4\frac 1{\sqrt{- \epsilon}} + \Theta (1).
\eeq
For $p=2/5$, one of the two $\mathrm{arctan}$ vanishes for all $\alpha$, 
and the  first $\ln$ brings another singularity ($\epsilon<0$):
\beq
-\ln \Psuccess \left[ \frac 53 (1 + \epsilon) ,\frac 25 \right] =
 \frac \pi8\frac 1{\sqrt{- \epsilon}} - \frac{1}{8} \ln(-\epsilon) +
 \Theta (1).
\eeq
And for $p<2/5$, the two $\mathrm{arctan}$ have opposite signs so that
the first term of equation (\ref{expression_Psuccess_UC}) has no 
singularity while crossing the `limiting' resolution trajectory $\alpha 
= 24 p / (2+p)^2$ (thin short-dashed line of the inset of
Fig.~\ref{figure_traj}). 
%For $\alphainit$ between this trajectory and the $\adeux=1$
%line, (\ref{expression_Psuccess_UC}) reads
%\bea
% \label{expression_Psuccess_UC_audessus_trajcritique}
% -\ln \Psuccess(\alphainit) & = &
% \frac{1}{4 \sqrt{1 - 24 p/(2+p)^2 / \alphainit}}
% \left\{ - \mathrm{arccotanh} \frac{1}{\sqrt{1 - 24 p/(2+p)^2/\alphainit}} +
%        \mathrm{arccotanh}
%          \left[1 + \frac{24p(1-p-1/\alphainit)}{(5p-2)^2}\right]^{-1/2}
% \right\} \nonumber \\ & &
% -\frac{1}{8} \ln \left( \frac{1}{1-p} - \alphainit \right)
% -\frac{1}{8} \ln(1-p) + \frac{\alphainit (p-4)}{16}
%\eea
A singularity is found when $\alphainit$ reaches the $\adeux=1$ line,
$\alphainit=1/(1-p)$ (thick long-dashed line of the inset of
figure~\ref{figure_traj}), with the outcome ($\epsilon<0$)
%: the argument of the second $\mathrm{arccotanh}$
%above tends to 1, and the argument of the first $\ln$ above vanishes. 
%This yields two singularities in $\ln$ with respective coefficients 
%$-(p+2)/(2-5p)/8$ and $-1/8$ that add up:
\beq
 \label{singularite_psuccess_alphafixe_famille2sat}
 -\ln \Psuccess \left[(1 + \epsilon) \alphaR(p),p \,\right] =
 - \frac{1-p}{2(2-5p)} \ln(-\epsilon) + \Theta (1).
\eeq
The difference of nature of the singularities between the 2-SAT and
3-SAT families corresponds to different divergences of $- \ln \Psuccess$
with $N$, as will be computed in the next section.

For completeness, let us check that the above calculation is compatible
with our approximation (\ref{principe_approximation_concentration_C_j}).
The first term of the l.h.s. there is equal to $\Psuccess$ plus the
$1/(N-T)$ correction from
equation~(\ref{developpement_G1_succes_ordre1enx1}). The second term
there, in $\mathcal{O}(N^\zeta)$, corresponds here to the fluctuations
of $C_2$: $\mathcal{O}(N^{\delta-1})$ in
equation~(\ref{eq1succ_avec_erreurs}), thus $\zeta=\delta-1$. If we take
for $\expdeviationcjs$ any value on the allowed interval $]1/2, 1[$, the
two approximation terms in
(\ref{principe_approximation_concentration_C_j}) vanish as $N \to
+\infty$. Therefore, as long as $\Psuccess$ is finite for large $N$,
these approximation terms are actually negligible.

\subsection{Large $N$ scalings in the critical regime}
\label{section_choix_echelles_critiques}

Our previous study of the success regime breaks down when
$\adeux(t)$  reaches 1 during operation of the algorithm. 
%Under such circumstances, the average
%$\langle C_1 \rangle$ could not be shown to be finite for large $N$ (or
%$S$) any more. 
Indeed, consider  the infinite-$N$ generating
function for $C_1$, $G_1$, given by  eq.~(\ref{expression_pi0_succes}).
 As a function of $X_1$, $G_1$ vanishes uniformly on any compact interval
$[0,1-\eta]$, $\eta > 0$, as $\adeux \to 1$. The $X_1=1$
point is singular since normalization enforces $\pi_0(1,t)=1$ (assuming
$X_0=1$): all useful information is concentrated in a small region
around $X_1=1$. Expanding eq.~(\ref{expression_pi0_succes}) for
$X_1 \to 1$ yields
\bea
 \label{expression_pi0_succes_critique}
\pi_0(X_1,t) & = & \left\{1 + \adeux(t) \frac{1-X_1}{1-\adeux(t)} -
  \frac{X_1 \adeux^2(t)}{2} \frac{1-X_1}{1-\adeux(t)} +
    \right. \nonumber \\ & & \left. \ \,
  \mathcal{O}\left[ \frac{(1-X_1)^2}{1-\adeux(t)} \right] \right\}^{-1}.
\eea
%(if $X_0 \neq 1$, we get the same expression for the normalized quantity
%$\pi_0(X_0, X_1, t) / \pi_0(X_0, 1, t)$). 
Non-trivial results are obtained when $1-X_1$ and $1-\adeux(t)$ are of
the same (vanishingly small) order of magnitude --- let us call it
$\Delta$. We suspect that $\Delta$ is some negative power of $N$, to be
determined below. Let us define
\beq
 \label{scaling_x1_et_e2}
 X_1 =: 1 - x_1 \Delta, \quad e_2(t):= \Delta^{-1} (1-\adeux(t))
\eeq
so that eq.~(\ref{forme_de_C2_impose}) now reads
\beq
 \label{forme_de_C2_impose_critique}
 C_2[\alphainit,T=tN] = [1 - e_2(\alphainit, t) \Delta ] (N-T) +
   \mathcal{O}\left[ (N-T)^\expdeviationcjs \right]
\eeq
where, as previously, the exponent $\expdeviationcjs$ will be tuned
later according to the framework
eq.~(\ref{principe_approximation_concentration_C_j}).

 We assume that, in the thermodynamic limit $N \to \infty, \Delta \to 0$
but at fixed $x_1$, $e_2(t)$, and time $t$, the normalized generating
function of $C_1$ conditioned to the typical value of $C_2(t)$,
\beq 
  \label{definition_pibarre}
  \frac{G_1(X_1=1-x_1\Delta, T=t N|C_2)}{ G_1(1, T=t N|C_2)}
  \tendsto_{N \to +\infty} \pib(x_1,t)
\eeq
where the limit $\pib$ is a smooth function of $x_1$ and $t$ (which also
depends on $X_0$ and $e_0$). $\pib(x_1,t)$ is the generating function
for the stochastic variable $c := C_1/\Delta$, conditioned to the
success of the algorithm. Eq.~(\ref{expression_pi0_succes_critique})
gives information about the $e_2 \to +\infty$ limit of $\pib$.
Furthermore, eq.~(\ref{expression_pi0_succes}) shows that
\beq
 \label{definition_sigma}
 \Delta^{-1} G_1(0, T|C_2) / G_1(1, T|C_2)
 \tendsto_{N \to +\infty} \sigma (t)\ ,
\eeq
a well defined limit of the order of unity for $T=t N$ and $C_2$ given
by eq.~(\ref{forme_de_C2_impose_critique}).
 We now plug the previous conventions and assumption into the evolution
equation for the \emph{conditional}, normalized generating function of
$C_1$. This equation is formed by dividing
eq.~(\ref{equation_evolution_fonctgen_c1}) by $G_1(1, T|C_2)$, which
yields, in a formal way,
\bea
 \label{equation_evolution_fonctgen_c1_normalisee}
 \mathrm{LHS} & = & \mathrm{RHS}_1 + \mathrm{RHS}_2 \\
 \mathrm{LHS} & := &
  G_1(X_1, T+1 | C_2) / G_1(1, T | C_2) \nonumber \\
 \mathrm{RHS}_1 & := &
  \left(1+\frac{X_1-1}{N-T}\right)^{C_2(\alphainit,T)}
  \frac{1}{f_1}
  \frac{G_1(f_1;T|C_2)}{G_1(1, T | C_2)} \nonumber \\
 \mathrm{RHS}_2 & := &
  \left(1+\frac{X_1-1}{N-T}\right)^{C_2(\alphainit,T)}
  \left(1 - \frac{1}{f_1} \right)
  \frac{G_1(0;T|C_2)}{G_1(1, T | C_2)} \ . \nonumber
\eea
 From this equation we get, in the following subsections, all results
relevant to the critical behaviour of the success probability of the
greedy algorithm.

\subsubsection{Analysis of the RHS terms}

 The two contributions of the r.h.s. of the evolution
equation~(\ref{equation_evolution_fonctgen_c1_normalisee}) have the
detailed expression (for $X_0=0$):
\bea
 \mathrm{RHS}_1 & = &
 \left\{ 1 + x_1 e_2(t) \Delta^2 + \frac{1}{2 N (1-t)} +
 \frac{x_1^2}{2} \Delta^2 +
   \right. \nonumber \\ & & \  
 \mathcal{O}( (x_1^3 + x_1^2 e_2(t) ) \Delta^3 ) +
 \mathcal{O}( x_1 \frac{\Delta}{N(1-t)} ) +
   \nonumber \\ & & \ 
 \mathcal{O}\left[x_1 \Delta (N-T)^{\delta-1}\right] \bigg\} \times
  \nonumber \\ & &
\left\{
 \pib(x_1,t) + \frac{1}{2N(1-t) \Delta} \partial_{x_1} \pib(x_1,t) +
  \right. \nonumber \\ & & \ \,
 \mathcal{O}\left[(N-T)^{-2} \Delta^{-2}\right] \bigg\} \quad
\eea
and
\bea & &
 \mathrm{RHS}_2 =
 - \sigma(t) \Delta
 \left\{ x_1 \Delta + \frac{1}{2N(1-t)} +
   \mathcal{O}(\frac{x_1\Delta}{N(1-t)}) +
     \right. \nonumber \\ & & \quad \nonumber
   \mathcal{O}(x_1^3 \Delta^3) + \mathcal{O}\left[ (N-T)^{-2} \right] +
   \mathcal{O}\left\{x_1^2 \Delta^2 (N-T)^{\delta-1}\right] \bigg\} 
\eea
where, as usual, $T = t N$.

 Apart from the dominant term $\pib(x_1,t)$ that cancels with the
dominant term of the l.h.s., the first terms have order $\Delta^2$ and
$1/(N\Delta)$ (we here assume that $t<1$ in the critical regime, which
will be the case in all subsequent situations). Then, $\Delta^3$, $1/N$,
$\Delta/N$ and so on are negligible for large $N$ (because $\Delta$
vanishes, but slower than $1/(N-T)$ since the integer $C_2$ may not vary
by less than unity). So do the terms stemming from fluctuations of $C_2$
around its typical value, $\Delta^2 (N-T)^{\delta-1}$ and $\Delta
(N-T)^{\delta-1}$, if we choose $\delta$ carefully (see below). Choosing
$\Delta$ such that $\Delta^2 = 1/(N \Delta)$ allows us to gather a
maximal number of terms in the equation for $\pib$. Other choices are
possible but trivial, in that they would correspond to either the
success or the failure regimes, but not to the critical case. From now
on, $\Delta := N^{-1/3}$.

 Then, the fluctuations of $C_2$ in the above expansion are of the order
of $\Delta (N-T)^{\delta-1} = (N-T)^{\delta-4/3}$. Using the notations
from eq.~(\ref{principe_approximation_concentration_C_j}),
$\zeta=\delta-4/3$. These fluctuations are negligible with respect to
$N^{2/3}$ if $\delta<2/3$. Remember that the range of possible values
for $\delta$ was $]1/2, 1[$; in the critical situation here, we may
choose $\delta\in ]1/2, 2/3[$. It will be checked later that the third
term of the l.h.s. of
eq.~(\ref{principe_approximation_concentration_C_j}) is also negligible
w.r.t. the first one.
% With this choice of scale for $\Delta$ and $\delta$, the two terms
%above yield the total r.h.s. of the evolution
%equation~(\ref{equation_evolution_fonctgen_c1_normalisee}) in the
%thermodynamic limit (when $x_1=0$, the term stemming from fluctuations 
%of $C_2$ vanishes like in eq.~(\ref{eq1succ_avec_erreurs})):
Finally,
\bea & &
 \label{equation_evolution_fonctgen_c1_critique_md}
 \mathrm{RHS}_1 + \mathrm{RHS}_2 =
 \pib(x_1,t) +  N^{-2/3}
   \times \nonumber \\ & & \quad
 \left[ \frac{1}{2(1-t)} \partial_{x_1} \pib(x_1,t) +
 \big(e_2(t)\; x_1 + \frac{x_1^2}{2} \big)  \pib(x_1,t)
   \right. \nonumber \\ & & \quad \ \,
 - \sigma(t) x_1 \bigg] +
 \mathcal{O}\left[(N-T)^{-1} , x_1 (N-T)^{\delta-4/3}\right] .
\eea

\subsubsection{Analysis of the LHS terms}

LHS in eq.~(\ref{equation_evolution_fonctgen_c1_normalisee}) has to do
with time evolution. The values of $\adeux$ or $e_2$ are given by the
average value of $C_2$ calculated in Section \ref{secavecla}. $e_2$ is
of the order of unity when $N \to +\infty$ if the resolution trajectory
comes close to $\adeux=1$ \ie if the initial clauses-per-variable ratio
$\alphainit$ is close to its threshold value
eqs.~(\ref{alpha_critique_famille2sat} -
\ref{alpha_critique_famille3sat}). We zoom in around the time, say,
$\tst$ where $\adeux$ is closest to 1 (or equal to 1 if we are exactly
on the borderline between the success and failure cases) and let
\beq
 \label{scaling_temps}
 T = \tst \, N + t_0 N \Delta^\expDeltatemps
   = \tst \, N + t_0 N^{1-\expDeltatemps/3} .
\eeq
$\expDeltatemps$ will be fixed later for each family of near-critical
trajectories so that $1-\adeux(t)$ is indeed of order $\Delta$ on a
finite interval of rescaled times $t_0$. We now assume that $\pib$ and
$G_1(X_1=1)$ have, when $N \to\infty$ and time $T$ is given by
(\ref{scaling_temps}), well defined limits, regular w.r.t. $t_0$
\footnote{This assumption could presumably be demonstrated using the
same technique as for Wormald's theorem~\cite{wormald-theoreme}. If we
introduce ex nihilo the functions $\pib$ and $G_1$ that satisfy the
equations (\ref{edp_pi})--(\ref{edo_pib}), we could show that the
difference between the sequences, for $T$ from 1 to $N$, of the discrete
quantities $G_1(X_1, T|C_2)$ and of the approximate quantities
\mbox{$\pib[(1-X_1) N^{1/3},$} $(T - N \tst) N^{-1+\expDeltatemps/3}]$
$G_1[1,$ \mbox{$t_0=(T - N\tst) N^{-1+\expDeltatemps/3}]$}, vanishes
when $N \to +\infty$. The reason is that the difference between two
consecutive terms of each of the two sequences is the same up to a small
quantity that yields a negligible difference at the final date $T=N$,
and the initial conditions for both sequences are equal.}.
 The l.h.s.\ of eq.~(\ref{equation_evolution_fonctgen_c1_normalisee})
may be written for large $N$
\beq
 \label{equation_evolution_fonctgen_c1_critique_mg_brut}
 \mathrm{LHS} = G_{1}( 1, T+1|C_2) / G_{1}( 1, T|C_2)
                \ \pib(x_1, T+1)
\eeq
where $T+1 = t_0 + N^{-1} \Delta^{-\expDeltatemps} = t_0 +
N^{-1+\expDeltatemps/3}$.
 As time $t_0$ goes on, the shape of the distribution of $C_1/\Delta$
(encoded into $\pib$) and the probability of success (given by $G_1(1,
T|C_2)$) both vary. Fix first $x_1$ at 0 so that $X_1=1$ and $\pib=1$,
then eqs.~(\ref{equation_evolution_fonctgen_c1_critique_mg_brut}) and
(\ref{equation_evolution_fonctgen_c1_critique_md}) read
\bea
 \label{equation_evolution_fonctgen_c1_critique_mg_normalis}
 \mathrm{LHS} & = & 1 +
   N^{-1+\expDeltatemps/3} \partial_{t_0} \ln[ G_{1}( 1, t_0|C_2) ] \\
 \label{equation_evolution_fonctgen_c1_critique_md_x1_nul}
 \mathrm{RHS}_1 + \mathrm{RHS}_2 & = &
   1 + \frac{1}{2(1-t)} N^{-2/3} \partial_{x_1} \pib(1,t) +
     \nonumber \\ & &
   \mathcal{O}\left[(N-T)^{-1}\right] .
\eea
 Comparing the two members,
eq.~(\ref{equation_evolution_fonctgen_c1_critique_mg_normalis}) and
eq.~(\ref{equation_evolution_fonctgen_c1_critique_md_x1_nul}), of
eq.~(\ref{equation_evolution_fonctgen_c1_normalisee}) shows that $\ln[
G_1(1, t_0|C_2) ]$ is of the order of \linebreak
$N^{(1-\expDeltatemps)/3}$, with subleading terms of the order of
$N^{-\expDeltatemps/3}$ at most.  Defining $\expproba:=
(1-\expDeltatemps)/3$, we have in the large $N$ limit
\beq
 \label{definition_mu}
 - N^{-\expproba} \ln G_1(1, T|C_2) \to \mu (t_0) ,
\eeq
a regular function of $t_0$. 
Eqs.~(\ref{equation_evolution_fonctgen_c1_critique_mg_normalis}) and
(\ref{equation_evolution_fonctgen_c1_critique_md}) with $x_1=0$ yield
\beq
 \label{lien_mu_pi}
 \partial_{t_0} \mu(t_0) =
 - \frac{1}{2(1-\tst)} \partial_{x_1} \pib(0, t_0) .
\eeq
As $\pib$ is the generating function of $c = C_1/\Delta$ (conditioned 
to success of the algorithm),
\beq
\cbarre (t_0):= - \partial_{x_1} \pib(0, t_0)
\eeq 
is the conditional average of the rescaled number $c$ of unit-clauses.

 For general $x_1$ now, the previous assumptions lead from the
expression (\ref{equation_evolution_fonctgen_c1_critique_mg_brut}) of
the l.h.s. of eq.~(\ref{equation_evolution_fonctgen_c1_normalisee}) to
\bea
 \label{equation_evolution_fonctgen_c1_critique_mg}
 \mathrm{LHS} & = & \pib(x_1, t_0) +
 N^{-1+\expDeltatemps/3} \partial_{t_0} \pib(x_1, t_0)
   \nonumber \\ & &
 - N^{-2/3} \pib(x_1, t_0) \partial_{t_0} \mu(t_0) +
 \mathcal{O}(N^{-2+2\expDeltatemps/3}) . \quad
\eea

\subsection{Critical evolution equations}

 Comparing the two sides of the evolution equation,
eq.~(\ref{equation_evolution_fonctgen_c1_critique_md}) and
eq.~(\ref{equation_evolution_fonctgen_c1_critique_mg}), we are left with
two situations.
\begin{itemize}
\item If $\expDeltatemps=1$: $\expproba=0$ and it is convenient to use
the non-normalized (non-conditional) generating function \[ \pi(x_1,t_0)
:= \exp[-\mu(t_0)] \pib(x_1,t_0).\] The total probability here, $\pi(0,
t_0)$, is not 1 as for $\pib$ but the success probability of the greedy
algorithm. $\pi$ satisfies the following PDE:

\bea
 \label{edp_pi}
 \partial_{t_0} \pi(x_1, t_0) & = &
 \frac{1}{2(1-\tst)} \partial_{x_1} \pi(x_1,t_0) +
    \\ \nonumber & &
  \left[ e_2(t_0) x_1 + \frac{x_1^2}{2} \right] \pi(x_1,t_0)
  - \sigma(t_0) x_1 .
\eea

\item If $\expDeltatemps < 1$: the probability of success has the
scaling relationship $\Psuccess \propto \exp[- \Theta(N^{(1-a)/3})]$. 
The time-derivative term $\partial_{t_0} \pib(x_1, t_0)$ is negligible
w.r.t. other terms, and $\pib$ satisfies the ODE:

\bea
 \label{edo_pib}
 0 & = & \frac{1}{2(1-\tst)}
   \left[ \partial_{x_1} \pib(x_1,t_0) + \cbarre(t_0) \pib(x_1,t_0) \right] +
    \nonumber \\ & &
  \left[ e_2(t_0) x_1 + \frac{x_1^2}{2} \right] \pib(x_1,t_0)
  - \sigma(t_0) x_1 \ .
\eea

\end{itemize}
The third possibility, $\expDeltatemps > 1$, leads to inconsistencies
and has to be rejected~\footnote{We would have either subdominant terms
of the order of $N^{-1+a/3}$, larger than the dominant term (of the
order of 1) $\pib(x_1,t_0)$ if $\expDeltatemps \ge 3$, or the two
equations $\partial_{t_0} \pib(x_1, t_0) =0$ and an ODE for $\pib$ at
fixed $t_0$ but with coefficients $e_2(t_0)$ and $\sigma(t_0)$. In the
latter case, since $\pib$ would be constant with time, $e_2$ and thus
$\adeux$ should also be constant, which is impossible in the context of
our algorithm (see Eq.~(\ref{equation_evolution_c_j})).}.

\medskip

In the previous section, we classified the critical resolution
trajectories into two families: those of the 2-SAT family ($p<2/5$)
start from the $\adeux=1$ line but are secant to it, and those of the
3-SAT family ($p>2/5$) do not start on this line but get tangent to it
(`parabola situation'). As we will see in the next sections, these two
families correspond, respectively, to values of the exponent
$\expDeltatemps$ equal to 1 and $1/2$, making successively
eq.~(\ref{edp_pi}) and (\ref{edo_pib}) relevant. 

%To summarize, we have changed, in the purpose of capturing
%the critical phenomenon between success and failure, the scales of $C_1$
%(through $X_1$) and of time. We also required that $\adeux(t)$ should be
%close to 1, and this change of scale implies that we choose accordingly
%an initial clause-per-variable ratio $\alphainit$ close to its
%threshold value. But this last rescaling depends, just like the value of 
%$\expDeltatemps$, on the shape of the critical resolution trajectory 
%(secant or tangent) and has to be deferred to the next Subsections.

\section{The 2-SAT class (power law class)}
\label{section_famille_2sat}

\subsection{Equations for 2-SAT and its family}

When $p<2/5$, the critical resolution trajectory starts on the $\adeux=1$
line and is secant to it (at time $\tst=0$) with slope
\beq
 \label{definition_beta}
 \beta(p):= \frac{2-5p}{2(1-p)} .
\eeq
The threshold value of $\alphainit$ is $\alphaR(p)=1/(1-p)$
(eq.~\ref{alpha_critique_famille2sat}). On this resolution trajectory,
$1-\adeux(t-\tst) \propto t-\tst$. Therefore, this resolution trajectory
is at distance $\Delta = N^{-1/3}$ of the $\adeux=1$ line as long as
$t-\tst$ is of order $\Delta^1$: the exponent $\expDeltatemps$ equals 1
here and the relevant equation is eq.~(\ref{edp_pi}).

 The critical regime is realized when $\alphainit$ is close to the value
$\alphaR(p)=1/(1-p)$.
The relevant scaling is
\beq
 \label{scaling_alpha_2sat}
 \alphainit = \alphaR(p) (1 + \epsalpha N^{-1/3})
\eeq
with finite $\epsalpha$ since, if $\alphainit$ is less than $\alphaR(p)$
by more than $\Delta = N^{-1/3}$, at the initial date $t=0$, $\adeux(0)$
is already out of the critical region $1-\Delta$ (remember that
$\adeux(t)$ decreases with time if $p<2/5$ as can be seen in
Fig.~\ref{figure_traj}). Conversely, if $\alphainit$ is above
$\alphaR(p)$ by a distance much greater than $\Delta$ at time $t=0$,
$1-\adeux(0)$ is an order of magnitude higher than the critical distance
$\Delta$ and an infinite duration, on the scale of $\tvt$ in
eq.~(\ref{scaling_temps}) with $\expDeltatemps=1$, is needed until this
critical distance is reached. Notice that the \emph{critical window}
here coincides with the critical window of the static phase transition
for 2-SAT~\cite{bollobas-borgs-chayes-2sat-scaling-window}. % 
remember, deals with the mere existence of satisfying % 
rather than with the ability of some algorithm to find them)

 Finally, $\tst$ is such that $\adeux(t)=1$, therefore \[ \tst =
\epsalpha/\beta(p) N^{-1/3} + \mathcal{O}(N^{-2/3}) \] from
eq.~(\ref{solution_c2c3_UC_c2}), and the relevant scaling for time is
\beq
 \label{scaling_temps_2sat}
 T = \left[ \tvt + \epsalpha/\beta(p) \right] N^{2/3}
\eeq
according to eq.~(\ref{scaling_temps}), where we replaced the notation
$t_0$ with $\tvt$ to emphasize that this scaling is proper to the 2-SAT
family.

 We have to solve eq.~(\ref{edp_pi}) with this choice of scales and with
proper initial and boundary conditions. Define
\beq
 \psuccess(\to \tvt) := \pi(x_1=0, \tvt) .
\eeq
This is the probability that the algorithm detects no contradiction from
$T=0$ up to the (rescaled) time $\tvt$. We shall send 
$\tvt \to +\infty$ at the end. In the case of the greedy UC algorithm,
$1/[2(1-\tst)] = 1/2 + \mathcal{O}(N^{-1/3})$ and $e_2(\tvt) = \beta(p)
\tvt + \mathcal{O}(N^{-1/3})$, so that eq.~(\ref{edp_pi}) reads:
\bea & &
 \label{edp_pi_2sat}
 \partial_{\tvt} \pi(x_1, \tvt) =
 \frac{1}{2} \partial_{x_1} \pi(x_1,\tvt) +
    \nonumber \\ & & \qquad
  \left[ \beta(p) \tvt x_1 + \frac{x_1^2}{2} \right] \pi(x_1,\tvt)
  - \sigma(\tvt) x_1 .
\eea

 In practice, we find it easier to perform an inverse Laplace transform
of eq.~(\ref{edp_pi}) before solving it; this amounts to work with
probability density functions (PDFs) rather than with generating
functions. In particular, the difficulty
of computing $\sigma$ that appears in eq.~(\ref{edp_pi}) is turned 
into a boundary condition on the PDF that is easier to deal with. 
%One might ask why we
%used generating functions so far, instead of working directly with PDFs.
%The reason is that the fundamental equation,
%eq.~(\ref{equation_evolution_fonctgen_c1}), is much easier to write with
%generating functions than with PDFs, thanks to the absence of
%correlations between time steps 1, 2, ..., $N$: up to the fluctuations
%of $C_2$, which we may neglect as shown previously,
%eq.~(\ref{equation_evolution_fonctgen_c1}) is just a linear recursion
%relation with explicit time-dependence.

 If we plug the critical scaling of $X_1$, eq.~(\ref{scaling_x1_et_e2}),
into the definition of the generating function $G_1(X_1)$ of $C_1$:
\bea
  G_1(X_1) & := & \sum_{C_1=0}^{+\infty} X_1^{C_1} P(C_1)
    \nonumber \\
  & = & \sum_{C_1=0}^{+\infty} e^{-x_1 C_1 N^{-1/3} + \mathcal{O}(N^{-2/3})} 
      P(C_1)
\eea
and (in an heuristic way) change the discrete sum on $C_1$ into an
integral on $c := C_1 N^{-1/3}$, letting $N$ go to $+\infty$:
\beq
 \label{expression_pi_comme_transformee_de_Laplace}
 \pi(x_1) = \int_{0}^{+\infty} e^{-x_1 c} \rho(c) \, \dd c
\eeq
where $\rho(c)$ is the probability density function (PDF) of $c$, we see
that $\pi(x_1)$ is the Laplace transform of $\rho(c)$ with respect to
$c$. Here we have dropped the time dependence, but $\rho(c,\tvt)$ is
actually a function of $c$ and $\tvt$ and the Laplace transform is taken
at fixed time.

 In terms of $\rho(c,\tvt)$, eq.~(\ref{edp_pi_2sat}) translates into
\beq
 \label{edp_rho_2sat}
 \partial_{\tvt} \rho(c,\tvt) = \frac{1}{2} \partial_c^2 \rho(c,\tvt) +
  \beta(p)\; \tvt \;\partial_c \rho(c, \tvt)
  - \frac{1}{2} c \;\rho(c, \tvt) \ .
\eeq
This inverse Laplace transform can be performed only if the limit when
$x_1 \to +\infty$ of the r.h.s. of eq.~(\ref{edp_pi_2sat}) is zero. 
Writing from (\ref{expression_pi_comme_transformee_de_Laplace}) the
asymptotic expansion for $\pi(x_1,\tvt)$ in terms of the density of
clauses and its derivatives at the origin $c=0$, 
\beq
 \pi(x_1, \tvt) = \rho(0, \tvt)/x_1 +
   \partial_c \rho(0, \tvt) / x_1^2 + o(1/x_1^2)
\eeq
and plugging it into eq.~(\ref{edp_pi_2sat}), we find that
$\sigma(\tvt) = 1/2 \; \rho(0, \tvt)$ and 
$\rho(c, \tvt)$ satisfies the boundary condition:
\beq
 \label{condition_bord_2sat}
 \frac{1}{2} \partial_c \rho(0,\tvt) + \beta(p)\; \tvt \; \rho(0,\tvt) = 0 .
\eeq
Conversely, one verifies that eq.~(\ref{edp_rho_2sat}) supplemented with
the boundary condition eq.~(\ref{condition_bord_2sat}) leads by direct
Laplace transform to eq.~(\ref{edp_pi_2sat}) where $\sigma(\tvt)$ is
replaced with $\rho(0, \tvt)/2$.

Eq.~(\ref{edp_rho_2sat}), supplemented with eq.~(\ref{condition_bord_2sat}), 
is a reaction-diffusion equation on the semi-infinite axis of $c =
C_1/(N-T)^{1/3} > 0$. At the initial time step $T=0$, \ie $\tvtinit = -
\epsalpha / \beta(p)$, there are no 1-clauses, so that $\pi(x_1,
\tvtinit)=1$ for all $x_1$ and $\rho(c, \tvtinit)$ is a Dirac $\delta$
distribution centered on $c=0$: the diffusing particles all sit on the
$c=0$ point. Then, they start diffusing (second-derivative term in
Eq.~(\ref{edp_rho_2sat})) because new 1-clauses are produced randomly
from 2-clauses when variables are assigned by the algorithm. This
diffusion is biased: the drift term $\partial_c \rho(c, \tvt)$ comes
from the tendency of the algorithm to make 1-clauses disappear (to
satisfy them). A picture of this process may be found in the upper-right
inset of Figure~\ref{graphe_rhobarre_2sat}, where the PDF $\rho$ is
shown \emph{after normalization}. The total number of particles is not
conserved: the absorption term $c \rho(c, \tvt)$ results from the
stopping of some runs of the algorithm, those where a contradiction (a
0-clause) is detected.
The probability that no contradiction has been encountered till time
$\tvt$, $\psuccess(\to \tvt) = \pi(0, \tvt) =
\linebreak \int_{c=0}^{+\infty} \rho(c, \tvt) \dd c$, is a decreasing
function of $\tvt$, smaller than unity.

\subsection{Results for 2-SAT and its family}

 Unfortunately, we were not able to solve analytically
eq.~(\ref{edp_rho_2sat}). Our study relies on an asymptotic expansion of
the solution of this equation for large times $\tvt$ and on a numerical
resolution procedure to get results at finite times $\tvt$. This
numerical resolution was in turn helped with an asymptotic expansion at
small times $\tvt$.

 Details about the large times $\tvt$ expansion may be found in 
Appendix~\ref{appendice_2sat_critique_developpement_grand_t23}. In 
short, we find that the probability that the greedy algorithm does not 
stop till time $\tvt$ decays algebraically at large times,
\beq
 \label{decroissance_algebrique_Psucces_2sat}
 \psuccess(\to \tvt) = \int_{0}^{+\infty} \rho(c, \tvt) \dd c
  \quad \propto \quad \tvt^{-\frac{1}{4\beta}} \ .
\eeq
The leading order of the probability of success
at the final time step $T=N$ can be guessed by replacing $\tvt$ with 
$N^{1/3}$:
\beq
 \label{annonce_Psucces_2sat}
 \Psuccess[\alphaR(p),p] \propto N^{-\frac{1}{12\beta}} \ ,
\eeq
an intermediate behaviour between the success
(finite \linebreak $\Psuccess$) and failure ($-\ln \Psuccess \propto N$)
situations defined in Section~\ref{suctofaisec}.

The proportionality factor in
eq.~(\ref{annonce_Psucces_2sat}) can be calculated through a numerical
resolution of eq.~(\ref{edp_rho_2sat}) for finite values of $\tvt$.

\subsubsection{Numerical resolution of eq.~(\ref{edp_rho_2sat}) at 
finite times $\tvt$}

We have solved the reaction-diffusion-like eq. (\ref{edp_rho_2sat}) 
thanks to a standard
numerical resolution scheme (the Crank-Nicholson method) after some
preliminary steps.
 First we discretized both time and `space' (the semi-infinite axis of
$c$). It is convenient to consider finite-support functions, 
and we have tried the changes of
variables $b=1/(c+1)$ and $b = \exp(-c)$; the latter turned out to be
better. The range $]0,1]$ for $b$ was
discretized into $\mathcal{N}$ points. The Crank-Nicholson method allows
us to take a time step $1/\mathcal{N}$ for the numerical resolution
(quite efficient as compared to the time step $1/\mathcal{N}^2$
for Euler's method), provided that the Courant condition is
respected. With eq.~(\ref{edp_rho_2sat}) this is not the case, since the
coefficient of the drift term, $\beta(p) \tvt$,
is not bounded with growing $\tvt$. We actually consider $\rhot(c, \tvt) =
\exp(-\beta c \tvt - \beta^2 \tvt^3/6) \rho(c, \tvt)$ rather than
$\rho(c, \tvt)$ so that the Courant condition is satisfied. What we have
to solve now is
\bea & &
 \label{edp_rhot_de_b}
 \partial_{\tvt} \rhot(b, \tvt) =
 \frac{b^2}{2} \partial_b^2 \rhot(b, \tvt) +
 \frac{b}{2} \partial_b \rhot(b, \tvt)
  \nonumber \\ & & \qquad \quad
 - \left[ \beta(p) - \frac{1-X_0}{2} \right] \ln(b) \rhot(b, \tvt)
\eea
with $X_0=0$ and for $0 < b \le 1$, and with the boundary condition
\beq
 \partial_b \rhot(1, \tvt) - \beta(p) \tvt \rhot(1, \tvt) = 0 \ .
\eeq

At initial time ${\tvt}_\mathrm{init} = -\epsalpha/\beta(p)$, $\rho(c,
{\tvt}_\mathrm{init}) = \delta (c)$. 
The most relevant term in
eq.~(\ref{edp_rho_2sat}) is the diffusion term, and we expect $c$ to grow 
like $\sqrt{\tvt-{\tvt}_\mathrm{init}}$ typically~\footnote{See also
eq.~(\ref{developpement_cbarre_tvt_petit}).}. Therefore, we
start our numerical resolution at time ${\tvt}_\mathrm{init} + 
\mathcal{N}^{-1/2}$ so
that a finite number of discretization points (instead of just
the point on the $b=1$ boundary) share the support of $\rhot$. 
The initial condition is given by a short-time series expansion of the
solution of eq.~(\ref{edp_rho_2sat}). Details about this expansion are
found in Appendix~\ref{appendice_2sat_critique_developpement_petit_t23}.

\begin{figure}[htbp]
\begin{center}
\includegraphics[width=\largeurgraphes]{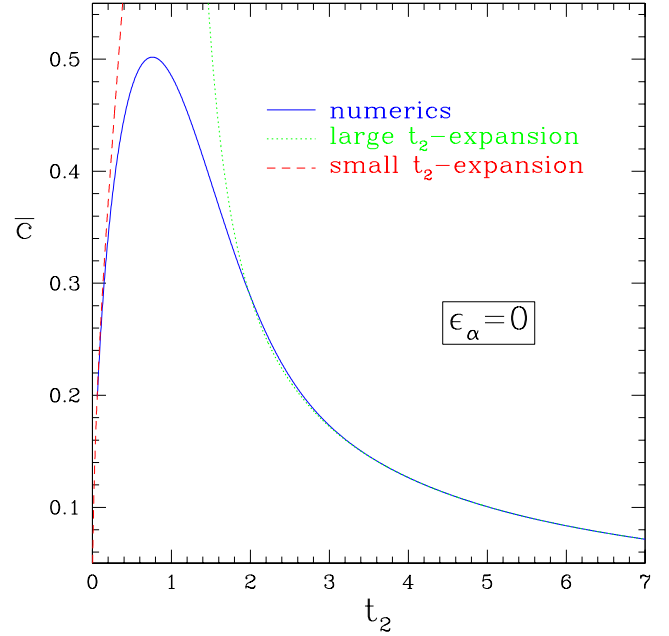}
\end{center}
\caption{\label{figure_cbarre_2sat} Numerical data for $\cbarre(\tvt)$ 
in the case of critical 2-SAT ($\beta=1$ and $\epsalpha=0$), obtained by 
numerically solving eq.~(\ref{edp_rho_2sat}). Comparisons with the
analytical small- and large-times asymptotic expansions are shown.}
\end{figure}

 We first tested our program by studying the linear equation
eq.~(\ref{edp_rho_2sat}), or more precisely eq.~(\ref{edp_rhot_de_b})
with $X_0$ set to 1. In this case, the distribution $\rho$ is normalized
to unity. We observed that this conservation rule is fulfilled by our numerical
resolution scheme up to a small
deviation of order $1/\mathcal{N}$, diverging with the simulation
time $\tvt$.  Therefore, we must be careful in our choice for  the
final time of the simulation. Moreover, when we plotted the conditional
average $\cbarre(\tvt)$, we found a very good agreement with the
analytical expansions at small and large times $\tvt$. This agreement
was also observed for the non-linear equation eq.~(\ref{edp_rhot_de_b})
--- see Figure~\ref{figure_cbarre_2sat}. Therefore, we think that our 
numerical results are quite reliable, at least on finite time ranges.

\subsubsection{Probability of success in the critical time regime and
the scaling function $H$}

 As the numerical precision on $\cbarre(\tvt)$ is greater than on the
total probability~\footnote{For instance, in the case of $X_0=1$, with
$\mathcal{N}=100$ and at $\tvt=6.5$, the relative error on the total
probability equals 6.8\%, whereas it is only 2.5\% on $\cbarre$.}, we
have calculated the probability of success through the numerical results
for $\cbarre$. The probability $\pib(0, \tvt)$ is indeed related to the
values of $\cbarre(\tvt)$ by integrating eq.~(\ref{edp_rho_2sat}) over
$c$ from 0 to $+\infty$, see
 eq.~(\ref{lien_mu_pi}):
\beq
 \partial_{\tvt} \ln \psuccess(\to \tvt) = - \cbarre(\tvt)/2
\eeq
where we have used the boundary condition
eq.~(\ref{condition_bord_2sat}). In practice, we integrated $\cbarre$
numerically from $\tvtinit$ (which depends on $p$ and $\epsalpha$) to
$\tvt \approx 5$, and used our large-$\tvt$ expansion (see
Eq.~(\ref{decroissance_algebrique_Psucces_2sat}) and
Appendix~\ref{appendice_2sat_critique_developpement_grand_t23}) for
$\tvt > 5$, which yields
\beq
 \label{critical_2sat_expression_pi_de_t23}
 - \ln \psuccess(\to \tvt; \epsalpha, p) = \frac{\ln \tvt}{4 \beta(p)} +
  H[\epsalpha,\beta(p)] + \mathcal{O}\left(\tvt^{-3}\right)
\eeq
with the following values for $H$ at criticality ($\epsalpha=0$):
$0.24371 \pm 10^{-5}$ for $p=0$, $0.24752 \pm 10^{-5}$ for $p=1/7$,
$0.20157 \pm 10^{-5}$ for $p=1/4$, $-0.208 \pm 10^{-3}$ for $p=1/3$.
These values are extrapolations to $\mathcal{N}=\infty$ of numerical
results for a number of discretization points $\mathcal{N}$ up to 1600. 
We checked that changing the end time of numerical integration from
$\approx 5$ to $\approx 10$ did not change this extrapolation (although
it notably affects the numerical integral for values of $\mathcal{N} \ll
1000$).

 The behaviour of $\psuccess$ in the critical time range is illustrated
in the inset of Figure~\ref{graphe_mu_2sat} in the case of 2-SAT. For
$N=+\infty$ (continuous line),
eq.~(\ref{critical_2sat_expression_pi_de_t23}) yields the large-$\tvt$
asymptote while data for small $\tvt$ come from numerical results for
eq.~(\ref{edp_rho_2sat}). This compares well with results for finite
sizes $N$ from 25 to 1000 (points), even though the finite-size effects
in $N^{-1/3}$ are large; for fixed $\tvt$, finite-size data converge to
the $N=+\infty$ result but, on a series of data for fixed $N$, there is
a cross-over from the time regime $T = \tvt N^{2/3}$ to the time regime
$T= t N$ (where the correct scaling is illustrated in the main plot). 
The finite-$N$ data were computed by direct solution of the evolution
equation~(\ref{equation_evolution_fonctgen_c1}) for the generating
function $G_1$ of $C_1$, thanks to the technique exposed in
Appendix~\ref{appendice_solution_equation_fonctgen_C1}. They have no
Monte-Carlo error but don't take into account the Gaussian fluctuations
of $c_2$.

\begin{figure}[htbp]
\begin{center}
\includegraphics[width=\largeurgraphes]{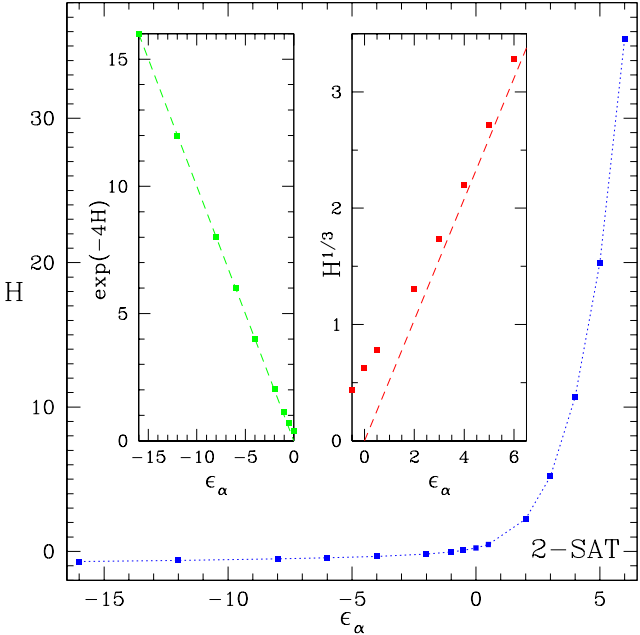}
\end{center}
\caption{\label{figure_constante_de_ea_2sat} Numerical results (dots)
for the scaling function $H$ of the probability $\psuccess$ in the
critical time scale  as a function of the relative distance
$\epsalpha$ to the critical constraint-per-variable ratio, in
the case of 2-SAT ($p=0$, $\beta=1$). The error bars are smaller than
the symbol's size. The dotted line is a guide for the eye. \textbf{Left
inset:} $y:=\exp(-4 H)$ is plotted vs. $\epsalpha$ to show that, when
$\epsalpha \to -\infty$, $H \sim -\ln(-\epsalpha)/4$. The straight
line has equation $y=-\epsalpha$. \textbf{Right inset:} $H^{1/3}$ is
plotted vs. $\epsalpha$ to show that, when $\epsalpha \to +\infty$, $H
\propto \epsalpha^3$. The straight line is a tentative linear fit.}
\end{figure}

 $H$ may be viewed as a scaling function of the parameter $\epsalpha$
for the probability not to find a contradiction in the time scale 
$T = \Theta(N^{2/3})$. 
We computed along the same lines several values of $H$ for various
$\epsalpha \ne 0$ at fixed $p=0$. Results are shown in
Figure~\ref{figure_constante_de_ea_2sat}.

% The result eq.~(\ref{critical_2sat_expression_pi_de_t23}) is an
%asymptotic expansion for large $N$ with fixed $\tvt$. Although it
%doesn't give readily the success probability of the algorithm at the
%very final time step $T=N$ (see the next paragraph), the leading order
%in $N$ in the expression of $-\ln \Psuccess$ may be found by setting
%$\tvt = t N^{1/3}$:
%\beq
% \label{proba_succes_2sat_0_a_1_extrapolation_regime_t23}
% -\ln \psuccess(\to t N^{1/3}) =
% \frac{1}{12 \beta(p)} \ln N + \frac{\ln t}{4 \beta(p)} +
% H[\epsalpha, \beta(p)] + \mathcal{O}\left(\frac{1}{t^3 N}\right)
%\eeq
%which shows that $\Psuccess$ decays algebraically with $N$.

 Let us check heuristically that the success and failure cases are
recovered from the critical results when $\epsalpha$ tends to $-\infty$
and $+\infty$ respectively; this amounts to precise the
large-$\epsalpha$ behaviour of the scaling function $H$. $\epsalpha$
fixes the initial date $\tvtinit = -\epsalpha/\beta(p)$ in the time
scale of $\tvt$ and this in turn influences the value of $H$.

 For $\alphainit < \alphaR(p)$ \ie $\epsalpha = -\delta N^{1/3}$ for
some fixed $\delta>0$, $\cbarre(\tvt)$ reaches very quickly its
asymptotic regime for large $\tvt$: $\cbarre(\tvt) \sim 1/(2 \beta
\tvt)$, and we obtain
\bea
 - \ln \psuccess(\tvtinit = \delta N^{1/3} \to t N^{1/3}) \approx
   & & \nonumber \\
 \int_{\delta/\beta N^{1/3}}^{t N^{1/3}} \frac{\dd\tvt}{4 \beta \tvt} =
 \frac{1}{4 \beta} \left( \ln t - \ln \delta \right) . & &
\eea
$\psuccess$ is finite, as expected in the success case. 
This computation shows that the scaling function $H$
should behave like $-\ln(-\epsalpha)/(4 \beta)$ for large negative
$\epsalpha$; this is confirmed numerically in the left inset of
Figure~\ref{figure_constante_de_ea_2sat}.

Above the critical threshold, 
for $\epsalpha = + \delta N^{1/3}$ with $\delta>0$,
$c$ is driven away from 0 at speed $\approx \delta/\beta N^{1/3}$ from
time $\tvt = - \delta/\beta N^{1/3}$ to time 0 (see
eq.~(\ref{edp_rho_2sat})), hence $c\approx \delta^2/\beta^2
N^{2/3}$ for a duration $\approx \delta/\beta N^{1/3}$. Hence,
\bea
 - \ln \psuccess(\tvtinit = -\delta N^{1/3} \to \tvt = t N^{1/3}) \approx
   & & \nonumber \\
 \frac{1}{12 \beta} \ln N + \frac{\delta^3}{\beta^3} N \ . & &
\eea
$\psuccess$ vanishes exponentially with $N$ as expected. $H$ should
behave like $\epsalpha^3$ for large positive $\epsalpha$; this is
confirmed numerically in the right inset of
Figure~\ref{figure_constante_de_ea_2sat}.

\subsubsection{Matching together critical and non-critical time scales
--- final result for $\Psuccess$}

% Formally and forgetting for the moment the perturbations due to the
%fluctuations of $C_2$ around its typical value (see
%eq.~(\ref{principe_approximation_concentration_C_j})), the result
Equation~(\ref{critical_2sat_expression_pi_de_t23}) may be written as,
setting $\tvt + \epsalpha/\beta = t N^{1/3}$,
%\beq
% \label{proba_succes_2sat_0_a_t_selon_t23}
% -\ln \Psuccess(T=0 \to (\tvt + \epsalpha/\beta) N^{2/3}) =
%  -\ln \psuccess(\to \tvt) + N^{-1/3} Q(\tvt,N)
%\eeq
%or, setting $\tvt + \epsalpha/\beta = t N^{1/3}$ and using the results
%in the critical time scale,
%eq.~(\ref{critical_2sat_expression_pi_de_t23}),
\bea
 \label{proba_succes_2sat_0_a_t_selon_t}
 -\ln \Psuccess(T=0 \to t N) =
 \frac{1}{4 \beta(p)} \left(\frac{1}{3} \ln N + \ln t \right) +
   & & \nonumber \\
 H[\epsalpha, \beta(p)] + \mathcal{O}\left(\frac{1}{t^3 N}\right) +
 N^{-1/3} Q(t N^{1/3}, N) \quad & &
\eea
where $Q(\tvt, N)$ is bounded when $N \to +\infty$ with fixed
$\tvt$~\footnote{$Q$ is like the sum of the asymptotic expansion of
$-\ln \Psuccess$ in powers of $N^{-1/3}$ without the leading term. Our
aim is to precise how $Q(\tvt, N)$ behaves when both $\tvt$ and $N$ go
to $+\infty$.}.

\begin{figure}
\begin{center}
\includegraphics[width=\largeurgraphes]{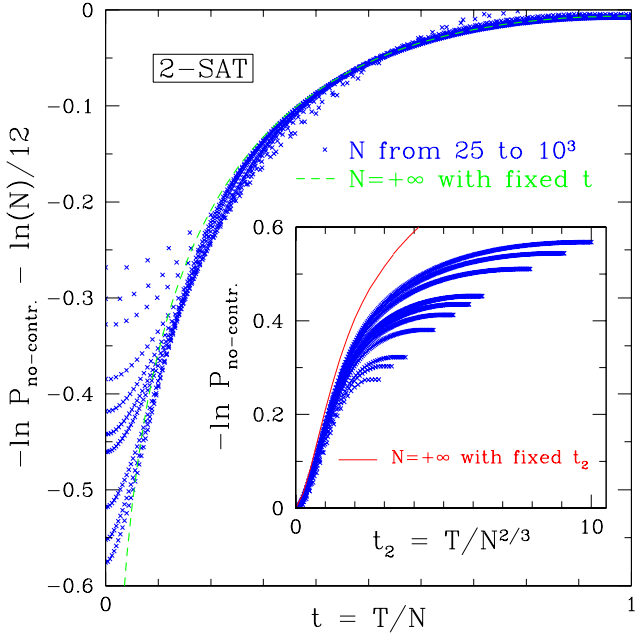}
\end{center}
\caption{Illustration of the two relevant time regimes for the
probability $\psuccess$ that no contradiction is found between $T=0$ and
some step $T$. Points are data for the exact solution of
eq.~(\ref{equation_evolution_fonctgen_c1}) in the case of 2-SAT at
criticality ($\alphainit=1$) for sizes $N=25$, 37, 51, 101, 151, 201,
251, 501, 751 and 1001 (increasingly close to the $N=+\infty$ asymptotic
lines). \textbf{Main plot:} $- \ln \psuccess - \frac{1}{12} \ln N$ is
plotted vs. $t=T/N$. This quantity has a well-defined continuous limit
when $N \to +\infty$ with fixed $t>0$ (dashed line); see text for the
computation of this limit. Because of the cross-over from the critical
time-regime, convergence is non-uniform; finite-size effects are huge
except if $T \gg N^{2/3}$. \textbf{Inset:} $- \ln \psuccess$ is plotted
vs. $\tvt=T/N^{2/3}$; it has a well-defined continuous limit when $N \to
+\infty$ with fixed $\tvt > 0$ (solid line). Convergence is non-uniform;
finite-size effects are huge in the cross-over regime $T \gg N^{2/3}$.
Data for the solid line are from numerical solution of
eq.~(\ref{edp_rho_2sat}) and the asymptotic
expression~(\ref{critical_2sat_expression_pi_de_t23}).}
\label{graphe_mu_2sat}
\end{figure}

 The behaviour of $\psuccess$ for times $T$ of the order of $N$ is
illustrated in the main part of Figure~\ref{graphe_mu_2sat} (for 2-SAT),
where $- \ln \psuccess - \frac{1}{12} \ln N$ is plotted as a function of
$t$. The dashed line is the $N=+\infty$ result $\frac{1}{4}(\ln t -
t)+H(0,1)$ (see below for the expression of $Q$). Data for finite-sizes
$N$ (points) compare well with this result if $T \gg N^{2/3}$; otherwise
there are strong finite-size effects and the critical time regime
results are relevant (see inset).

 Outside the critical regime, the probability that no contradiction is
found can be calculated along the lines of Section
\ref{section_success_case}. In the $N \to +\infty$ limit with fixed
ratio $T/N$, the probability that no 0-clause is found between times
$0<t<1$ and $1$ satisfies
\bea & &
 \label{proba_succes_2sat_t_a_1_selon_t1}
 -\ln \psuccess(T = t N \to N) =
    \nonumber \\ & & \qquad
  \int_{t}^{1} \frac {d\tau}{4 (1-\tau)}
   \, \frac{\adeux(\tau)^2}{1-\adeux(\tau)}
  +\mathcal{O} \left(N^{-1/3}\right) =
    \nonumber \\ & & \qquad
  -\frac{\ln t}{4 \beta(p)}  + \int_t^1 d\tau f(\tau)
  +\mathcal{O} \left(N^{-1/3}\right)
\eea
since the reasoning that led to
eqs.~(\ref{developpement_G1_succes_ordre1enx1}) and
(\ref{expr_proba_succes}) is still valid here: if $t_\mathrm{init}>0$
and $\alphainit = \alphaR(p) (1 + \epsalpha N^{-1/3})$ according to
eq.~(\ref{scaling_alpha_2sat}), we know that $\adeux(t)$ is bounded away
from 1 when $N \to +\infty$. Notice that the subdominant term in
eq.~(\ref{proba_succes_2sat_t_a_1_selon_t1}) is not of order $1/N$ like
in eq.~(\ref{developpement_G1_succes_ordre1enx1}) because we approximate
$\alphainit$ with $\alphaR$. The expression for function $f$,
\beq
 f(t) := - \frac{1}{4} - \frac{3 p t}{8(1-p)}
         - \frac{9 p^2}{4(2-5p)(3 p t + 2 - 5 p)} \ .
\eeq
is found from eqs.~(\ref{definition_adeux}) and
(\ref{solution_c2c3_UC_c2}).

% Eq.~(\ref{proba_succes_2sat_t_a_1_selon_t1}) may also be written with
%$Q$, using eq.~(\ref{proba_succes_2sat_0_a_t_selon_t}) at times $1$ and
%$t$ and taking the difference:
%\beq
% \label{proba_succes_2sat_t_a_1_selon_t23}
% -\ln \Psuccess(T = t N \to N) =
%  \frac{1}{4 \beta(p)} \ln \frac{1}{t}
%  + \mathcal{O}\left(N^{-1}\right)
%  + N^{-1/3} \left[ Q(N^{1/3}, N) - Q(t N^{1/3}, N) \right]
%\eeq
 Comparing eq.~(\ref{proba_succes_2sat_t_a_1_selon_t1}) and
%eq.~(\ref{proba_succes_2sat_t_a_1_selon_t23})
eq.~(\ref{proba_succes_2sat_0_a_t_selon_t}) yields
\beq
 N^{-1/3} Q(t N^{1/3}, N) = F(t) + \mathcal{O}\left( N^{-1/3} \right)
  + \mathcal{O}\left( t^{-3} N^{-1} \right)
\eeq
where $F$ is a primitive of $f$. Using this expression for $Q$ in
eq.~(\ref{proba_succes_2sat_0_a_t_selon_t}), setting $t = N^{-1/3}(\tvt
+ \epsalpha/\beta)$ and letting $\tvt$ go to 0 shows that $F(0)=0$.
%In order to find out which
%primitive is relevant, substitute this expression for $Q$ into
%eq.~(\ref{proba_succes_2sat_0_a_t_selon_t23}):
%\beq
% -\ln \Psuccess[T = 0 \to (\tvt + \epsalpha/\beta) N^{2/3}] =
%  - \ln \psuccess(\to \tvt) + F(\tvt N^{-1/3}) +
%  \mathcal{O}\left( N^{-1/3} \right)
%\eeq
%and let $\tvt$ go to its initial value $-\epsalpha/\beta(p)$. Of course,
%at the very initial time $T=0$, no 0-clause was produced yet and
%$\Psuccess = \psuccess = 1$, hence $F(0)=0$. Finally, the total
%probability of success of the greedy UC=UP+R algorithm is given by (this 
%is the central result of this subsection):
Finally, the probability of success $\Psuccess$ is
\beq
 \label{moinslnPsuccess_famille_2sat}
 -\ln \Psuccess = \frac{\ln N}{12 \beta(p)} +
  H[\epsalpha, \beta(p)] + F(1) + \mathcal{O}\left( N^{-1/3} \right)
\eeq
with
\beq
 F(1) = \frac{\ln \beta(p)}{\beta(p)} \frac{3p}{8(1-p)}
        - \frac{4-p}{16(1-p)} .
\eeq

 The corrections due to the fluctuations of $C_2$, temporarily left
aside, are of the order of, from
eq.~(\ref{principe_approximation_concentration_C_j}), \[
\mathcal{O}(N^{\expdeviationcjs-4/3} \ln N)\] (without the $\ln$ factor
if $X_0=1$) and \[ \mathcal{O} \left[ N^{3/2-\expdeviationcjs} \exp
\left( -N^{2\expdeviationcjs-1}/2 \right) \right] \] where $\delta$ has
to be in the range $]1/2, 2/3[$. They are negligible w.r.t. all other
terms of eq.~(\ref{moinslnPsuccess_famille_2sat}). Notice that
$\expdeviationcjs$ has opposite effects on the two corrections, as was
anticipated in the discussion following
eq.~(\ref{principe_approximation_concentration_C_j}).

 Let us give precise values for some special cases, to illustrate the
predictive power of our computation, although we have no analytical
formula for $H$. For the greedy algorithm ($X_0=0$) at the critical 
point ($\alphaR(p)$), $-\ln \Psuccess$ equals, up to $\mathcal{O}\left( 
N^{-1/3} \right)$,
\bea
 \label{resultats_lnp_famille_2sat_p0}
 & & \ln N /12 + 0.24371 \pm 10^{-5} - 1/4 \\
 & & \ln N /9 + 0.24752 \pm 10^{-5} -9/32 +\ln(3/4)/12 \quad \\
 & & \ln N /6 + 0.20157 \pm 10^{-5} -5/16 -\ln(2)/4 \\
 \label{resultats_lnp_famille_2sat_p13}
 & & \ln N /3 - 0.208 \pm 10^{-3} -11/32 - 3\ln(2)/2
\eea
for 2-SAT, 2+1/7-SAT, 2+1/4-SAT and 2+1/3-SAT respectively.

\begin{figure}[htbp]
\begin{center}
\includegraphics[width=\largeurgraphes]{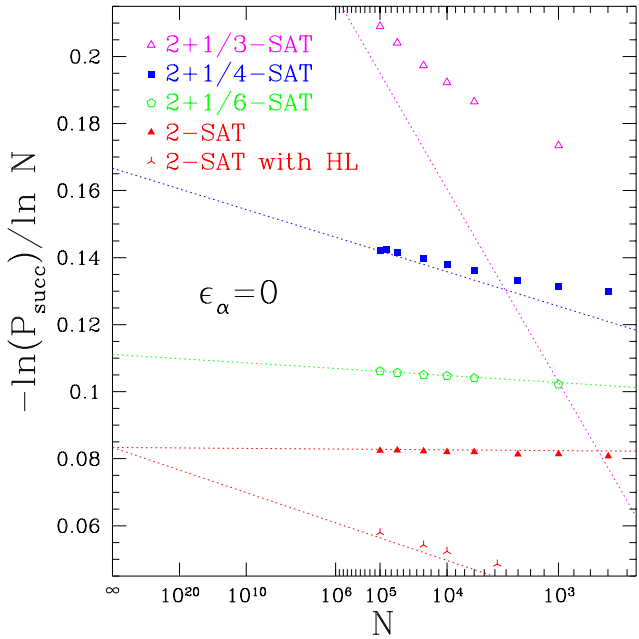}
\end{center}
\caption{\label{figure_comparaison_psucces_ea0_2sat} Comparison of
eqs.~(\ref{resultats_lnp_famille_2sat_p0})--(\ref{resultats_lnp_famille_2sat_p13})
(lines) with empirical estimates (symbols) for the probability of
success of the greedy UC algorithm on instances of the random $2+p$-SAT
problem at the critical initial constraint-per-variable ratio
$\alphaR(p)$ (\ie $\epsalpha=0$) for p=0, $p=1/7$, $p=1/4$ and $p=1/3$,
and data for the probability of success of the greedy HL
algorithm~\cite{kaporis-kirousis-lalas-hl-cl} on instances of critical
2-SAT. The error bars are smaller than the symbols' sizes. Data for
$-\ln \Psuccess/\ln N$ are plotted against $1/\ln N$, and the straight
lines come from our asymptotic analytical results,
eqs.~(\ref{resultats_lnp_famille_2sat_p0}--\ref{resultats_lnp_famille_2sat_p13}) 
(ignoring the $\mathcal{O}(N^{-1/3})$ terms), except for HL where it is 
a tentative extrapolation to $N=+\infty$.}
\end{figure}

 Figure~\ref{figure_comparaison_psucces_ea0_2sat} compares these results
with empirical success probabilities, obtained by running the greedy UC
algorithm on a large number ($4.10^5$ to $3.10^6$) of instances of
random 2+p-SAT at critical initial clauses-per-variable threshold
$\alphaR(p)$ with sizes up to $N=10^5$. Interpretation of the data could
be difficult because finite-size effects are strong. But if we take into
account the finite corrections to the $\ln N$ terms in
eqs.~(\ref{resultats_lnp_famille_2sat_p0})--(\ref{resultats_lnp_famille_2sat_p13}),
a very good agreement is found.

\subsubsection{The critical distribution of $C_1$}
\label{section_distribution_C1_2sat}

\begin{figure}[htbp]
\begin{center}
\includegraphics[width=\largeurgraphes]{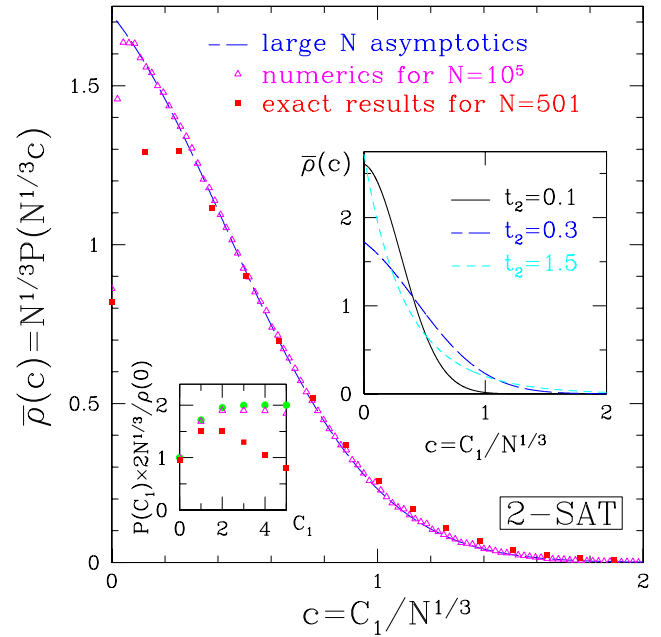}
\end{center}
\caption{\label{graphe_rhobarre_2sat} The normalized PDF $\rhob$ of
$c=C_1/N^{1/3}$ for 2-SAT at the critical initial clauses-per-variable
ratio $\alpha=1$ ($\epsalpha=0$) and at $\tvt=0.3$. Long-dashed line:
limit result for $\rhob$ at $N=+\infty$. Dots: numerical results from
$8.10^5$ runs of the UC algorithm for $N=10^5$ ($\triangle$; the error
bars are smaller than the symbol's size) and results from the direct
solution of eq.~(\ref{equation_evolution_fonctgen_c1}) for $N=501$
($\blacksquare$) --- see
appendix~\ref{appendice_solution_equation_fonctgen_C1}.
\textbf{Lower-left inset:} probability $P$ (expressed in $N^{-1/3}
\rho(0)/2$ units) that $C_1$ takes finite values at the $N=+\infty$
limit ($\bullet$). This shows how the discontinuity, at the scale of
$c$, between $N^{1/3} P(C_1=0)=\sigma$ and $\lim_{c \to 0+} \rhob(c) = 2
\sigma$ is resolved at the scale of $C_1$. Numerical data (same as for
the main plot) converge to $P$ when $N \to +\infty$. \textbf{Upper-right
inset:} Normalized distributions $\rhob$ for $N=+\infty$ at times
$\tvt=0.1$ (solid line), $\tvt=0.3$ (long dashes) and $\tvt=1.5$ (short
dashes).}
\end{figure}

 In the critical time regime ($T$ of the order of $N^{2/3}$), the PDF of
$c=C_1/N^{1/3} > 0$ is the solution
$\rho$ of eq.~(\ref{edp_pi}). As a special case, the probability that no
1-clause is present is $\sigma \sim
N^{-1/3} \rho(0)/2$. Convergence to this distribution, for $c>0$ on one
hand and for $c=0$ on the other hand, is observed numerically --- see
Figure~\ref{graphe_rhobarre_2sat}. The convergence is not uniform in the
neighbourhood of $c=0$, which is expected since the distribution $\rho$
is singular in $c=0$. There is rather a cross-over from the regime where
$C_1 \ll N^{1/3}$ to the regime where $C_1$ is of the order of
$N^{1/3}$.
%To precise the probability that $C_1$ takes a finite value,
%write the generating function $G_1$ of $C_1$ with a correction term:
%\beq
% \label{expression_G1_2sat_avec_correction_U}
% G_1(x_1) = \pi(x_1) + N^{-1/3} U(x_1,N)
%\eeq
%after eq.~(\ref{definition_pibarre})~\footnote{The order of magnitude of
%the correction, $N^{-1/3}$ is known after the neglected terms of
%eq.~(\ref{equation_evolution_fonctgen_c1}) in the study of
%Section~(\ref{section_choix_echelles_critiques}).}, with $U$ bounded
%when $N \to +\infty$ at fixed $x_1$. Setting $x_1 = (1-X_1) N^{1/3}$
%into eq.~(\ref{expression_G1_2sat_avec_correction_U}) and using
%$\pi(x_1) = \rho(0)/x_1 + \mathcal{O}(1/x_1^2)$ for large $x_1$ from
%eq.~(\ref{expression_pi_comme_transformee_de_Laplace}),
%\beq
% \label{expression_G1_2sat_X1_dapres_x1}
% G_1(X_1) = \rho(0)/(1-X_1) N^{-1/3} +
% N^{-1/3} U\left[(1-X_1) N^{1/3}, N \right] +
% \mathcal{O}\left(N^{-2/3}\right) \ .
%\eeq
%On the other hand,
Eq.~(\ref{eq1suc}) yields, going to the $\adeux \to 1$ limit
(well-defined if $X_1>0$),
\beq
 \label{expression_G1_2sat_X1_dapres_X1}
 \pi_0(X_1) = \sum _{C_1} P(C_1) X_1 ^{C_1} 
 = \frac{1-X_1}{1-X_1 \exp(1-X_1)} \pi_0 (0) \ .
\eeq
%Expressing the leading order of $U$ 
%from eqs.~(\ref{expression_G1_2sat_X1_dapres_x1}) and 
%(\ref{expression_G1_2sat_X1_dapres_X1}) and substituting the result in 
%eq.~(\ref{expression_G1_2sat_avec_correction_U}) yields
%\beq
% G_1(x_1) = \pi(x_1) -\rho(0)/x_1 + 2 G(0) N^{1/3}/x_1 -4G_1(0)/3 + o(1)
%\eeq
%hence $G_1(0)=N^{-1/3} \rho(0)/2$ to prevent the divergence of $G_1$ at
%$x_1=0$. This is consistent with eq.~(\ref{raccord_sigma_rho}). Finally,
%for large $N$ with fixed $X_1>0$,
Hence (for large $N$ with fixed $X_1$)
\beq
 G_1(X_1) = N^{-1/3} \frac{\rho(0)}{2} \frac{1-X_1}{1-X_1 e^{1-X_1}} +
  o\left( N^{-1/3} \right) \ .
\eeq
The probabilities that $C_1$ takes the values 0, 1, 2, ... are given by 
the coefficients of the Taylor expansion, in $X_1=0$, of $G_1$ above. It 
is observed that these probabilities converge very quickly to $N^{-1/3} 
\rho(0)$, which coincides with the $c \to 0$ limit of the distribution 
$\rho$ of $c=C_1/N^{1/3}$. These probabilities are plotted in the 
lower-left inset of Figure~\ref{graphe_rhobarre_2sat}, together with 
numerical data for finite sizes. A good agreement is found.

\subsubsection{Universality}
\label{section_universalite_famille_2sat}

 For a given $p<2/5$, all algorithms that use the UP rule fall into the
same universality class (which depends on $p$). They share the result
eq.~(\ref{moinslnPsuccess_famille_2sat}) with common $H$ (but $F(1)$ is
a non-universal correction), and the critical distribution of $C_1$
studied in Section~\ref{section_distribution_C1_2sat}. 

 The reason is two-fold: first, the analysis done so far in
Section~\ref{section_famille_2sat} is still valid for another heuristic
than R, run on random instances of the $2+p$-SAT problem, provided the
critical trajectory starts on the $\adeux=1$ line and is secant to it
with some slope $\beta>0$. Second, the value of $\beta$ \emph{at
criticality} is universal and depends on $p$ only, because, at
criticality, the heuristic is almost never used and UP alone fixes the
slope: even if the resolution trajectories of several heuristics may be
quite different in general (compare \eg Eqs.~(\ref{solution_c2c3_UC_c2})
for R and (\ref{solution_c2c3_GUC_c2}) for GUC), in the critical regime,
the probability that $C_1=0$ and the heuristic is used is of the order
of $N^{-1/3}$ only. Most of the time, the UP rule is used, and the
resulting evolution of $c_2$ and $c_3$ is common to all algorithms: the
slope of the critical trajectory is $\beta(p)=(2-5p)/[2(1-p)]$ as for
UC. We verified this by direct computation from
eq.~(\ref{solution_c2c3_UC_c2}) for R, eq.~(\ref{solution_c2c3_GUC_c2})
for GUC and the corresponding equations of
reference~\cite{kaporis-kirousis-lalas-hl-cl} for HL and CL.

Figure~\ref{figure_comparaison_psucces_ea0_2sat} shows the agreement of
empirical data for the HL heuristic, used on random 2-SAT instances,
with the scaling of $\Psuccess$ that we derived for the R heuristic.

\section{The 3-SAT class (stretched exponential class)}
\label{section_classe_3sat}

\subsection{Equations and results for 3-SAT and its class}

 We now address the case $p>2/5$. Here, the critical resolution
trajectory starts \emph{below} the $\adeux=1$ line and gets tangent to
it, at point $(p=2/5, \alpha=5/3)$ at a finite time, $\tst \in ]0,1[$. 
From eq.~(\ref{solution_c2c3_UC_c2}), $\adeux(t)$ is locally a parabola
around $\tst$ : $1-\adeux(t-\tst) \propto (t-\tst)^2$.  The critical
resolution trajectory is at distance $\Delta = N^{-1/3}$ of the
$\adeux=1$ line as long as $t-\tst$ is of order $\Delta^{1/2}$: the
exponent $\expDeltatemps$ equals 1/2 here and the relevant equation is
eq.~(\ref{edo_pib}), not eq.~(\ref{edp_pi}) as for 2-SAT. The
computation is easier here (at least for the leading order) since we
have an ordinary differential equation (the time enters into play only
as a parameter of the coefficients of this ODE) rather than a partial
derivatives equation. The relevant scaling for time is
\beq
 \label{scaling_temps_3sat}
 T = \tst N + \tcs N^{5/6}
\eeq
according to eq.~(\ref{scaling_temps}) where we replaced the notation
$t_0$ with $\tcs$ to emphasize that this scaling is proper to the 3-SAT
class.

 As for 2-SAT, the critical regime extends to a non-empty range of
values of $\alphainit$. This critical window is the same: we set
\beq
 \label{scaling_alpha_3sat}
 \alphainit = \alphaR(p) (1 + \epsalpha N^{-1/3})
\eeq
with finite $\epsalpha$. Indeed, if $\alphainit$ is less than
$\alphaR(p)$ by more than $\Delta = N^{-1/3}$, because $\adeux(t)$ is
increasing proportionally with $\alphainit$ (see
Eqs.~(\ref{solution_c2c3_UC_c3}) and (\ref{solution_c2c3_UC_c2})), the
resolution trajectory will be out of the critical region in particular
at the time $\tst$ where $\adeux(t)$ is maximal since $1 - \adeux(t) \gg
\Delta$, and therefore at all times. Conversely, if $\alphainit$ is
above $\alphaR(p)$ by a distance much greater than $\Delta$, at time
$\tst$ $1-\adeux$ is an order of magnitude higher than the critical
distance $\Delta$, which implies that the resolution trajectory would
stay for an infinite duration, on the scale of $\tcs$ in
eq.~(\ref{scaling_temps}) with $\expDeltatemps=1/2$, above the
$\adeux=1$ line -- this would yield numerous contradictions (0-clauses)
and let the probability of success be exponentially small.

\subsubsection{Results for the critical time regime}

 We now have to solve eq.~(\ref{edo_pib}). As for the 2-SAT family, we
prefer to do computations on the (here normalized, or conditioned to
success of the greedy algorithm) PDF $\rhob$ of the stochastic variable
$C_1/N^{1/3}$ rather than on its generating function $\pib$. Performing
an inverse Laplace transform on eq.~(\ref{edo_pib}) yields
\beq
 \label{edo_rhob_3sat}
 0 = \frac{1}{2} \partial_c^2 \rhob(c,\tcs) +
  e_2(\tcs) \partial_c \rhob(c, \tvt) +
  \frac{1}{2(1-\tst)} [\cbarre(\tcs) - c] \rhob(c, \tcs)
\eeq
with the boundary condition
\beq
 \label{condition_bord_3sat}
 \frac{1}{2} \partial_c \rhob(0,\tvt) + e_2(\tcs) \rhob(0,\tvt) = 0 .
\eeq
 The parameters in eqs.~(\ref{edo_rhob_3sat}) and
(\ref{condition_bord_3sat}) are $\tst = 5/6-1/(3p)$ and $e_2(\tcs)= 36
\tcs^2 p^2/(p+2)^2 - \epsalpha$.

 Here, the initial condition is mostly irrelevant: the initial step of
the algorithm, $T=0$ or $\tst \to -\infty$, is far out of the critical
time region (finite $\tst$). When the resolution trajectory enters this
region, the distribution of $C_1$ has already equilibrated to its
critical value and is only subject to `adiabatic' changes during the
crossing of the critical region (Eq.~(\ref{edo_rhob_3sat}) has no time
derivative). Solving the ODE~(\ref{edo_rhob_3sat}) brings out the
explicit critical distribution of 1-clauses.

\begin{figure}[htbp]
\begin{center}
\includegraphics[width=\largeurgraphes]{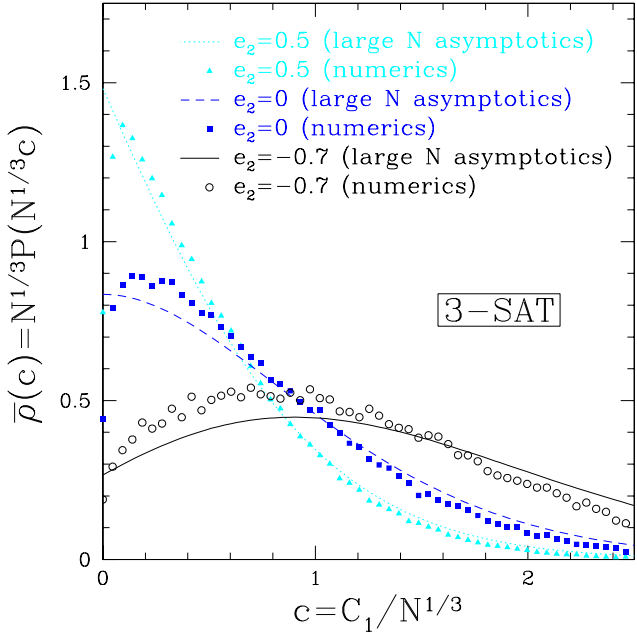}
\end{center}
\caption{\label{graphe_rhobarre_3sat} Critical distributions $\rhob$ for
$c=C_1/N^{1/3}$ in the case of 3-SAT with drifts $e_2=0.5$ (solid line,
$\bigcirc$), $e_2=0$ (dashed line, $\blacksquare$) and $e_2=-0.7$
(dotted line, $\triangle$). Points are from numerical estimates with
size $N=10^4$.}
\end{figure}

 Let $u(c, \tcs) := \rhob(c, \tcs) \exp[e_2(\tcs) c]$,
$k(p)=[6p/(2+p)]^{-1/3}$ and
\beq
 \label{definition_z_3sat}
 z := \left[c -\cbarre(\tcs) + \frac{2+p}{6p} e_2(\tcs)^2 \right] / k(p) \ .
\eeq
Eq.~(\ref{edo_rhob_3sat}) is recast into an equation that admits Airy's
$\Ai$ and $\mathrm{Bi}$ functions as linearly independent solutions:
\beq
 \label{equation-qui-donne-airy}
 \partial_z^2 u(z, \tcs) - z u(z, \tcs) = 0 .
\eeq
See references in~\cite{albright-airy} for studies of
eq.~(\ref{equation-qui-donne-airy}) in the context of (semi-classical)
quantum mechanics or~\cite{groeneboom-airy} in the context of Brownian
motion (similar to our situation). Since $u$ has to vanish for large $z$
(because $\rhob(c,\tcs) \to 0$ for large $c$) whereas $\mathrm{Bi}(z)$
is not bounded for large $z$, $u(z, \tcs) = A(\tcs) \Ai(z)$ where $A$ is
a normalization coefficient. The boundary condition
eq.~(\ref{condition_bord_3sat}) reads
\beq
  \label{condition_bord_3sat_qui_donne_z0}
  \Ai'(z_0)/\Ai(z_0) = - k(p) e_2(\tcs)
\eeq
where $z_0$ is expressed from eq.~(\ref{definition_z_3sat}) with $c=0$. 
Let $\recdifflnai$ be the reciprocal function of $\Ai'/\Ai$. Inverting
eq.~(\ref{condition_bord_3sat_qui_donne_z0}) yields an expression for
$\cbarre(\tcs)$:
\beq
 \label{expression_cbarre_3sat}
 \cbarre(\tcs) = k(p)^3 e_2(\tcs)^2 -
   k(p) \recdifflnai\left[ - k(p) e_2(\tcs) \right]
\eeq
and
\beq
 \rhob(c, \tcs) = A(\tcs) e^{-e_2(\tcs) c}
  \Ai \left\{ k(p) c + \recdifflnai \left[ - k(p) e_2(\tcs)
  \right] \right\} \ .
\eeq
We did not compute explicitly the normalization constant $A(\tcs)$. The
critical distribution $\rhob$ is plotted in
Figure~\ref{graphe_rhobarre_3sat} for 3-SAT ($p=1$) and several values
of $e_2$, to show the influence of the drift on its shape. The agreement
with numerics is good; the same phenomenon in $c=0$ as for the 2-SAT
family is observed: at $N=\infty$, $\rhob$ is singular, and for finite
$N$ there is a cross-over from the regime $C_1 \ll N^{1/3}$ to the
regime $C_1 = c N^{1/3}$ (see Sect.~\ref{section_distribution_C1_2sat}).

 The probability that the greedy algorithm doesn't find a contradiction
in the critical regime from time $\tcs^{(1)}$ up to time $\tcs^{(2)}$
satisfies, according to eq.~(\ref{definition_mu}) and the discussion
preceding it,
\bea & &
  - \ln \psuccess \left( \tcs^{(1)} \to \tcs^{(2)} \right) =
      \nonumber \\ & & \qquad
    N^{1/6} \left[  \mu\left( \tcs^{(2)} \right)
                  - \mu\left( \tcs^{(1)} \right) \right] +
    \mathcal{O}\left(N^{-1/6}\right) \qquad
\eea
where, from eq.~(\ref{lien_mu_pi}) and the initial condition
$\psuccess(T=0)=1$, $\partial_{\tcs} \mu = k(p)^{-3} \cbarre(\tcs)/2$
and $\mu(\tcs) \to 0$ as $\tcs \to -\infty$. In the interesting
situation where $\tcs^{(1)} < 0 < \tcs^{(2)}$, using the $y$ variable
such that $\Ai'(y)/\Ai(y) := - k(p) e_2(\tcs)$ rather than $\tcs$,
\bea & &
 \label{expression_mu_3sat}
 \mu\left(\tcs^{(2)}\right) - \mu\left(\tcs^{(1)}\right) =
 \frac{k(p)^{1/2}}{4} \times
   \nonumber \\ & & \qquad
 \left( \int_{\recdifflnai[k(p) \epsalpha]}^
             {\recdifflnai[- k(p) e_2( -\tcs^{(1)} )]}
      + \int_{\recdifflnai[k(p) \epsalpha]}^
             {\recdifflnai[- k(p) e_2( \tcs^{(2)} )]} \right) \times
   \nonumber \\ & & \qquad
 \frac{\dd y}{\sqrt{k(p) \epsalpha - \Ai'(y)/\Ai(y)}}
 \left[ \frac{\Ai'(y)^2}{\Ai(y)^2} - y \right]^2 . \quad
\eea
Define (the integral is finite)
\beq
 \label{definition_Phi}
 \Phi(x) := \frac{1}{4} \int_{\recdifflnai(x)}^{+\infty}
   \frac{\dd y}{\sqrt{x - \Ai'(y)/\Ai(y)}}
   \left[ \frac{\Ai'(y)^2}{\Ai(y)^2} - y \right]^2 \ .
\eeq
For large positive $\tcs^{(2)}$ (and similarly for large negative 
$\tcs^{(1)}$),
\beq
 \label{expression_mu_3sat_avec_tcspos_grand}
 \mu\left(\tcs^{(2)}\right) - \mu(0) =
 \sqrt{k(p)} \Phi[k(p) \epsalpha] - \frac{k(p)^3}{4 \tcs^{(2)}}
 + \mathcal{O}\left({\tcs^{(2)}}^{-5}\right)
% \\
% \label{expression_mu_3sat_avec_tcsneg_grand}
% \mu(0) - \mu\left(\tcs^{(1)}\right) & = &
% \sqrt{k(p)} \Phi[k(p) \epsalpha] + \frac{k(p)^3}{4 \tcs^{(1)}}
% + \mathcal{O}\left({\tcs^{(1)}}^{-5}\right)
\eeq
where we have used a large-$\tcs$ expansion of
eq.~(\ref{expression_cbarre_3sat}). From eq.~(\ref{scaling_temps_3sat}),
$\tcs$ is of order $N^{1/6}$ at most, thus
eq.~(\ref{expression_mu_3sat_avec_tcspos_grand})
% -- \ref{expression_mu_3sat_avec_tcsneg_grand})
allows one to express $\mu$ in terms of $\Phi$ only, up to corrections
of the order of $N^{-1/6}$. Anticipating that the non-critical time
regime, like the success case in Section~\ref{section_success_case},
brings contributions to $\ln \Psuccess$ of order $\mathcal{O}(1)$ in
$N$, the total probability of success (at time $T=N$) of the greedy UC
algorithm reads
\bea & &
 -\ln \Psuccess[(1+\epsalpha)\alphaR(p),p] =
    \nonumber \\ & & \qquad \qquad
  N^{1/6} 2 \sqrt{k(p)} \Phi[k(p) \epsalpha] + \mathcal{O}(1) \qquad
\eea
where the function $\Phi$ is closely related~\footnote{To lighten
notations here, we have rescaled the argument $x$ and the value of
$\Phi$ by constant coefficients w.r.t. 
reference~\cite{deroulers-monasson-universalite-up-lettre}. Moreover,
the new $\Phi$ is more universal: it is exactly the scaling function at
the tricritical point $p=2/5$, $\alphainit=5/3$ for all heuristics (both
$r_H^\Phi$ and $r_H^{\epsalpha}$ equal one in this case) --- see
section~\ref{section_2.4sat}.} to the universal function introduced in
reference~\cite{deroulers-monasson-universalite-up-lettre}. It may
easily be computed with a mathematical software and is plotted in
Figure~\ref{figure_comparaison_Phi}.

 Let us heuristically check that the success and failure cases are
recovered from the critical results when $\epsalpha$ tends to $-\infty$
and $+\infty$ respectively. Using the asymptotic expansions of $\Ai$ and
$\Ai'$~\cite{abramowitz+stegun,sommabilite-developpement-asymptotique-airy},
the asymptotic behaviour of $\Phi$ is found:
\bea
 \Phi(x) \sim \frac{\pi}{8 \sqrt{-x}}
  & & \text{when } x \to -\infty \nonumber \\
 \Phi(x) = \frac{4}{15} x^{5/2} - a \frac{\sqrt{x}}{2} + o(\sqrt{x})
  & & \text{when } x \to +\infty \nonumber
\eea
where $a \approx -2.338107410$ is the greatest zero of $\Ai$ on the real
axis. Taking now $\epsalpha = N^{1/3} \epsilon$ shows that $-\ln
\Psuccess \approx \pi/(4\sqrt{-\epsilon})$ for $\epsilon<0$, in
agreement with eq.~(\ref{singularite_psuccess_alphafixe_famille3sat}),
and $-\ln \Psuccess \approx 4\epsilon N/15$, in agreement with the
expected failure behaviour.

\subsubsection{Matching critical and non-critical time scales --- final
result for $\Psuccess$}

Here we use a heuristic reasoning, based on what we learned from the
study of the 2-SAT family. On the one hand, the results from the previous
paragraph show that the probability no to find a contradiction between
times $\tst N$ and $T = \tst N + \tcs N^{5/6}$ equals (with an obvious 
convention for negative $\tcs$)
%\bea
% \label{proba_succes_3sat_t56_a_tst_selon_t56}
%  - \ln \psuccess( T= \tst N + \tcs N^{5/6} \to \tst N ) & = &
%    N^{1/6} \left[ \mu(0) - \mu(\tcs) \right] + R(\tcs, N)
%    \text{ if } \tcs<0 \\
\bea & &
 \label{proba_succes_3sat_tst_a_t56_selon_t56}
  - \ln \psuccess( T= \tst N \to \tst N + \tcs N^{5/6} ) =
      \nonumber \\ & & \qquad
    N^{1/6} \left[ \mu(\tcs) - \mu(0) \right] + R(\tcs, N)
%    \text{ if } \tcs>0
\eea
where the function $R(\tcs, N)$ (assumed to be regular) is bounded when
$N \to +\infty$ with fixed $\tcs$. Setting $\tcs =$ \mbox{$(t - \tst)
N^{1/6}$} and using eq.~(\ref{expression_mu_3sat_avec_tcspos_grand}),
% -- \ref{expression_mu_3sat_avec_tcsneg_grand}))
%eqs.~(\ref{proba_succes_3sat_t56_a_tst_selon_t56} --
%\ref{proba_succes_3sat_tst_a_t56_selon_t56}) read, for $t<\tst$ and 
%$t>\tst$ respectively,
%\bea
% \label{proba_succes_3sat_t_a_tst_selon_t56}
% -\ln \psuccess(T = t N \to \tst N) & = &
% N^{1/6} \sqrt{k(p)} \Phi[k(p) \epsalpha]
% - \frac{k(p)^3}{4} \frac{1}{\tst-t}
% + \mathcal{O}\left( N^{-1/6} \right) + R[(t - \tst) N^{1/6}, N] \\
\bea & &
 \label{proba_succes_3sat_tst_a_t_selon_t56}
 -\ln \psuccess(T = \tst N \to t N) =
 N^{1/6} \sqrt{k(p)} \Phi[k(p) \epsalpha]
   \ \  \nonumber \\ & & \,
 - \frac{k(p)^3}{4} \frac{1}{t-\tst}
 + \mathcal{O}\left( N^{-1/6} \right) + R[(t - \tst) N^{1/6}, N] .
\eea

 On the other hand, out of the critical time regime, we may modify
eq.~(\ref{proba_succes_2sat_t_a_1_selon_t1}) (with the
expression of $\adeux(t)$ for $\alphainit=5/3$) to compute the lost of
success probability in the large $N$ limit between given times on the
scale of $T/N$. For $t < \tst$ and $t > \tst$ respectively,
\bea & &
 \label{proba_succes_3sat_0_a_t_selon_t1}
 -\ln \psuccess(T = 0 \to t N) =
 -\frac{k(p)^3}{4} \left( \frac{1}{t-\tst} + \frac{1}{\tst} \right) +
   \nonumber \\ & & \qquad
 \frac{1}{4} \ln \left( \frac{\tst-t}{\tst} \right) +
 \int_0^t d\tau f(\tau) + \mathcal{O} \left(N^{-1/3}\right) \\ & &
 \label{proba_succes_3sat_t_a_1_selon_t1}
 -\ln \psuccess(T = t N \to N) =
   \nonumber \\ & & \qquad
 -\frac{k(p)^3}{4} \left( \frac{1}{1-\tst} - \frac{1}{t - \tst} \right) +
   \nonumber \\ & & \qquad
 \frac{1}{4} \ln \left( \frac{1-\tst}{t-\tst} \right) +
 \int_t^1 d\tau f(\tau) + \mathcal{O} \left(N^{-1/3}\right)
\eea
where $f$ is the function
\beq
 f(t) := - \frac{3p}{2(p+2)}
         - \frac{9p^2}{(p+2)^2} (t-\tst) \ .
\eeq

 Eqs.~(\ref{proba_succes_3sat_0_a_t_selon_t1} --
\ref{proba_succes_3sat_t_a_1_selon_t1}) share with
eq.~(\ref{proba_succes_3sat_tst_a_t_selon_t56}) a divergence in 
$1/(t-\tst)$, but they also have a logarithmic divergence that does not
appear in
eq.~(\ref{proba_succes_3sat_tst_a_t_selon_t56}). We speculate that, if 
we pushed the asymptotic expansion for large $N$ that led to
eq.~(\ref{proba_succes_3sat_tst_a_t56_selon_t56}) one step further, we
would find
\beq
 \label{hypothese_sur_R_3sat}
 R(\tcs, N) = g(\tcs) + S(\tcs, N)
\eeq
with $g(\tcs) \sim \ln(|\tcs|)/4$ when $\tcs \to \pm \infty$ and with
$S(\tcs, N)$ regular and bounded in the two limits, first $N \to
+\infty$ with fixed $\tcs$, then $\tcs \to \pm \infty$. At this new
order, the time-derivative term $\partial_{t_0} \pib(x_1, t_0)$ that was
canceled to write eq.~(\ref{edo_pib}) becomes relevant. It yields a
correction to the probability of success that originates physically from
the slow, `secular', evolution of the shape of the probability
distribution $\rhob$ of $c=C_1/N^{1/3}$: the solution of
eq.~(\ref{edo_pib}) is the distribution of $c$ in a true stationary
state, but here we have only a quasi-stationary situation ($c$ is slowly
driven) and the actual distribution is always delayed w.r.t. the
perfectly equilibrated solution of eq.~(\ref{edo_pib}).

 With the assumption~(\ref{hypothese_sur_R_3sat}),
eq.~(\ref{proba_succes_3sat_tst_a_t_selon_t56}) reads, for $t \gtrless 
\tst$,
\bea & &
% \label{proba_succes_3sat_t_a_tst_selon_t56_selon_hypothese}
% -\ln \psuccess(T = t N \to \tst N) & = &
% N^{1/6} \sqrt{k(p)} \Phi[k(p) \epsalpha] - \frac{1}{24} \ln N
% - \frac{k(p)^3}{4} \frac{1}{\tst-t} - \frac{1}{4} \ln|t-\tst| + g(0)
%   \nonumber \\ & &
% + \mathcal{O}\left( N^{-1/6} \right) + S[(t - \tst) N^{1/6}, N]
%   \text{ if } t<\tst \\
 \label{proba_succes_3sat_tst_a_t_selon_t56_selon_hypothese}
 -\ln \psuccess(T = \tst N \to t N) =
   \nonumber \\ & & \qquad
 N^{1/6} \sqrt{k(p)} \Phi[k(p) \epsalpha] \pm \frac{1}{24} \ln N
 \mp \frac{k(p)^3}{4} \frac{1}{t-\tst}
   \nonumber \\ & & \qquad
 \pm \frac{1}{4} \ln|t-\tst|
 \mp g(0) + \mathcal{O}\left( N^{-1/6} \right) +
   \nonumber \\ & & \qquad
 S[(t - \tst) N^{1/6}, N]
%   \text{ if } t>\tst
\eea
and comparing
%eqs.~(\ref{proba_succes_3sat_t_a_tst_selon_t56_selon_hypothese} --
eq.~(\ref{proba_succes_3sat_tst_a_t_selon_t56_selon_hypothese}) with
eqs.~(\ref{proba_succes_3sat_0_a_t_selon_t1} --
\ref{proba_succes_3sat_t_a_1_selon_t1}) yields
\beq
 \label{expression_S_3sat}
 S[(t-\tst) N^{1/6}, N] =
 \left\{ \begin{array}{ll}
   F_<(t) + \mathcal{O}\left( N^{-1/6}\right) & \text{if }t<\tst \\
   F_>(t) + \mathcal{O}\left( N^{-1/6}\right) & \text{if }t>\tst \\
 \end{array} \right.
\eeq
where $F_<$ and $F_>$ are two primitives of $f$. Replacing 
expressions~(\ref{hypothese_sur_R_3sat}) and (\ref{expression_S_3sat}) 
with $t = \tst + \tcs N^{-1/6}$ in 
eq.~(\ref{proba_succes_3sat_tst_a_t56_selon_t56}) and using the
regularity of $S$ in $\tcs=0$ shows that
\beq
 \label{expression_F_3sat}
 F_<(t) = F_>(t) = \int_{\tst}^t f(\tau) \dd \tau \ .
\eeq

 Finally, adding
%eqs.~(\ref{proba_succes_3sat_t_a_tst_selon_t56_selon_hypothese}) and
eq.~(\ref{proba_succes_3sat_tst_a_t_selon_t56_selon_hypothese}) for the
two cases $t \gtrless \tst$ after making the substitution
eq.~(\ref{expression_F_3sat}) yields the total probability of success of
the greedy algorithm UC=UP+R for $p>2/5$:
\beq
  \label{moinslnPsuccess_famille_3sat}
  -\ln \Psuccess =
  N^{1/6} 2 \sqrt{k(p)} \Phi[k(p) \epsalpha] + E(p) +
  \mathcal{O}\left( N^{-1/6}\right)
\eeq
with
\beq
  E(p) = \frac{3 p (p-4)}{2 (p+2)^2} - \frac{3p}{2(5p-2)} +
         \frac{1}{4} \ln \left( \frac{6p}{5p-2} - 1 \right) .
\eeq
While the $N^{1/6}$ divergences of the two cases of
%eqs.~(\ref{proba_succes_3sat_t_a_tst_selon_t56_selon_hypothese}) and
eq.~(\ref{proba_succes_3sat_tst_a_t_selon_t56_selon_hypothese}) add up,
the $\ln N$ divergences cancel out. Similarly, if we set $t=\tst - \tau$
in eq.~(\ref{proba_succes_3sat_0_a_t_selon_t1}) and $t = \tst + \tau$ in
eq.~(\ref{proba_succes_3sat_t_a_1_selon_t1}), add the results and let
$\tau \to 0$, the $\ln |t-\tst|$ divergences cancel out. This seems
reasonable since the slow adaptation of the shape of $\rhob$ is
symmetric w.r.t. the time $\tst$: before $\tst$, the driving term in
eq.~(\ref{edo_pib}) pushes $c$ away from 0 and the equilibration delay
of the distribution of $c$ makes the actual $c$ smaller than the
perfectly equilibrated $c$. Hence the $\ln N$ term in
eq.~(\ref{proba_succes_3sat_tst_a_t_selon_t56_selon_hypothese}) for $t <
\tst$ has a negative contribution to the probability of finding two
contradictory 1-clauses. Conversely, after $\tst$, the driving pulls $c$
towards 0 back. The delay of $c$ makes it larger than what the perfectly
equilibrated $c$ would be. This yields a positive $\ln N$ correction in
eq.~(\ref{proba_succes_3sat_tst_a_t_selon_t56_selon_hypothese}). The
balance of the two slow adaptations is null for symmetry reasons.

 The corrections due to the fluctuations of $C_2$, temporarily left
aside, have order, after
eq.~(\ref{principe_approximation_concentration_C_j}),
\[ \mathcal{O}[N^{\expdeviationcjs-4/3} \exp(-N^{1/6} \Phi)] \text{ and 
} \mathcal{O} \left[ N^{3/2-\expdeviationcjs} \exp \left(
-N^{2\expdeviationcjs-1}/2 \right) \right] \] with possibles choices of
$\delta$ in the range $]1/2, 2/3[$. Taking $\delta$ in the range $]7/12,
2/3[$ ensures that both corrections are negligible w.r.t. all terms of
eq.~(\ref{moinslnPsuccess_famille_3sat}).

\begin{figure}[htbp]
\begin{center}
\includegraphics[width=\largeurgraphes]{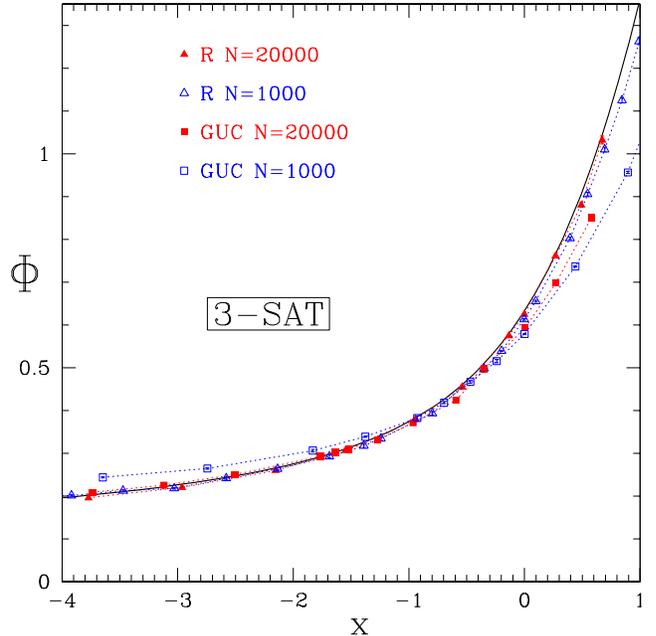}
\end{center}
\caption{\label{figure_comparaison_Phi} Empirical data for the
probability of success of the greedy algorithms with R and UC heuristics
on $10^3$ to $7 \times 10^5$ instances of the random 3-SAT problem, for
several initial clauses-per-variable ratios. Error bars are smaller than
the symbol's sizes. Dotted lines are guides for the eye. Data are
plotted after rescaling their $x$ and $y$ axis with coefficients
$r^\Phi$ and $r^\epsalpha$ to compare them with universal scaling
function $\Phi$ (solid line), according to
eq.~(\ref{moinslnPsuccess_famille_3sat_heuristique_quelconque}).}
\end{figure}

 The result (\ref{moinslnPsuccess_famille_3sat}) is compared to
empirical success probabilities of the greedy UC=R+UP algorithm on a
large number (2000 to $7 \times 10^5$) of instances of the random 3-SAT
problem with sizes up to $N=20000$ in Figure
\ref{figure_comparaison_Phi}. In spite of strong finite-size effects (in
$1/N^{1/6}$), there is an excellent agreement because
eq.~(\ref{moinslnPsuccess_famille_3sat}) provides also the first
subdominant term.

\subsubsection{Universality}

 Any heuristic H run on a set of random instances with self-averaging
$C_2$ and a typical $C_2$ such that, for a given initial
constraint-per-variable ratio $\alphainit$, the resolution trajectory
becomes tangent to the $\adeux=1$ line at a finite time $\tst \in ]0,1[$
with
\beq
 \label{condition_tangence_comme_parabole}
 1-\adeux(t - \tst) \propto (t-\tst)^2
\eeq
has the same critical behaviour as $R$. Indeed, in such a case, the
generating function for $C_1$ satisfies
eq.~(\ref{equation_evolution_fonctgen_c1}) and one may use critical
scalings for the quantities $C_1$, $T$, $\alphainit$ and $\Psuccess$ to
derive eq.~(\ref{edo_rhob_3sat}) from
eq.~(\ref{equation_evolution_fonctgen_c1}). In these scalings, the
exponents are independent of H because the geometric situation expressed
by eq.~(\ref{condition_tangence_comme_parabole}) is the same as for
heuristic R. Solving eq.~(\ref{edo_rhob_3sat}) yields the same scaling
function $\Phi$ as for the R heuristic, \ie there exists numbers
$\alpha_H$, $r_H^{\Phi}$ and $r_H^{\epsalpha}$ (that depend on H and p)
such that~\footnote{For heuristic R, $r_\mathrm{R}^{\Phi} = 2
\sqrt{k(p)}$ and $r_\mathrm{R}^{\epsalpha} = k(p)$.}
\bea & &
  \label{moinslnPsuccess_famille_3sat_heuristique_quelconque}
  -\ln \Psuccess[\alpha_H (1+\epsalpha N^{-1/3})] =
    \nonumber \\ & & \qquad
  N^{1/6} \; r_H^{\Phi} \ \Phi \left( r_H^{\epsalpha}\;  \epsalpha \right) +
  E_H + \mathcal{O}\left( N^{-1/6}\right) . \quad
\eea
$E_H$ is a non-universal correction (even the contribution from a
primitive of the universal term $\ln |t-\tst| /4$ in
eq.~(\ref{proba_succes_3sat_tst_a_t_selon_t56_selon_hypothese}) to $E_H$
is not universal because $\tst$ depends on H). Scaling
relation~(\ref{moinslnPsuccess_famille_3sat_heuristique_quelconque}) is
expected to hold for most, if not all, algorithms using UP on random
$2+p$-SAT instances with $p>2/5$. For GUC we performed analytic
computations on the basis of eq.~(\ref{solution_c2c3_GUC_c2}). The
values of the numbers above are, in the case of $p=1$, \ie random 3-SAT,
with GUC heuristic:
\bea & &
  \alpha_\mathrm{GUC} \approx 3.003494331 \nonumber \\ & &
    \qquad \text{ is such that }
    3 \alpha_\mathrm{GUC}/2 - \ln(3 \alpha_\mathrm{GUC}/2) = 3
      \nonumber \\ & &
  r_\mathrm{GUC}^{\Phi} = \alpha_\mathrm{GUC}^{-1/12} \alphaR(1)^{1/12}
    2^{5/6} \approx 1.764223038 \nonumber \\ & &
  r_\mathrm{GUC}^{\epsalpha} =
    (3\alpha_\mathrm{GUC}/4 - 1/2) \alpha_\mathrm{GUC}^{-1/6} \alphaR(1)^{1/6}
    2^{-1/3} \nonumber \\ & &
    \qquad\ \, \approx 1.363750542 \nonumber \\ & &
  F_\mathrm{GUC}(1) \approx -1.2438849 . \nonumber
\eea
Empirical data for the probability of success of the UP+ GUC algorithm
are compared with the universal function $\Phi$ in
Figure~\ref{figure_comparaison_Phi} --- as for heuristic R, the
agreement is very good, despite strong finite-size effects.

 Notice also that the point where the critical resolution trajectory
gets tangent to the $\adeux=1$ line is universal. At this point, the
residual 2-clauses-per-unassigned-variables ratio $\adeux= \alphainit
(1-p) = 1$ and the residual 3-clauses-per-unassigned-variables ratio
$\alphainit p = 3/2$ so that each affectation of variable through UP
produces, in average, a new 2-clause from the remaining 3-clauses ---
this is why $\adeux(t)$ has a vanishing derivative and the trajectory
does not cross the $\adeux=1$ line. Moreover, the resolution trajectory
(\eg its curvature) is locally the same for all heuristics since almost
all time steps use UP; the chances that the heuristic rule is used in
one step during the critical regime scale like $P(C_1=0) \sim N^{-1/3}$.
Therefore, improving some heuristic may only affect the pre- and
post-critical time regimes. A good heuristic is one that does its best
to avoid the critical region, or to delay entering it as much as
possible.

\subsection{The special case of 2+2/5-SAT}
\label{section_2.4sat}

\subsubsection{The tricritical point $(p=2/5, \alphainit=5/3)$}

 For $p=2/5$, the critical window for $\alphainit$ is the same as for 2-
and 3-SAT, $\alphainit = 5/3 (1 + \epsalpha N^{-1/3})$.
The critical resolution trajectory is tangent to the $\adeux=1$ line so
that $-\ln \Psuccess$ scales like $N^{1/6}$ like in the 3-SAT class. In
addition, the delay of the actual distribution $\rhob$ of
$c=C_1/N^{1/3}$ w.r.t. the fully equilibrated distribution that solves
eq.~(\ref{edo_pib}) contributes to the success probability with a
non-vanishing subdominant $\ln N$ term. This is because, instead of
reversing its direction, the driving of $c$ is directed towards 0 during
the whole algorithm's run, for the critical resolution trajectory starts
\emph{on} the $\adeux=1$ line.

 For times $T$ of the order of $N^{5/6}$, eq.~(\ref{edo_rhob_3sat}) is
relevant (with $\tst=0$ and $e_2(\tcs)=\tcs^2 - \epsalpha$). In the
expression~(\ref{expression_mu_3sat}), $k(p)=1$ and $\tcs^{(1)}$ has to
be 0. This yields
\beq
 -\ln \Psuccess \left[\frac 53 (1+\epsalpha) , \frac 25\right] 
 = \Phi(\epsalpha) N^{1/6} + o(N^{1/6}) .
\eeq
The critical distribution $\rhob$ of $C_1/N^{1/3}$ is the same as for the
3-SAT family, up to scaling factors.

\begin{figure}
\begin{center}
\includegraphics[width=\largeurgraphes]{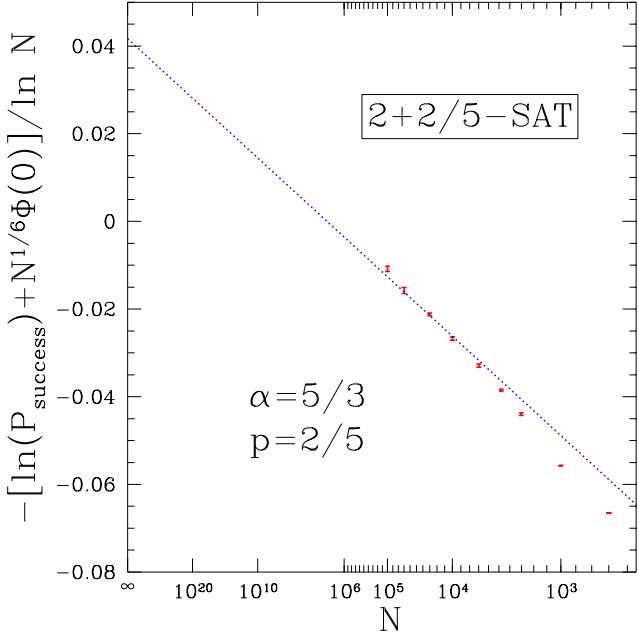}
\end{center}
\caption{\label{figure_comparaison_psucces_ea0_2.4sat} Comparison of
empirical estimates of $\Psuccess$ with prediction
eq.~(\ref{moinslnPsuccess_2.4sat}) for $\epsalpha=0$. We plot $-[\ln
\Psuccess + \Phi(0)]/\ln(N)$ as a function of $1/\ln(N)$.
The straight line is the analytical prediction, $1/24-5/8/\ln(N)$.}
\end{figure}

 As for 3-SAT, we did not compute directly the correction due to secular
evolution of $\rhob$~\footnote{This would be possible by keeping a
further order in the expansion of
eq.~(\ref{equation_evolution_fonctgen_c1}) and supplementing
eq.~(\ref{edo_pib}) with a PDE where $\rhob$ appears as a driving
term.}, but we deduced its contribution to the final result by
comparison between the time scales of $\tcs=T/N^{5/6}$ and of $t=T/N$.
For fixed $t$ and large $N$,
eq.~(\ref{proba_succes_3sat_tst_a_t_selon_t56_selon_hypothese}) reads
here
\bea & &
 \label{proba_succes_2.4sat_0_a_t_selon_t56}
 -\ln \psuccess(T = 0 \to t N) =
 N^{1/6} \Phi(\epsalpha) + \frac{1}{24} \ln N - \frac{1}{4\, t}
   \nonumber \\ & & \qquad
 + \frac{1}{4} \ln t - g(0) + \mathcal{O}\left( N^{-1/6} \right)
 + S(t N^{1/6}, N)
\eea
while eq.~(\ref{proba_succes_3sat_t_a_1_selon_t1}) reads
\bea & &
 \label{proba_succes_2.4sat_t_a_1_selon_t1}
 -\ln \psuccess(T = t N \to N) =
 \frac{1}{4} \left( \frac{1}{t} - 1 \right) - \frac{1}{4} \ln t
   \nonumber \\ & & \qquad
 -\frac{3}{8} + \frac{1}{4} t + \frac{1}{8} t^2
 + \mathcal{O} \left(N^{-1/3}\right) .
\eea
Thus $S(t N^{1/6}, N) = -t/4 - t^2/8 + g(0)$, and
\beq
 \label{moinslnPsuccess_2.4sat}
 -\ln \Psuccess =
 N^{1/6} \Phi(\epsalpha) + \ln(N)/ 24 -5/8 + o(1) \ .
\eeq
This expression compares well with numerical estimates in the critical
$\epsalpha=0$ case, see
Figure~\ref{figure_comparaison_psucces_ea0_2.4sat}.

 As a side remark, in the range of time steps of the order of $N^{2/3}$,
eq.~(\ref{edp_rho_2sat}) with vanishing $\beta(p)$ is relevant.
Numerical evidence shows that its solution $\rho$, once normalized,
converges to the PDF $\rhob$ that satisfies eq.~(\ref{edo_rhob_3sat})
with $e_2=\tst=0$, which is natural since $\rhob$ is the stationary
solution of eq.~(\ref{edp_rho_2sat}). This equilibration process takes a
finite range of time $T/N^{2/3}$, but a vanishing range of time $\tcs$:
this is why the solution of eq.~(\ref{edo_rhob_3sat}) yields a finite
value for $\cbarre(\tcs=0)$, whereas $C_1$ at $T=0$ is 0.

 For $p=2/5$, the scaling function $\Phi$ is truly universal, in the
sense that $r^{\epsalpha}=r^\Phi=1$ for all heuristics. Indeed, the
resolution starts already in the critical regime where UP is used at
almost every step and the heuristic becomes irrelevant; the trajectory
$\adeux(t)$ is then locally universal.

\subsubsection{Matching the 2-SAT family with the 3-SAT class}

 The refined scaling $p = 2/5(1+ \epsp N^{-1/6})$ allows us to blow up
the transition between the 2-SAT family where $\Psuccess$ decays
algebraically with exponent $\expprobadeuxsat(p)$ and the 3-SAT class
where it decays as a stretched exponential. Now, the critical window for
$\alphainit$ is $\alphainit = \frac{5}{3} (1 + \frac{2}{3} \epsp
N^{-1/6} + \epsalpha N^{-1/3})$, $e_2(\tcs) = (\tcs-\frac{5}{6}\epsp)^2
-\frac{1}{4}\epsp^2 - \epsalpha$ and \[ -\ln \Psuccess = N^{1/6}
\Phi(\epsalpha,\epsp) + \ln(N)/24 + \mathcal{O}(1) \] with
\bea
 \Phi(\epsalpha,\epsp) & = &
 \frac{1}{4} \int_{\recdifflnai[\epsalpha-\frac{4}{9}\epsp^2]}^{+\infty}
 \frac{\dd y}{\sqrt{\epsalpha + \frac{1}{4} \epsp^2 - \Ai'(y)/\Ai(y)}}
   \times \nonumber \\ & & \qquad \qquad
 \left[ \frac{\Ai'(y)^2}{\Ai(y)^2} - y \right]^2
\eea
for $\epsalpha \le \frac{4}{9} \epsp^2$ and $\epsp \le 0$. If we send
$\epsp$ to $-\infty$ (as \eg $-N^{1/6} q$) with $\epsalpha \ll
\frac{4}{9}\epsp^2$, $\Phi$ behaves like $1/\epsp$ and $\Psuccess$ is
finite. This was expected since, in this case, we dive into the success
region below the $\adeux=1$ line. However, if we follow the $\adeux=1$
line and set $\epsalpha = \frac{4}{9} \epsp^2$, $\Phi$ behaves like $-
\ln(\epsp)/\epsp$ and $-\ln \Psuccess = \ln(N) [1/24-1/(20q)]$ to the 
leading order in $N$, which matches the
singularity~(\ref{singularite_psuccess_alphafixe_famille2sat}).

\subsection{Case of K-SAT with $K \ge 4$}
\label{section_ksat_k_quelconque}

 For general $K$, that is for random instances with initially $N$
variables and $C_2$ 2-clauses, $C_3$ 3-clauses, ..., $C_K$ $K$-clauses,
$\adeux(t)$ may become tangent to the $\adeux=1$ line with an exponent
greater than 2: $1-\adeux(t-\tst) \propto (t-\tst)^n$ with $n<K$, and
the scaling exponent $\expproba$ for $-\ln \Psuccess$ may take the value
$\frac{1}{3}(1-\frac{1}{n})$. This happens when reduction of $j$-clauses
into $j-1$-clauses compensates exactly the lost of $j-1$-clauses for $j
\ge 4$, so that $\adeux(t)$ stays longer close to the critical
$\adeux=1$ line. $n$ is necessarily integer because $c_2$ is computed
after solving a triangular system of equations
like~(\ref{equation_evolution_c_j}). When $n$ is odd, the critical
resolution trajectory crosses (with vanishing slope) the $\adeux=1$
line, coming from the failure region ($\adeux>1$) into the success one.
Thus the critical behaviour may be reached only if the trajectory starts
\emph{on} this line (initial $C_2=N$; otherwise, the trajectory stays
for a number of time steps $0 \le T \le N$ of the order of $N$ in the
failure region $\adeux>1$ and the probability of success vanishes
exponentially), and $-\ln \Psuccess$ is necessary accompanied with a
$\ln N$ subleading term (because of the secular equilibration of the
critical distribution of $C_1$, like in the $2+2/5$-SAT case).

 The critical behaviour of the 2-SAT family is recovered if the initial
clauses-per-variable ratios $\alpha_j = C_j/N$ are low enough, with
initial $\alpha_2=1$, so that the resolution trajectory is secant to the
$\adeux=1$ line at time $t=0$. The results for the R heuristic are
universal (see Sect.~\ref{section_universalite_famille_2sat}) and
eq.~(\ref{solution_c2_UC}) leads to
\beq
  \Psuccess \propto N^{\frac{1}{12 \left( 1- \frac{3 C_3}{2 C_2} \right)} }
\eeq
provided $\alpha_2=1$, $\alpha_3/\alpha_2 < 2/3$, $\alpha_4/\alpha_3<1$, 
$\alpha_5/\alpha_4<6/5$, $\ldots$, $\alpha_K/\alpha_{K-1} < 2(K-2)/K$.

 As for $2+p$-SAT, the critical initial clauses-per-variable ratios that
yield a stretched-exponential behaviour are not universal, but the
position where the resolution trajectory meets the $\adeux=1$ hyperplane
(with exponent $n$) in the space of the $\alpha_j$'s is. It lies on the
boundary of the region that inequalities above define: if $n=2$,
$\alpha_2=1$, $\alpha_3=2/3$ and $\alpha_4<2/3$. If $n=3$,
$\alpha_4=2/3$ additionally but $\alpha_5<4/5$, and so on. The
trajectory can't reach tangentially the boundary twice (because, in the
neighbourhood of the points where the boundary is hit tangentially, the
flow is going away; coming for a second time to one of these points
would need to re-increase some clauses-per-variable ratios, which is
impossible precisely because of the inequalities above); thus $- \ln
\Psuccess$ can have at most one power-law divergence (\ie only one power
of $N$) and one $\ln N$ divergence (always present when $n$ is odd).

 In the situation where the greatest power of $N$ that appears in $- \ln
\Psuccess$ is $(1-1/n)/3$, the technique of computation based on
eq.~(\ref{principe_approximation_concentration_C_j}) may still be
applied, provided that $\expdeviationcjs \in ]2/3-1/(6n), 2/3[$. The
shrinking of this interval as $n \to +\infty$ probably means that large
deviations of $c_2$ for finite size $N$ have more and more influence (on
the statistics of $C_1$ and on $\Psuccess$) as $n$ is increased.

\section{Discussion and perspectives}
\subsection{How well does UP estimate the ground state energy of
$2+p$-SAT formulas?}
\label{section_energie_fondamental_2sat}

 At the critical initial constraint-per-variable ratio $\alphainit$, the
probability $\Psuccess$ that an algorithm using the UP rule finds a
satisfying assignment to a random formula vanishes algebraically with
large $N$. In contrast, for 2-SAT and $\alphainit=1$, such a satisfying
assignment exists with finite
probability~\cite{bollobas-borgs-chayes-2sat-scaling-window}, that we
numerically estimated to $0.907 \pm 10^{-3}$ -- see Appendix
\ref{appendice_proba_2sat_au_seuil}. To this respect, even though the
dynamic threshold of algorithms using the UP rule is \emph{equal} to the
static threshold $\alpha_C(p)$ for $p<2/5$, they perform not very well
\emph{at} the threshold (at least for $p=0$). But we argue here that they
don't overestimate too much the minimum number of clauses that can be
satisfied simultaneously in an instance of SAT. This number is also
commonly referred to as the ground state energy of the instance. Finding
it amounts to solve the so-called MAX-SAT optimization problem, and, like
the SAT problem~\cite{papadimitriou-steiglitz-livre}, this problem is
classified as difficult by the computer scientists. It is known to grow
like $(\alphainit-\alphaR)^3$ for $\alphainit >
\alphaR$~\cite{monasson-zecchina-statistical-mechanics-of-random-K-sat}.

\begin{figure}[htbp]
\begin{center}
\includegraphics[width=\largeurgraphes]{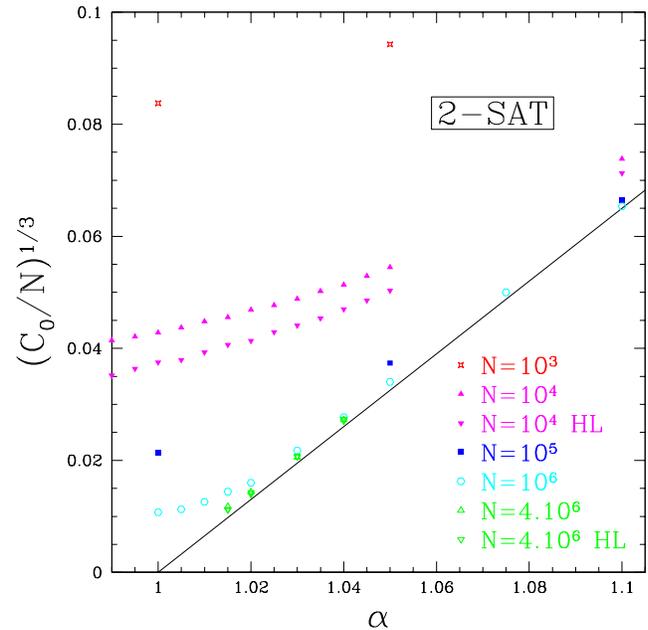}
\end{center}
\caption{\label{figure_C0_2sat} Numerical test of the scaling
relationship~(\ref{expression_moyenne_C0_2sat}) for the UC=UP+R and
UP+HL algorithms on 2-SAT (data are for heuristic R when not otherwise
stated). The rescaled average number of 0-clauses $(\langle C_0
\rangle/N)^{1/3}$ is plotted vs. the initial clauses-per-variable ratio
$\alphainit$. The error bars are smaller than the symbol's sizes. The
solid line is a tentative fit with slope $0.65 \approx 0.27^{1/3}$. For
$\alpha=1$, finite-size effects are strong since $\langle C_0 \rangle$
decreases with $N$ as $N^{-1/12}$, which is consistent with
eq.~(\ref{expression_moyenne_C0_2sat}). Using the HL heuristic rather
than the fully random R heuristic gives some improvement for finite
sizes, but no improvement in the thermodynamic limit because the
heuristic plays no role in the critical regime.}
\end{figure}

 If we allow the greedy algorithm to keep running even if 0-clauses are
found, the final number $C_0$ of 0-clauses is an upper bound to the
ground state energy. The average $C_0$ may be computed from the
generating function $G_{01}$ that satisfies
eq.~(\ref{equation_evolution_fonctgen_c1}) with $X_0=1$. Using non-zero
values of $X_0$ doesn't modify substantially the computations of $G_1$
that we did so far. Eq.~(\ref{edp_rho_2sat}) reads now
\beq
 \label{edp_rho_2sat_X0nonnul}
 \partial_{\tvt} \rho(c,\tvt) = \frac{1}{2} \partial_c^2 \rho(c,\tvt) +
  \beta(p) \tvt \partial_c \rho(c, \tvt)
  - \frac{1-X_0}{2} c \rho(c, \tvt)
\eeq
and eq.~(\ref{proba_succes_2sat_0_a_t_selon_t}) becomes
\bea & &
 \label{expression_G01_2sat_X0nonnul}
 G_{01}(\to t N^{1/3}; X_0, X_1=1) =
   \\ \nonumber & & \quad
 \frac{1-X_0}{12 \beta(p)} \ln N + \frac{\ln t}{4 \beta(p)} +
 H[X_0, \epsalpha, \beta(p)] + \mathcal{O}\left(\frac{1}{t^3 N}\right)
\eea
where $H$ depends also on $X_0$. Deriving
eq.~(\ref{expression_G01_2sat_X0nonnul}) w.r.t. $X_0$ and setting
$X_0=1$ yields
\bea & &
 \label{expression_moyenne_C0_2sat_tempscritique}
 \langle C_0 \rangle(\to t N^{1/3}) =
   \\ \nonumber & & \quad
 \frac{1}{12 \beta(p)} \ln N +
 \partial_{X_0} H[X_0, \epsalpha, \beta(p)] \big|_{X_0=1} +
 \mathcal{O}\left(\frac{1}{t^3 N}\right) .
\eea
We know that, if $X_0=0$, $H[X_0, \epsalpha, \beta(p)]$ behaves like
$\epsalpha^3$ for large positive $\epsalpha$. Assuming that the
behaviour is the same for general $X_0$, the average number of 0-clauses
at the end of the critical time regime is
\beq
 \label{expression_moyenne_C0_2sat}
 \langle C_0 \rangle(\to t N^{1/3}) \approx
 h(p) [\alphainit-\alphaR(p)]^3 N + \frac{1}{12 \beta(p)} \ln N +
 \mathcal{O}(1)
\eeq
for some universal number $h(p)$ (as in
Section~\ref{section_universalite_famille_2sat}, in the critical regime
the heuristic is almost never used, only UP is used, hence $h(p)$ is
universal, or, in other words, independent of the heuristic). The
subsequent non-cri\-ti\-cal time regime $N^{1/3} \lesssim T \le N$
brings only a finite contribution to $\langle C_0 \rangle$. Therefore,
algorithms using UP give an upper bound to the ground state energy of
$2+p$-SAT problems with $p<2/5$ that grows with $\alphainit$ like $h(p)
(\alphainit-\alphaR)^3$. This speculation agrees well with numerical
results for the average $C_0$ at the end of runs of UC=UP+R or UP+HL on
instances of 2-SAT, see Figure~\ref{figure_C0_2sat}. The scaling
$\langle C_0 \rangle \propto N^{1/12}$ is also numerically correct (not
shown on the figure).

 The exponent 3 in $h(p) (\alphainit-\alphaR)^3$ agrees with the
(non-rigorous) results for the ground state energy of random 2-SAT
computed in Section VI
of~\cite{monasson-zecchina-statistical-mechanics-of-random-K-sat}.  
However, the numerical estimate $h(0) \approx 0.27$ (see
Fig.~\ref{figure_C0_2sat}) is clearly different from the coefficient
0.15 computed in Section VI
of~\cite{monasson-zecchina-statistical-mechanics-of-random-K-sat}: the
bound that UP provides is not tight~\footnote{It might be that the exact
ratio of $h(p)$ to the coefficient of the ground state energy is 2.}. As a
side remark, general rigorous
results~\cite{hastad-inapproximabilite-max-2sat-conference,hastad-inapproximabilite-max-2sat-journal}
show that, using techniques quite different from UP, one can design an
algorithm running in polynomial time that finds, for any instance of
random 2-SAT, an assignment which satisfies at least 93.1\% of the number
of clauses that can be satisfied simultaneously in that instance (let us
call this number optimum of the instance; it is the difference between the
number of clauses and the ground state energy). Conversely, it is proved
impossible to find a polynomial time-algorithm that would, for each
instance, satisfy simultaneously a number of clauses of at least $21/22
\approx$~95.5\% of the optimum (for 3-SAT, these two bounds become
$7/8$). For 2-SAT, our results state that, \emph{in average}, the
polynomial time greedy algorithms that use UP outperform these worst-case
bounds for $\alphainit$ close to $\alphaR$ (they satisfy in average a
number of clauses that tends to the optimum as $\alphainit$ tends to
$\alphaR$, although on some rare instances they may perform badly). It
might be interesting to know the average behaviour of the algorithms
of~\cite{hastad-inapproximabilite-max-2sat-conference,hastad-inapproximabilite-max-2sat-journal}
and see whether they get closer to the ground state energy as algorithms
based on UP.

\subsection{Interpretation as a random graph percolation phenomenon}
\label{section_graphe_aleatoire}

 Introduction of the oriented graph $\gG$ representing 1- and 2-clauses
allows us to re-interpret some of the scalings that we found in the
previous sections. $\gG$ is made of $2\, (N-T)$ vertices (one for each
variable $x_i$ and its negation $\bar x_i$), $C_1$ marked vertices (one
for each 1-clause $z_i$), and $2\, C_2$ edges ($z_i\vee z_j$ is
represented by two oriented edges $\bar z_i \to z_j$ and $\bar z_j \to
z_i$)
\cite{aspvall-plass-tarjan-temps-lineaire,bollobas-borgs-chayes-2sat-scaling-window}.
$\adeux$ is simply the average (ingoing or outgoing) degree of vertices
in $\gG$.

 A step of UP removes a marked vertex, say $z$, and its attached
outgoing edges, after having marked its descendants (2-clauses with
$\bar z$ are reduced). If $\bar z$ appears in some 3-clauses, these
clauses become 2-clauses, \ie new pairs of edges in $\gG$. UP-steps are
repeated until no vertex is marked, then some vertex is marked according
to the heuristic of the algorithm and another \emph{round} of UP starts.
During a round, there is a competition between the elimination of edges
and vertices and the creation of new edges. All vertices in the
(outgoing-edges) connected component of the initial $\gG$ that started
with the first marked vertex are removed. In addition, the vertices that
have been linked to this component through the new edges are also
removed. A contradiction arises (a 0-clause appears) when two conjugate
vertices $z$ and $\bar z$ are marked. When $\gG$ percolates (this
happens for $\adeux \ge
1$~\cite{janson-luczak-rucinski-livre,bollobas-borgs-chayes-2sat-scaling-window}),
there exist many oriented loops going from one literal $z$ to its
conjugate and back to it~\cite{aspvall-plass-tarjan-temps-lineaire}.
Marking a vertex of such a loop results sooner or later in the marking
of both $z$ and $\bar z$, and the algorithm fails. Conversely, if $\gG$
doesn't percolate, there is no such loop with finite probability. The
success/failure transition of the algorithms using UP is related to the
percolation transition of $\gG$; if the value of $\adeux$ is bounded
away from 1, $\gG$ never percolates and the probability of success if
finite. Although the results for percolation of random graphs were
obtained in a static context, they apply readily to the graph $\gG$ that
is kinetically built, because of conditional uniformity (see
Sect.~\ref{section_cinetique_fonctions_generatrices}).

 Notice that the percolation phenomenon of random graphs is very robust. 
For instance, taking random graphs conditioned to a certain degree
distribution of the
vertices~\cite{newman-strogatz-watts-gral-avec-distribution-degres} does
not change the universality class if the probability of the large
degrees decays not too slowly. Such (directed) graphs $\gG$ appear in
the context of algorithms with the HL or CL heuristics that select
literals according to their degrees. We have checked numerically for HL
that the degree distribution in $\gG$ is far from the Poisson
distribution that we find for non-conditioned $\gG$'s (\eg for R and
GUC). Still, the probabilities of finding vertices of high degrees
decrease fast and $\gG$ has the same percolation critical behaviour as a
random graph. A parallel can be drawn between the robustness of the
critical behaviour of random graphs and, in our computation, the weak
dependence of the probability law for $C_1$ on that of $C_2$ (only the
average of $C_2/N$ matters thanks to self-averaging).

 In the percolation critical window $|\adeux -1| \sim
N^{-1/3}$~\cite{janson-luczak-rucinski-livre,bollobas-borgs-chayes-2sat-scaling-window},
the probability that a vertex belongs to a component of size $S$ is
$Q(S)\sim
S^{-3/2}$~\cite{newman-strogatz-watts-gral-avec-distribution-degres},
with a cut-off equal to the largest size, $N^{2/3}$. From
Figure~\ref{figure_traj}, departure ratios $\alpha$ have to differ from
$\alphaflat$ by $N^{1/3}$ for resolution trajectories to fall into the
critical window. Hence the critical window in $\alphainit$ has width
$N^{-1/3}$ for both the 2-SAT family and the 3-SAT class.

 When the resolution trajectory is tangent to the $\adeux=1$ line (3-SAT
class), it spends the time $\Delta t \sim \sqrt {|\adeux -1|} \sim
N^{-1/6}$ in the critical window, corresponding to $\Delta T = N \,
\Delta t \sim N^{5/6}$ eliminated variables. Let $S_1, S_2, \ldots ,
S_J$ be the sizes of components eliminated by UP in the critical window;
we have $J \sim \Delta T/ (\int \! \dd S \, Q(S) \, S) \sim N^{1/2}$.
During the $j^{th}$ elimination, the number of marked vertices `freely'
diffuses, and reaches $C_1 \sim \sqrt {S_j}$. The probability that no
contradiction occurs is $[(1-q)^{C_1}]^{S_j} \sim \exp(- S_j^{3/2}/N)$
where $q\sim \frac 1N$ is the probability that a marked vertex is the
negation of the one eliminated by UP. Thus $-\ln \Psuccess \sim J \!
\int \! \dd S \, Q(S) \, S^{3/2}/N \sim N^{1/6}$, giving $\expproba =
\frac 16$. Notice that, while the average component size is $S\sim
N^{1/3}$ (and thus the probability that $C_1$ vanishes in the critical
time regime is $\sim N^{-1/3}$, consistently with
Eq.~(\ref{definition_sigma})), the value of $\expproba$ is due to the
largest components with $S\sim N^{2/3}$ \ie $C_1 \sim N^{1/3}$ marked
vertices. 

 In the case of the 2-SAT family, the algorithm eliminates only $\Delta
T \sim N^{2/3}$ variables and $J \sim N^{1/3}$ connected components. The
estimation above yields a finite $- \ln \Psuccess$ and fails to predict
the correct answer because it doesn't take the $\ln N$ corrections into
account.

 For general $K$-SAT, one may attach to an instance a family of oriented
hypergraphs were vertices are the literals and the $l$-clauses are
$l$-hyperedges, for $2 \le l \le L$ and some fixed integer $L$ between 2
and $K$. The first graph of this family, for $L=2$, is $\gG$. The
critical behaviours exhibited in Section~\ref{section_ksat_k_quelconque}
appear when the first hypergraphs, $2 \le l \le L$ for some $3 \le L \le
K$, percolate. This happens~\cite{goldschmidt-percolation-hypergraphes}
for clauses-per-variable ratios $\alpha_l = 2^{l-1}/[l(l-1)]$ for $2 \le
l \le L$; then $- \ln \Psuccess \sim N^\expproba$ with $\expproba =
\frac{1}{3}(1-\frac{1}{L-1})$.

\subsection{Unit-Propagation in other problems}

 The Unit-Propagation rule that we defined for SAT problems may easily
be generalized to other contexts. In the graph coloring problem, one
wants to know if the vertices of a given graph may be colored with $K$
colors in such a way that no two vertices that share an edge have the
same color. A family of algorithms that deal with such graphs processes
the graph by maintaining the list of available colors on each vertex
(initially the list of all $K$
colors)~\cite{achlioptas-molloy-coloriage-par-liste,ein-dor-monasson-coloriage-symetrique}.
When a vertex is colored with some color, this color is removed from the
list of available colors of the neighbours of this vertex. The UP rule
would be defined as: ``When, on the partially colored graph, at least
one vertex has only one available color, color it with this color prior
to any other action''. We expect
eq.~(\ref{equation_evolution_fonctgen_c1}) to hold also for such graph
coloring algorithms, after choosing a relevant value for $C_2$. This
value may be computed by the technique of differential
equations~\cite{achlioptas-molloy-coloriage-par-liste,ein-dor-monasson-coloriage-symetrique}
like in eqs.~(\ref{equation_evolution_c_j}). The probability of success
of the randomized, greedy coloring algorithms should behave like
$\Psuccess$ for $K$-SAT; in particular, for the case of 3 colors, it
should have a success-to-failure transition with critical behaviour in
$N^{1/6}$.

 Similarly, the study of greedy algorithms using UP for other hard
\emph{decision} computational problems~\cite{garey-johnson-intractability}
should lead to the same critical behaviour (that is, problems where the
answer of the algorithm should be either ``yes -- satisfiable -- colorable
-- ...'' or ``no -- unsatisfiable -- uncolorable -- ...'').

 In the framework of error-correcting codes, the authors of
reference~\cite{amraoui-montanari-richardson-urbanke-taille-finie} also
found a critical behaviour that is governed by a Brownian motion with
parabolic drift~\cite{groeneboom-airy}, which leads to Airy distributions,
as we found for $C_1$ in Section~\ref{section_classe_3sat}.

 There is another important family of problems, called \emph{optimization}
problems, where the answer is a number. For example, in the MAX-SAT
problem, believed to be a hard computational problem, one wants to know,
for some instance of SAT, the maximal number of satisfied clauses or,
equivalently, the ground state energy, \ie the minimal number of violated
clauses over all assignments of variables of this instance. In the context
of graph coloring, one may look for the minimal number of colors needed to
color some graph, \etc Knowing whether their random formulation (where
instances are drawn uniformly at random) fall into the same universality
classes as the random SAT problem requires more investigation. However,
each optimization problem has a \emph{decision} version, which can be
related to this work. For instance, the decision version of random
MAX-$2+p$-SAT would be: ``is there an assignment of the variables of this
instance such that the number of violated clauses does not exceed some
fixed number $m$?'' We can deal with it by asking for the probability that
the greedy algorithm using UP ends, after having assigned all variables,
with a number $C_0$ of 0-clauses that does not exceed $m$. As long as $m$
is bounded in the large $N$ limit, this new success probability can be
computed by summing the 0th, 1st, 2nd, ..., $m$th derivatives of
$G_{01}(X_0,1;N|C_2)$ with respect to $X_0$ in $X_0=0$. A dynamic phase
transition between success and failure phases is found as for the random
$2+p$-SAT problem and, both for $p<2/5$ and $p \ge 2/5$, the largest term
in the expression for $-\ln\Psuccess$ at criticality is independent of
$m$: the critical exponents are unchanged, although subdominant terms now
depend on $m$ (in particular, for $p<2/5$, terms in $m \ln(\ln(N))$ appear
in the expression of $-\ln\Psuccess$ at criticality). We expect the
exponents to remain the same for algorithms using UP on the decision
version of other optimization problems. It would be interesting to know
what happens when $m$ is allowed to become large with $N$.

 Another interesting question is whether instances of SAT (or other
problems) with strong correlations, such that self-averaging of $C_2$ is
lacking, would lead to new universality classes. This could be the case
for instances of SAT such that $\gG$ can be embedded into a
finite-dimensional space, or with a power-law distribution of the number
of occurrences of variables in clauses, possibly built by preferential
attachment of new variables to highly-connected old variables. Graphs
with such features are observed in real-world applications such as the
`World Wide Web' or social networks; they might be helpful in modeling
industrial instances of SAT better than the random, `flat' distribution.

\bigskip
{\bf Acknowledgments.} We thank Andrea Montanari and Federico
Ricci-Tersenghi (who motivated the study in
Section~\ref{section_energie_fondamental_2sat}) for stimulating
discussions. Partial financial support from the`Statistical physics of
Information Processing and Combinatorial Optimization' (STIPCO) Network
of the Research Training Network programme of the European Commission
and from the French Ministry of Research through the `ACI Jeunes
Chercheurs' `Algorithmes d'optimisation et syst\`emes d\'esordonn\'es
quantiques' is acknowledged. We thank the La\-bo\-ra\-toi\-re de
Phy\-si\-que Th\'eo\-ri\-que in Stras\-bourg, where part of this work
was realized, for its hospitality.

\appendix  % 

 % Some tricks to comply with EPJB's policy for equation numbers in 
 % appendices, according to epjb_instructions.pdf, whereas the provided 
 % class "svjour.cls" doesn't comply with this policy:
\def\theequation{\thesection.\arabic{equation}}
\let\oldsection\section
\renewcommand{\section}[1]{\setcounter{equation}{0} \oldsection{#1}}

 % Trick to comply with EPJB's policy for section names in appendices,
 % according to epjb_instructions.pdf, whereas the provided class
 % "svjour.cls" doesn't comply with this policy:
\makeatletter
\def\@seccntformat#1{\appendixname\ \csname the#1\endcsname:\ }
\makeatother

\section{Numerical estimate of the probability of satisfiability for critical 2-SAT}
\label{appendice_proba_2sat_au_seuil}

\begin{figure}[htbp]
\begin{center}
\includegraphics[width=\largeurgraphes]{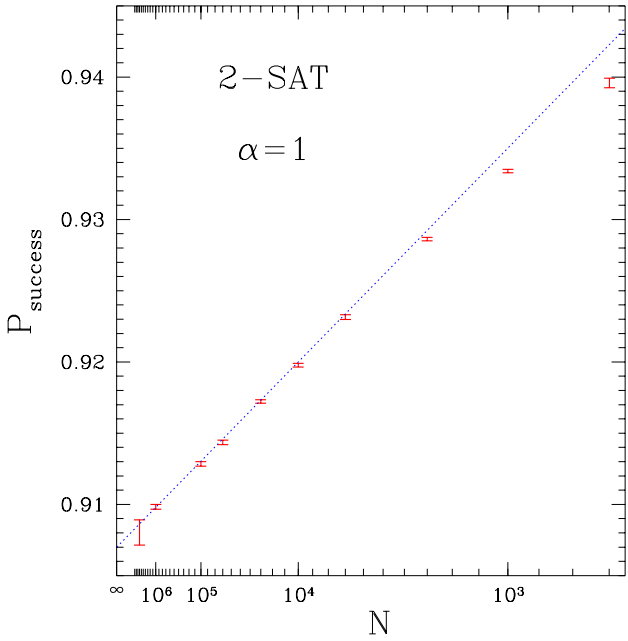}
\end{center}
\caption{\label{figure_proba_2sat_au_seuil}
Numerical estimates of the probability of satisfiability of critical 
($\alphainit=1$) random 2-SAT formulas for different sizes $N$, plotted 
vs. $N^{-1/3}$. The straight line is a guess for the tangent to the 
curve at the point where it cuts the $N=\infty$ axis.}
\end{figure}

 Reference~\cite{bollobas-borgs-chayes-2sat-scaling-window} rigorously
proves that the probability of a random 2-SAT formula of size $N$ with
initial constraint-per-variable $\alphainit=1$ is finite (whereas it
vanishes for large $N$ if $\alphainit>1$). A numerical estimate of this
probability was given on figure 5 of
reference~\cite{shen-zhang-max-2-sat-etude-empirique}, but for very
small sizes $N$ (up to 90). We give here more precise results, based on
numerical estimates of the probability of satisfiability for formulas of
sizes $N=500$ to $5 \times 10^6$. For each size $N$ up to $10^6$, we
drew at random more than $3 \times 10^6$ instances (for $N=5 \times
10^6$, we drew only $10^5$) and we determined if they were satisfiable
or not thanks to the well-known algorithm of
reference~\cite{aspvall-plass-tarjan-temps-lineaire}. This algorithm
finds, by depth-first exploration, the strongly connected components of
an oriented graph built from the 2-clauses, similar to the graph we
introduced in Section~\ref{section_graphe_aleatoire}. Results from the
simulations are plotted in Figure~\ref{figure_proba_2sat_au_seuil}. They
appear to be fully consistent with the finite-size scaling hypothesis
\beq
 \Psat(N) = {\Psat}_\infty + \Theta \left( N^{-1/3} \right)
\eeq
and yield
\beq
 {\Psat}_\infty = 0.907 \pm 10^{-3} .
\eeq

\section{Path-integral formalism for the kinetics of search}

 The standard, yet non-rigorous, technique of path integrals is another
tool to study the search process. Let us re-derive some results of
Sections~\ref{section_fonctions_generatrices} and
\ref{section_success_case} with it; this will show the correspondence
between, on one hand, K-SAT related quantities and physical quantities
such as moments, and, on the other hand, generating functions techniques
and path-integral techniques, hopefully bringing more insights to both
approaches.

We start from the evolution equation (\ref{equation_evolution_tousCs})
and write the probability $P(T)$ that the search process doesn't produce
any 0-clause from times $0$ to $T$ as an iteration (this quantity was
$G(0,1,1,\ldots,1;T)$ in Section~\ref{section_fonctions_generatrices}):
\begin{eqnarray} & &
\sum_{\vec{B}_T} P(\vec{B}_T;T) =
\sum_{\vec B_T} \sum_{\vec B_{T-1}} \ldots \sum _{\vec B_1}
  M(\vec{B}_T \gets \vec{B}_{T-1};T-1)
    \nonumber \\ & & \quad
  \times M(\vec{B}_{T-1} \gets \vec{B}_{T-2};T-2)
  \times \ldots \times M(\vec{B}_{1} \gets \vec C_0;0) \nonumber \\
& & = \sum _{\vec C_{1}, \ldots, \vec C_{T-1}}
 \sum_{\vec{B}_1 , \ldots , \vec{B}_T} \int_{-\pi}^\pi
  \frac{d\vec y_{1}}{(2 \pi)^K} \ldots \frac{d\vec y_{T}}{(2 \pi)^K}
    \times \\ \nonumber & & \quad
 \exp \left( \sum_{L=1}^T \imag \vec{y}_{L} .
   (\vec B_{L} - \vec C_{L})\right)
 \prod_{L=0}^{T-1} M(\vec B_{L+1} \gets \vec{C}_L;L)
\end{eqnarray}
with $\vec C_0 := (0, 0, \ldots, 0, \alpha N)$ and $\imag^2=-1$. All
clause vectors are, in this Appendix, of dimension $K$ instead of
$K+1$ since the number of 0-clauses is always zero. The $\vec{y}_L$'s
constrain the $\vec{C}_L$'s to mimic the $\vec{B}_L$'s, so that we can
write the formally quadratic expression $M(\vec{B}_{L+1} \gets
\vec{B}_L;L)$ as the uncoupled expression $M(\vec{B}_{L+1} \gets
\vec{C}_L;L)$. Carrying out the sums over the $\vec B_L$'s first, we
obtain
\begin{eqnarray} & &
P(T) := \sum_{\vec{B}_T} P(\vec{C}_T;T) \\ \nonumber & & \quad
= \sum_{\vec C_{1}, \ldots, \vec C_{T}} 
 \int_{-\pi}^\pi \frac{\dd\vec y_1}{(2 \pi)^K} \ldots
   \frac{\dd\vec y_T}{(2 \pi)^K}
 \exp \left( -\sum_{L=1}^T \imag \vec{y}_{L} . \vec{C}_{L}\right)
 \\ \nonumber & & \qquad
\times \exp \left( \sum_{L=0}^T
 \ln \left[ \vec{f}(\vec{X}_L;L) \right] . \;\vec{C}_L \right)
 \\ \nonumber & & \qquad
\times \prod_{L=0}^T
  \bigg[ (1-\delta_{(\vec C_L)_1,0} ) \; e^{-\imag (\vec y_{L})_1}
         + \delta_{(\vec C_L)_1,0}  \bigg] 
\end{eqnarray}
where $(\vec{X}_L)_i=\exp (\imag (\vec y_L)_i)$, the vector-valued
function $\vec{f}=(f_1, f_2, \ldots, f_K)$ is defined through
(\ref{deff}) and the dot denotes the usual scalar product of
$\mathbb{R}^K$.
 In the large $N$ limit, a continuous formulation for $P(T)$ can be
obtained. Let us define the reduced time, $\ell= L/N$ and the clause
densities $\vec c (\ell) = \vec C(L)/N$. Define for $i=1,\ldots,K$
\begin{equation} \label{defgamma}
\gamma_j( \vec y; t) := \frac j{1-t}
\bigg( \frac{ e^{-y_j} (1+e^{y_{j-1}})}2 - 1 \bigg)  
\end{equation}
with $y_0 := -\infty$, and $\vec \gamma := (\gamma _1, \ldots,
\gamma_K)$. The probability $P$ at time $T=tN$ can be written as a path
integral over the values of clause densities $\vec c$ and (from now on
complex) 'momenta' $\vec y$ between times 0 and $t$,
\bea & &
P(t N) =
\int_{\vec c(0)=(0,0,\ldots,0,\alpha)}^{\vec c(t)=(c_1,c_2,\ldots,c_K)}
 {\cal D}\vec c(\ell ) \,
 {\cal D}{\vec y (\ell)} \times
  \nonumber \\ & & \qquad \qquad \qquad
\exp \bigg( - N \;  {\cal S} \big[\{ \vec c(\ell),
 \vec y (\ell) \}\big] \bigg)
\eea
where the action reads
\bea
 \label{pip}
{\cal S} \big[\{ \vec c(\ell), \vec y (\ell)  \}\big] & = &
 \int _0^t d\ell \, \bigg\{
 \vec y (\ell) . \frac{d\vec c}{d\ell}(\ell) - \vec \gamma \big(
 \vec y(\ell);\ell \big) . \vec c(\ell )
   \nonumber \\ & &
 - \ln \big[ \rho _1 (\ell) \; e^{- y _1 (\ell) }
 +  1- \rho _1(\ell)  \big] \bigg\}
\eea
where $\rho _1(\ell)$ denotes the  probability that there is at least one
unit clause at time $\ell$ {\em i.e.} the  number of instants $L$ such 
that $C_1(L)\ge 1$ between $L=\ell N$
and $L=\ell N+ \Delta L$, divided by $\Delta L$, with $1\ll \Delta L \ll 
L$.
 Minimization of the action (\ref{pip}) yields the classical equations
of motion. Differentiating with respect to momenta, we find
\bea
 \label{eqacy}
\frac{\delta {\cal S}}{\delta  y _i (\ell)} = 0 & \rightarrow &
 \frac{d c _i}{d\ell}(\ell) =  \sum _j \frac{\partial \gamma _j \big(
 \vec y(\ell);\ell \big) }{\partial y_i(\ell)} \; c_j(\ell)
  \nonumber \\ & & \qquad
- \frac{\delta_{i,1}\; \rho_1 (\ell) }
 {\rho _1 (\ell) +  \big( 1- \rho _1(\ell) \big) e^{ y _1 (\ell) }} \quad
\eea
for $i=1,\ldots,K$. Some care has to be brought to the minimization of 
the action with respect to the clause densities since $\rho_1$ and $c_1$ 
are not independent. During the time 
interval $[\ell,  \ell+d\ell]$, the number $C_1$ of unit clauses is either
of the order of $N$ ($c_1>0$ and $\rho_1=1$) or of the order of unity
($c_1=0$ and $\rho_1 < 1$) \footnote{This statement is true outside the
critical regime where $c_1$ and $\rho_1$ are both vanishing as (negative) 
powers of $N$.}. With this caveat, we obtain
\begin{equation}
\frac{\delta {\cal S}}{\delta  c _i (\ell)} = 0 \rightarrow
\frac{d y _i}{d\ell}(\ell) =  - \gamma _i \big(
\vec y(\ell);\ell \big) 
\qquad (i=2,\ldots,K) 
\end{equation}
and
\begin{itemize}
\item if $c_1 (\ell) >0$, then $\rho_1(\ell)=1$ and  
\begin{equation}
\frac{\delta {\cal S}}{\delta  c _1 (\ell)} = 0 \rightarrow
\frac{d y _1}{d\ell}(\ell) =  - \gamma _1 \big(
\vec y(\ell);\ell \big)
\end{equation}
\item if $c_1(\ell)=0$, then $\rho_1(\ell)$ is given by equation
(\ref{eqacy}) for $i=1$,  
\begin{equation} \label{eqrho1}
\frac 1{\rho _1 (\ell)} = 1 - e^{ - y _1 (\ell) } +
e^{ - y _1 (\ell) } \bigg( 
\frac{\partial \gamma _2 \big(
\vec y(\ell);\ell \big) }{\partial y_1(\ell)} \; c_2(\ell) \bigg) ^{-1}.
\end{equation}
Eq.~(\ref{eqrho1}) makes sense if the r.h.s. is larger than unity. In 
the case where $y_1=0$, it agrees with eq.~(\ref{eq1asuc}).
\end{itemize}

\section{Large $\tvt$-expansion of the solution $\rho$ of the 
PDE~(\ref{edp_rho_2sat})}
\label{appendice_2sat_critique_developpement_grand_t23}

 In this appendix we explain the method we used to get an asymptotic
expansion of the solution of eq.~(\ref{edp_rho_2sat}) at large times
$\tvt$, and we quote the results. Eq.~(\ref{edp_rho_2sat}) is here
treated in the general case $X_0 \in [0,1]$ and reads
\beq
 \label{edp_rho_2sat_avec_X0}
 \partial_{\tvt} \rho(c,\tvt) = \frac{1}{2} \partial_c^2 \rho(c,\tvt) +
  \beta(p) \tvt \partial_c \rho(c, \tvt)
  - \frac{1-X_0}{2} c \rho(c, \tvt)
\eeq

\noindent First, we change the variables $c$ and $t$ for $u:= c \tvt$
and $s := 1/\tvt$ respectively, and define $g$ as $g(u,s) := \rho(c,
\tvt)$. This is motivated by the expectation that, at large times
$\tvt$, both the drift term (that pushes diffusing particles towards the
$c=0$ boundary) and the absorption term (that kills situations where
many particles are away from the $c=0$ boundary) constrain the PDF
$\rho$ to be concentrated around the $c=0$ boundary: $\rho(c, \tvt)$
will be non-vanishing only for values of $c$ that tend to 0 as $t \to
+\infty$. With this choice of scale, eq.~(\ref{edp_rho_2sat_avec_X0})
turns at the leading order in $s$ when $s \to 0$ into
\bea
 \partial_u^2 g(u,s) + 2 \beta \partial_u g(u,s) = 0 & & \\
 \partial_u g(0,s) + 2 \beta g(0,s) = 0
\eea
hence $g(u,s) = A(s) \exp(-2 \beta u)$: $c$ has at large times an
exponential statistical distribution, with average (conditioned to
success of the greedy algorithm) $\cbarre(\tvt) = 1/(2 \beta \tvt)$. 

 To actually compute the normalization factor $A(s)$, we need to
introduce formally the correction to this leading-order term. Let
$\rho(c, \tvt) =: A(s) \exp(-2 \beta u) + B(s) h(u,s)$ with the
requirement $B(s) \ll A(s)$ as $s \to 0$. The leading-order terms in
eq.~(\ref{edp_rho_2sat_avec_X0}) (after cancellation of the formerly
leading-order terms) satisfy
\bea & &
 \left[ -2 \beta s u A(s) - s^2 \partial_s A + u s (1-X_0) A(s)/2 \right]
 \exp(-2 \beta u)
   \nonumber \\ & & \qquad
 = s^{-2} B(s) \left( \partial_u^2 h + \beta \partial_u h \right)
\eea
with the boundary condition $\partial_u h(0,s) + 2 \beta h(0,s) = 0$.
This yields, after integration from $u=0$ to $+\infty$ (we assume that
$h(u,s) \to 0$ when $u \to +\infty$ just like $\rho$ does),
\beq
 A(s) \propto s^{-1+\frac{1-X_0}{4\beta}}
\eeq
and, for the leading order of $\rho(c, \tvt)$ at large times:
\beq
 \rho(c, \tvt) \propto \tvt^{1-\frac{1-X_0}{4\beta}}
  \exp(-2 \beta c \tvt) .
\eeq
This shows that the probability of success of the algorithm decays
algebraically at large times $\tvt$:
\beq
 \pi(0, \tvt) = \int_{c=0}^{+\infty} \rho(c, \tvt) \dd c
  \quad \propto \quad \tvt^{-\frac{1-X_0}{4\beta}} .
\eeq

\noindent Subleading orders in the expansion of $\rho$ at large $\tvt$
may be found iteratively by the same technique. For the sake of
completeness, let us quote here what we found:
\bea & &
 \rho(c, \tvt) \propto
 \tvt^{1-(1-X_0)/4/\beta} e^{-2 \beta c \tvt} \times \\ \nonumber & & \quad
  \left[ 1 +
   \left( 1-\frac{1-X_0}{4\beta} \right)
     \left(c^2 \tvt^2- \frac{6 \beta-1+X_0}{12 \beta^3}\right) \tvt^{-3} +
    \right. \nonumber \\ & & \quad \  \left.
   \left( 1-\frac{1-X_0}{4\beta} \right)
     \left\{ \frac{c^2 \tvt^2}{48 \beta^4}
       \left[ (6 \beta^3 X_0 + 24 \beta^4 - 6 \beta^3) c^2 \tvt^2 +
    \right. \right. \right. \nonumber \\ & & \left. \left. \left.
       \qquad \qquad \qquad \qquad \quad
              (48 \beta^3 + 8 \beta^2 X_0 - 8 \beta^2) c \tvt +
    \right. \right. \right. \nonumber \\ & & \left. \left. \left.
       \qquad \qquad \qquad \ 
              (-2 \beta + 2 X_0 \beta + 2 X_0 + 48 \beta^2-1-X_0^2)
       \right]
    \right. \right. \nonumber \\ & & \left. \left.
       \qquad \qquad \qquad \qquad \quad
       - \frac{6 \beta-1+X_0}{1152 \beta^7} \times
    \right. \right. \nonumber \\ & & \left.
         (384 \beta^2 - 50 \beta + 50 X_0 \beta-1+2 X_0-X_0^2)
     \bigg\} t^{-6}
   + o(t^{-6}) \right] \nonumber
\eea
and, for the conditional average of $c$:
\bea & &
 \label{developpement_cbarre_tvt_grand}
 \cbarre(\tvt) = \frac{1}{2 \beta \tvt} +
  \frac{4 \beta-(1-X_0)}{8 \beta^4 \tvt^4} +
    \\ \nonumber & & \quad
  \frac{5}{32 \beta^7 \tvt^7}
    \left[ 4 \beta-(1-X_0) \right] \left[ 5 \beta-(1-X_0) \right] +
  o(\beta^{-7} \tvt^{-7}) .
\eea

 Notice that the expansions here can only be asymptotic series, without
a finite radius of convergence (as a function of $s=1/\tvt$), because
they are independent of the initial conditions. Indeed, fix $\epsilon>0$
and suppose that the sum of the expansion for $\rho$ above is the right
solution, up to a uniform small difference $\epsilon$, on a finite
interval of $\tvt$, say $]A, +\infty[$. Let $\rho_\mathrm{approx}(c,
\tvt)$ be this sum of the expansion. Then ask for the solution of
eq.~(\ref{edp_rho_2sat}) that has initial condition $\rho(c, A+1) =
\rho_\mathrm{approx}(c, A+1) + 10 \epsilon$: it should also be
$\rho_\mathrm{approx}(c, A+1)$ up to a difference $\epsilon$ ---
contradiction! Therefore, the precision of the approximations obtained
with partial sums of the expansion above decreases as we try to get
values for smaller and smaller times $\tvt$, and we should keep that in
mind when we use the present results.

\section{Small $\tvt$-expansion of the solution $\rho$ of the
PDE~(\ref{edp_rho_2sat})}
\label{appendice_2sat_critique_developpement_petit_t23}

 In this Appendix we explain the method we used to get an asymptotic
expansion of the solution of eq.~(\ref{edp_rho_2sat_avec_X0}) at small 
times $\tvt$, and we quote the results.

 We use the exact solution of eq.~(\ref{edp_rho_2sat_avec_X0}) without
the non-linear term $c \rho(c, \tvt)$ (that is, for the special value
$X_0=1$) as a guide for a relevant choice of variables. Disregarding the
boundary condition eq.~(\ref{condition_bord_2sat}), we easily solve the
resulting linear PDE thanks to Fourier transform:
\bea & &
 \rho_\mathrm{linear}(c, \tvt) =
  \frac{\mathrm{constant}}{\sqrt{\tvt - \tvtinit}}
    \times \\ \nonumber & & \quad
  \exp \left\{ - \frac{1}{2(\tvt-\tvtinit)}
               \left[ c + \frac{\beta(p)}{2} (\tvt^2 - \tvtinit^2)
               \right]^2 \right\}
\eea
which suggests us to use the following variables for our expansion:
\[
 v:= \tvt -\tvtinit, \quad
 u:= \frac{c + \beta(p)/2 (\tvt^2-\tvtinit^2)}{\sqrt{\tvt-\tvtinit}}, 
\]
\[
 \rho(c, \tvt) =: \frac{1}{\sqrt{v}} g(u, v) \ .
\]
 The equations for $g$ are
\bea & &
 2 v \partial_v g = \partial_u^2 g + u \partial_u g +
    \\ \nonumber & & \quad
  \left[ 1 - (1-X_0) v^{3/2} u + (1-X_0) \beta(p)/2 v^2 (2 \tvtinit + v)
  \right] \\ & &
 \partial_u g[ \beta(p)/2 \sqrt{v} (2 \tvtinit + v), v] =
    \\ \nonumber & & \quad
  - 2 \beta(p) \sqrt{v} (\tvtinit + v)
    g[ \beta(p)/2 \sqrt{v} (2 \tvtinit + v), v ]
\eea

\noindent At $\tvt = \tvtinit$, \ie at $v=0$, $g$ is Gaussian, as
expected from the study of the linear equation (with $X_0=1$): $g(u, 0)
= \sqrt{2 \pi^{-1}} \exp(-u^2/2)$. Letting $v^\gamma h(u,v)$ be the
deviation, at positive times $v$, between the exact $g$ and this
Gaussian expression, and substituting this into the equations for $g$,
we see that the boundary condition constrains $\gamma$ to the value
$1/2$, and we are led to the ODE for $h(u,0)$
\beq
 \partial_u^2 h(u,0) + u \partial_u h(u,0) = 0
\eeq
with boundary condition $\partial_u h(0,0) = -2 \beta(p) \tvtinit /
\sqrt{2 \pi}$. Adding the physical requirement that $h(u,0) \to 0$ as $u
\to +\infty$, we find the unique solution $h(u,0) = \beta(p) \tvtinit [1
- \erf (u/\sqrt{2}) ]$. Going on with this iterative process, we find
that $\rho(c,v=\tvt-\tvtinit)$ has the following expansion at small $v$:
\bea & &
 \rho(c,v) =
  \sqrt{\frac{2}{\pi v}} \exp \left[- \frac{(c+\beta v w)^2}{2 v} \right]
   \bigg[ 1 + 2 \beta^2 \tvtinit^2 v
       \\ \nonumber & & \left. \quad
      - v (c+\beta v w)
      \left(\frac{3 \beta}{4} + \frac{1-X_0}{8} +2 \beta^3 \tvtinit^3\right)
   + o(v) \right] +
     \\ & & \quad
  \left( 1-\erf \frac{c+\beta v w}{\sqrt{2 v}} \right)
   \bigg\{ \beta \tvtinit - 2 \beta^2 \tvtinit^2 (c+\beta v w) +
       \nonumber \\  & & \quad
      v \left(\frac{3 \beta}{4}-\frac{1-X_0}{8}+2 \beta^3 
\tvtinit^3\right)
       \left[\frac{(c+\beta v w)^2}{v} +1\right]
         \nonumber \\  & & \qquad \qquad \qquad \qquad \qquad
      + o(v) \bigg\} 
\nonumber
\eea
where $w$ stands for $(2 \tvtinit+v)/2$. Hence the expression for the
conditional average of $c$:
\bea
 \label{developpement_cbarre_tvt_petit}
 \cbarre(\tvt) & = & \sqrt{\frac{2}{\pi}} \sqrt{\tvt-\tvtinit}
  - \frac{\beta(p)}{2} \tvtinit (\tvt - \tvtinit)
      \nonumber \\ & &
  + \frac{\beta(p)^2}{6} \sqrt{\frac{2}{\pi}} \tvtinit^2
    (\tvt-\tvtinit)^{3/2}
      \\ & &
  + (\tvt - \tvtinit)^2 \left[ \frac{9(1-X_0) + 10\beta(p)}{32}
     \right. \nonumber \\ & & \qquad \left.
     - \frac{2}{3\pi}(1-X_0) \right]
  + \mathcal{O}\left[ (\tvt-\tvtinit)^{5/2} \right] . \nonumber
\eea

\section{Direct numerical solution of the evolution
equation~(\ref{equation_evolution_fonctgen_c1}) for the generating
function $G_1$}
\label{appendice_solution_equation_fonctgen_C1}

 For limited sizes $N$ of the problem,
eq.~(\ref{equation_evolution_fonctgen_c1}) may be solved directly, with
exact or high precision computer arithmetic. It may be used to compute
the probability of success of the greedy algorithm from time steps 0 to
$T$, $G_1(1, T)$ or the distribution of $C_1$ from the generating
function $G_1$.

 We assume that $X_0=0$ and that $C_2$ is known (from \eg
Eq.~(\ref{solution_c2c3_UC_c2})). At $T=N$, $C_1$ is either 1 or 0 and a
useful set of values of $X_1$ in
eq.~(\ref{equation_evolution_fonctgen_c1}) is ${0,1}$ only. To compute
$G_1(X_1, N)$ with $X_1=0$ or 1, it is enough to know $G_1(X_1, N-1)$
with $X_1=0$, $1/2$ or 1 after eq.~(\ref{deff}) (we set $X_0=0$). This
in turns requires successively the values $G(X_1, T)$ for all $X_1 \in
{0, 1/[2(N-T)], 1/(N-T), \ldots, 1}$. Starting from the known initial
condition $G(X_1,0)=1$, it is possible to compute iteratively, from
$T=0$ to $N$, all values $G(k/[2(N-T)], T)$ for $0 \le k \le 2(N-T)$ (at
each step, only $2(N-T)+1$ numbers are stored). In practice, getting
accurate results requires either doing exact arithmetics (which may be
quite slow and/or memory consuming) or working with high precision
floating point arithmetic (roughly speaking, the number of decimals
digits has to be equal to $N$). This limited the range of $N$ to a few
hundreds on the computer we used (2.4GHz Pentium IV processor with 1GB
of RAM).

 Using rounded, integer values for $C_2$ rather than simply $N \times
c_2$ from eq.~(\ref{solution_c2c3_UC_c2}) makes computations much
faster. In the case where exact arithmetic is used, it is needed to
manage only rational numbers. For $N$ large enough, it brings only a
small error on $G_1$.

 If the number of 1-clauses at time $T$ is bounded by $2(N-T)$, $C_1$ is
a polynomial of degree at most $2(N-T)$. Knowing its $2(N-T)+1$ values
$G(k/[2(N-T)], T)$ is enough to compute it: $G_1$ may be expressed as a
weighted sum of Lagrange interpolating polynomials. Finally, the
coefficients of $G_1(X_1)$ indicate the probability law of $C_1$. We
used this technique to get the numerical results for the distribution of
$C_1$ for 2-SAT with $N=501$ that are plotted in
Figure~(\ref{graphe_rhobarre_2sat}). Contrary to estimates based on runs
of the very algorithms under study, it doesn't require averaging over
many series since there is no randomness in the computation.

 % Now redefine section (with 2 arguments) because it is used as 
 % section* for the bibliography :
\def\section#1#2{\oldsection#1#2}

\bibliographystyle{hunsrt}
\bibliography{ksat-compcrit-long}

\end{document}